\documentclass[letterpaper]{article}

\usepackage[T1]{fontenc}

\usepackage{geometry}
\usepackage{setspace}

\usepackage[
  backend=biber,
  style=chem-acs,
  sorting=none,
  natbib=true,
  articletitle=true
]{biblatex}
\usepackage{xpatch}

\xpatchbibdriver{inproceedings}
  {\usebibmacro{byauthor}}
  {\usebibmacro{byauthor}%
   \newunit
   \usebibmacro{title}}
  {}
  {\PackageWarning{main}{Could not patch inproceedings driver}}

\usepackage{changdefs}
\usepackage{graphicx}
\usepackage{subcaption}
\usepackage{float}
\newfloat{scheme}{htbp}{los}
\floatname{scheme}{Scheme}
\floatname{chart}{Chart}
\newfloat{graph}{htbp}{loh}
\usepackage{graphicx}
\usepackage{amsthm}
\usepackage{enumitem}
\usepackage[export]{adjustbox}
\usepackage[percent]{overpic}
\usepackage[theorems,most]{tcolorbox}
\newtcolorbox{mymath}[1][]{%
    nobeforeafter,  tcbox raise base,
    colframe=white!30!black,
    colback=olive!3,
    boxrule=0.2mm,
    boxsep=5pt,
    #1}
\usepackage{booktabs}
\usepackage{multirow}
\usepackage{amsthm}
\usepackage{amssymb}
\usepackage{xspace}
\usepackage{xcolor}
\theoremstyle{plain}
\usepackage{chemformula} 
\usepackage[version = 4]{mhchem} 
\usepackage[
  colorlinks=true,
  linkcolor=blue,
  citecolor=blue,
  urlcolor=blue
]{hyperref}

\newcommand{\ourmethod}{STEC\xspace}
\newcommand{\ourloss}{$\clL_{\rm \ourmethod}$\xspace}

\newcommand{\at}{{\rm AT}}
\newcommand{\cg}{{\rm CG}}
\newcommand{\dsm}{{\rm DSM}}

\newcommand{\dr}{{\rm d}\bfrr}

\newcommand{\kb}{k_{\rm B}}

\newcommand{\appxref}[1]{Appendix~\ref{appx:#1}}

\usepackage{authblk}

\author[1,2,3,\textdagger]{Yang Nie}
\author[1,\textdagger]{Lingyun Wang}
\author[2,3,*]{He Zhang}
\author[2,3,*]{Chang Liu}
\author[1,*]{Dan Hu}

\affil[1]{School of Mathematical Sciences, Institute of Natural Sciences, and MOE-LSC, Shanghai Jiao Tong University, Shanghai, 200240, China}

\affil[2]{Zhongguancun Academy}

\affil[3]{Zhongguancun Institute of Artificial Intelligence}

\date{}

\title{Spatiotemporal Energy Consistency Enables Reliable Coarse-Grained Molecular
Dynamics from Diffusion Models}

\begin{document}

\maketitle
\begingroup
\renewcommand{\thefootnote}{\fnsymbol{footnote}}

\footnotetext[1]{\normalfont\raggedright
E-mail: zhanghe@zgci.ac.cn (H.Z.);
liuchang@bza.edu.cn (C.L.);
hudan80@sjtu.edu.cn (D.H.).\par}

\footnotetext[2]{Yang Nie and Lingyun Wang
contributed equally to this work.}

\endgroup

\begin{abstract}

Coarse-grained molecular dynamics (CGMD) extends molecular simulations to length and time scales inaccessible to atomistic models, but building thermodynamically faithful CG potentials remains a longstanding challenge. Leveraging the connection between molecular equilibrium distributions and their underlying potentials, energy-based diffusion models (EBDMs) provide a promising route for building CG potentials through their zero-time energy, which mitigates training instability and limited scalability cost in prior approaches. However, previous studies have found that CGMD driven by diffusion-learned potentials can produce equilibrium ensembles that deviate substantially from the target distribution.
Here, we identify that this misalignment arises because the zero-time distribution error is jointly controlled by the spatial score error and temporal energy-derivative error, whereas conventional diffusion objectives directly supervise only the spatial component. 
Based on this finding, we introduce Spatiotemporal Energy Consistency (STEC) supervision, which complements spatial score matching by explicitly constraining temporal energy derivatives. 
Crucially, we derive an exact and tractable objective for this temporal component, avoiding the substantial computational overhead associated with regularization schemes in previous methods.
Across peptide and fast-folding protein systems, STEC-trained EBDMs accurately reproduce equilibrium distributions with free-energy errors within chemical accuracy, while supporting stable long-time simulations that recover key metastable transition patterns. 
These improvements further translate to strong generalization to unseen molecular systems. On held-out octapeptides, STEC reduces free-energy errors by more than 5-fold relative to the strongest competing baseline, while achieving a median production-cost reduction of approximately $1{,}700$-fold relative to atomistic molecular dynamics under a defined structural-coverage benchmark. These results establish STEC as a scalable framework for learning thermodynamically reliable CG force fields for increasingly complex biomolecular systems.

\end{abstract}


\section{Introduction}

\begin{figure*}[!t]
    \centering
     \includegraphics[width=\textwidth]{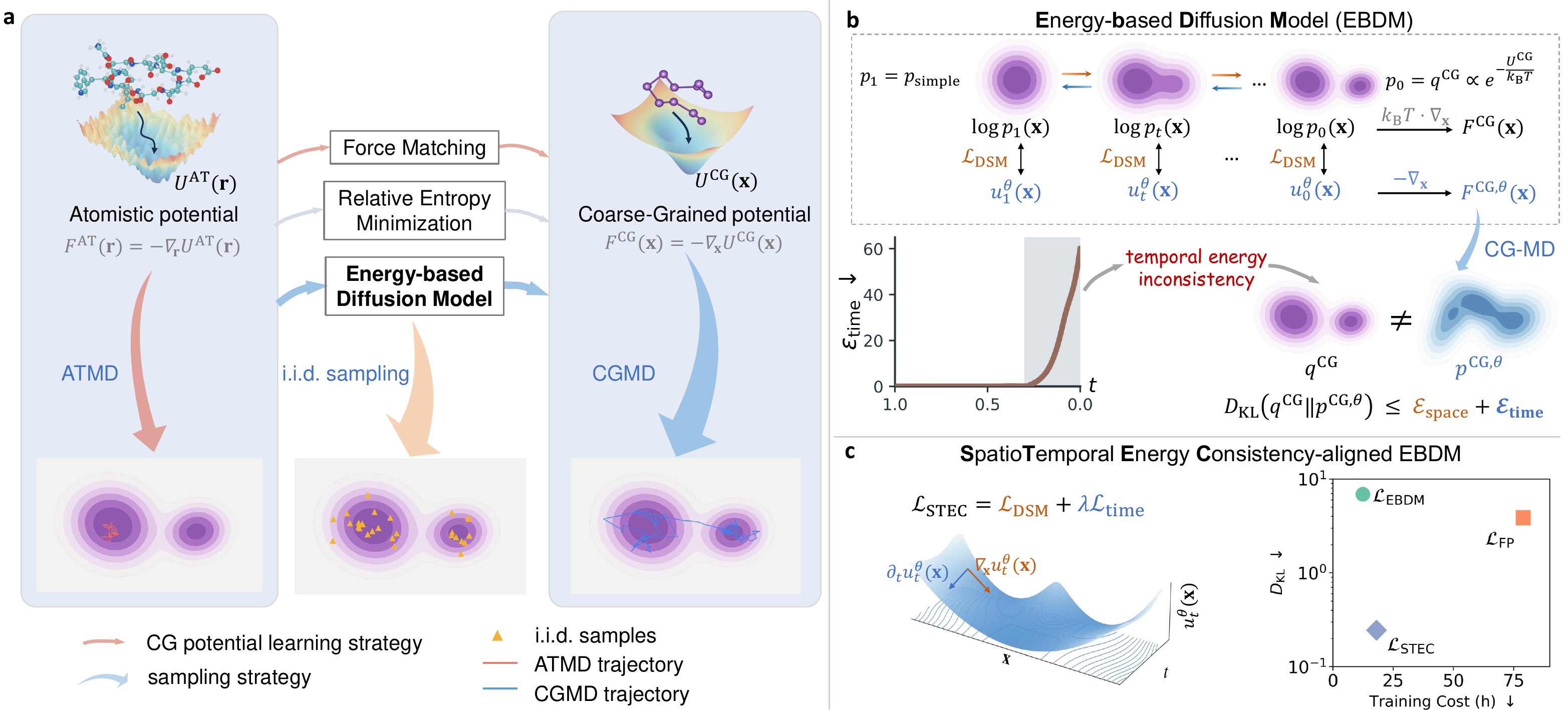}
    \caption{
    \textbf{Improving diffusion-based coarse-grained potential with spatiotemporal energy consistency.}
    \textbf{(a)} CGMD extends molecular simulations to longer length and time scales but requires thermodynamically accurate CG potentials. Force matching learns projected mean forces, relative entropy minimization matches equilibrium distributions through iterative CG sampling, and EBDMs learn distributions without model-driven sampling while also enabling i.i.d.\ sampling. Arrow widths qualitatively indicate the efficiency of the learning and sampling routes.
    \textbf{(b)} Standard EBDM training. A time-dependent energy $u^\theta_{t}(\bfxx)$ represents the diffusion path between a simple reference distribution and the target equilibrium distribution. Denoising score matching ($\clL_\dsm$) constrains spatial scores across diffusion times but not temporal energy evolution, potentially yielding a mismatch between the target distribution $q^\cg$ and the equilibrium distribution $p^{\cg,\theta}$ induced by the zero-time energy. Our theoretical analysis establishes that this zero-time distribution error is bounded by the sum of the spatial score error $\clE_{\rm space}$ and the temporal energy-derivative error $\clE_{\rm time}$. 
    \textbf{(c)} \ourmethod-aligned EBDM. \ourloss complements spatial score matching with explicit supervision of temporal energy derivatives. On the 35-residue Villin protein, \ourloss reduces distribution errors by approximately 10-fold relative to competing objectives, with modest training overhead.
    }
    \label{fig:overview}
\end{figure*}

Molecular dynamics (MD) simulations have become a central tool for connecting molecular structure to function, providing mechanistic insight into biomolecular processes and conformational pathways while increasingly informing drug discovery and materials design~\citep{dror2012biomolecular,devivo2016role,zong2020hydrogen,snow2002trp,Fu2022ForcesAN,hollingsworth2018molecular}. However, many biologically relevant processes remain beyond the accessible length and time scales of atomistic molecular dynamics (ATMD), owing to its large number of microscopic degrees of freedom (DoFs) and the small integration time steps required to resolve fast atomic motions~\citep{shaw2010atomic,kmiecik2016coarse,plattner2017complete,dominic2023memory}. Coarse-grained molecular dynamics (CGMD) alleviates these limitations by integrating out local DoFs, allowing larger integration time steps and more efficient exploration of conformational space~\citep{marrink2003vesicles,saunders2013coarse,zhang2016nucleosomes,souza2021martini,cullen2026integrating} (\figref{overview}a). Its reliability, however, critically depends on a CG potential that faithfully reproduces the equilibrium thermodynamics of the underlying atomistic ensemble~\citep{clementi2008coarse,noid2013perspective,noid2023perspective,noid2024rigorous}.

Given a molecular system, let $\Xi:\bbR^{3N}\rightarrow\bbR^{3n}$ denote a coarse-graining map from atomistic coordinates $\bfrr$ to CG variables $\bfxx$. The atomistic equilibrium ensemble then induces the CG distribution
$q^\cg(\bfxx)=\int q^\at(\bfrr)\delta(\bfxx-\Xi(\bfrr))\dr$,
where $q^\at(\bfrr)$ is the atomistic equilibrium distribution. The corresponding thermodynamically consistent potential is the potential of mean force (PMF), $U^\cg(\bfxx)$, defined up to an additive constant by $q^\cg(\bfxx)\propto\exp\left[-\frac{U^\cg(\bfxx)}{\kb T}\right]$, where $\kb$ is the Boltzmann constant and $T$ is the absolute temperature. Bottom-up~\citep{jin2022bottom,john2017manybody} coarse-graining seeks to recover this effective potential from atomistic-level potential, which is a challenging computational task as it requires high-dimensional marginalization over the eliminated DoFs. This has been primarily tackled through variational force matching (FM) and relative entropy minimization (REM). 
FM exploits the fact that $-\nabla_{\bfxx}U^\cg$ corresponds to the conditional mean of appropriately projected atomistic forces, allowing $U^\cg$ to be learned through supervised force regression~\citep{izvekov2005multiscale,izvekov2005multiscale,wang2019machine,zhang2018deepcg,husic2020coarse,durumeric2026learning,charron2025navigating,chaimovich2010relative}.
However, instantaneous projected forces can exhibit substantial variance around this conditional mean, leading to noisy supervision and potentially large errors in sparsely sampled transition regions~\citep{kramer2023statistically,park2026meanforce}. REM avoids this force-noise bottleneck by directly matching the target distribution $q^\cg$ with the Boltzmann distribution induced by the learned potential~\citep{shell2008relative,noid2008multiscale,durumeric2019adversarial,thaler2022deep}. Although this distribution-level objective can improve thermodynamic fidelity, its optimization requires expectations over the evolving model distribution and therefore repeated, computationally expensive CG sampling throughout training. 
Existing approaches thus face a fundamental trade-off between high-variance force supervision and expensive model-driven sampling (\appxref{related} provides more discussion).

Energy-based diffusion models (EBDMs) offer a promising route around this trade-off by retaining distribution-level learning without requiring model-driven sampling during training. 
Inspired by nonequilibrium thermodynamic annealing~\citep{sohl2015deep}, EBDMs construct a forward diffusion path $\{p_{t}\}_{t\in[0,1]}$ that progressively perturbs the target CG equilibrium distribution $q^\cg(\bfxx)$ toward a simple reference distribution $p_{\rm simple}$, recasting direct distribution learning as a sequence of denoising problems (\figref{overview}b, upper). 
The corresponding reverse diffusion process uses the spatial score $\nabla_{\bfxx}\log p_{t}(\bfxx)$ to guide samples from $p_{\rm simple}$ back toward the target distribution.
EBDMs represent this score as the negative spatial gradient of a time-dependent, dimensionless energy $u^\theta_{t}(\bfxx)$, yielding a conservative score field by construction. The model is trained through denoising score matching
(DSM)~\citep{song2019generative}
($\clL_\dsm$ in \figref{overview}b)
using CG configurations mapped from ATMD ensembles.
Crucially for CGMD, the energy function approaches the CG potential as the diffusion time approaches zero, $U^{\cg,\theta}(\bfxx)=\kb T u^\theta_{0}(\bfxx)$, whose negative gradient defines the conservative CG force field for molecular simulation.
However, directly using a zero-time energy function learned from conventional diffusion training has been observed inaccurate to produce a thermodynamic ensemble through CGMD~\citep{plainer2025consistent}. 
\citet{arts2023} partially mitigate this issue by extracting the force field at a finite diffusion time, while sacrificing fidelity to the clean zero-time potential associated with the target distribution (\figref{si_two_for_one}). Inspired by FP-Diffusion~\citep{lai2023}, which regularizes DSM by enforcing the score Fokker--Planck (FP) equation to promote consistency of the learned score fields with the underlying diffusion process, ScoreMD~\citep{plainer2025consistent} adapts FP-based regularization to EBDMs for CG potentials. Although this approach improves consistency near the clean-data limit, the full FP constraint involves higher-order spatial derivatives of the energy model and remains computationally expensive even with numerical approximations (\appxref{related-diffusion-consis}), limiting its scalability to larger molecular systems.

Here, we identify that the gap between successful diffusion training and thermodynamically faithful zero-time CG potentials originates from a fundamental difference in how molecular distributions are generated:
in conventional diffusion models, the target distribution is produced as the endpoint of the reverse process~\cite{song2021,chen2023sampling,lee2023convergence}, while for chemistry tasks, the Boltzmann distribution $p^{\cg,\theta}(\bfxx)\propto\exp[-u^\theta_{0}(\bfxx)]$ is produced by running MD using the learned zero-time potential~\citep{arts2023,plainer2025consistent}. Our analysis makes it more explicit: the distributional error in reverse-diffusion sampling can be bounded by the spatial score error $\clE_{\rm space}$~\cite{song2021maximum}, while the zero-time distributional error between $q^\cg$ and $p^{\cg,\theta}$ is bounded by the combination of $\clE_{\rm space}$ and the temporal energy-derivative error $\clE_{\rm time}$ (\figref{overview}b, bottom left).
Conventional DSM directly supervises spatial scores across diffusion times but does not explicitly constrain temporal energy evolution.
Consequently, strong denoising performance can coexist with errors in temporal density evolution and a distorted zero-time Boltzmann ensemble for CGMD (\figref{overview}b, bottom right).

Building on this theoretical insight, we introduce the \textbf{Spatiotemporal Energy Consistency (\ourmethod)} loss, which directly targets the spatial and temporal error components governing the zero-time distribution error (\figref{overview}c). To control the temporal component efficiently, we derive an analytical conditional target for the temporal energy derivative $\partial_tu^\theta_{t}(\bfxx)$, enabling efficient temporal matching without the higher-order spatial derivatives required by FP regularization.
Accordingly, this formulation substantially reduces the cost of enforcing temporal consistency and enables EBDM-based CG potentials to scale to larger molecules.

We evaluate \ourmethod across molecular systems of increasing complexity, examining thermodynamic fidelity, metastable dynamics, long-time simulation stability, transferability, and computational efficiency. Across four fast-folding proteins, \ourmethod achieves chemical accuracy in macrostate free energies. On the challenging Villin system, it recovers metastable basins missed by competing EBDMs, reduces distributional error by over 10-fold, and extends stable CGMD simulation times by two to three orders of magnitude relative to ScoreMD. Beyond equilibrium thermodynamics, STEC-driven CGMD more accurately reproduces metastable-state persistence and conformational interconversion, with transition probabilities closely matching atomistic references across all proteins. These capabilities extend to unseen peptide systems, where \ourmethod recovers key low-energy conformational basins and achieves more than 5-fold lower macrostate free-energy errors on held-out octapeptides. It also extends stable simulation times by three and two orders of magnitude relative to standard EBDM and ScoreMD, respectively. 
\ourmethod further enables rapid learning of accurate CG potentials with modest training overhead. More attractively, STEC-driven CGMD achieves substantial coverage of low-energy conformational regions at a median production cost approximately 1700-fold lower than that of generating the corresponding atomistic reference ensembles. 
Together, these results establish \ourmethod as an efficient approach to thermodynamically accurate, stable, and transferable CGMD.

\section{Methodology}
\label{sec:method}

\subsection{Diffusion-based Coarse-grained Potentials}
\label{sec:method-diffusion}

The goal of bottom-up coarse-graining is to learn a CG potential $U^{\cg,\theta}(\bfxx)$ whose Boltzmann distribution $p^{\cg,\theta}(\bfxx)
\propto\exp[\frac{-U^{\cg,\theta}(\bfxx)}{k_{\rm B}T}]$ reproduces the mapped equilibrium ensemble $q^\cg(\bfxx)$~\citep{jin2022bottom}.
In principle, the CG equilibrium distribution is obtained by marginalizing the atomistic distribution $q^\at(\bfrr)$ over unresolved DoFs. Nevertheless, evaluating this marginal density faces an intractable high-dimensional integration challenge. A simple and straightforward alternative is to marginalize at the sample level: CG equilibrium samples can be obtained by discarding unresolved DoFs from atomistic equilibrium configurations. 
EBDMs exploit this sample-level accessibility to efficiently learn CG potentials through a time-dependent, dimensionless energy $u_t^\theta(\bfxx)$, whose zero-time value defines the CG potential $U^{\cg,\theta}(\bfxx)=k_{\rm B}T\,u^\theta_0(\bfxx)$ for molecular simulation. Specifically, EBDMs construct a prescribed diffusion path $\{p_t\}_{t\in[0,1]}$ that progressively perturbs equilibrium configurations $\bfxx_0\sim q^\cg$, connecting the target distribution $p_0=q^\cg$ to a simple reference distribution $p_{\rm simple}$ (\figref{overview}b, upper panel).
Here, $t$ denotes diffusion time, which controls the perturbation level rather than the physical time of molecular dynamics. 
The forward process follows the prevailing stochastic differential equation
(SDE) formulation~\citep{oksendal2003,song2021,ho2020denoising},
$\mathrm d\bfxx_t
=-\frac{1}{2}\beta(t)\bfxx_t\,\mathrm dt
+\sqrt{\beta(t)}\,\mathrm d\bfww_t$,
where $\beta(t)>0$ is the noise schedule,
and $\bfww_t$ is a standard Wiener process.
The generation process follows the reverse SDE formulation $\mathrm d\bfxx_t=\left[-\frac{1}{2}\beta(t)\bfxx_t-\beta(t)\nabla_{\bfxx_t}\log p_{t}(\bfxx_t)\right]\mathrm dt+\sqrt{\beta(t)}\,\mathrm d\bar{\bfww}_t$ to transform samples from $p_{\rm simple}$ toward $q^\cg$, with $\bar{\bfww}_t$ as a reverse-time Wiener process~\citep{anderson1982reverse,song2021}. EBDMs approximate the spatial score $\nabla_{\bfxx_t}\log p_{t}(\bfxx_t)$ using the conservative field $\bfss^\theta_{t}(\bfxx)=-\nabla_{\bfxx}u^\theta_{t}(\bfxx)$. 
As $p_{\rm simple}$ is chosen to enable independent sampling, reverse diffusion can generate independent and identically distributed (i.i.d.) configurations approximating the target equilibrium distribution. Further theoretical details are provided in \appxref{method-diffusion}. 

\subsection{Spatiotemporal Characterization of Zero-Time Distribution Error}
\label{sec:method-stec}

Since the spatial score $\nabla_{\bfxx}\log p_{t}(\bfxx)$ is generally unavailable in closed form, EBDMs learn it by minimizing the DSM objective, 
\begin{gather}
\clL_\dsm(\theta)
=
\bbE_{t,\,\bfxx_0\sim p_0,\,\bfxx_t\sim p_t(\cdot\mid\bfxx_0)}
\left[
w(t)
\left\|
-\nabla_{\bfxx_t}u^\theta_t(\bfxx_t)
-\nabla_{\bfxx_t}\log p_t(\bfxx_t\mid\bfxx_0)
\right\|_2^2
\right],
\label{eq:methods_dsm}\\[2pt]
\text{where}\quad p_t(\bfxx_t\mid\bfxx_0)
=
\clN\left(\bfxx_t;\alpha(t)\bfxx_0,\sigma^2(t)\mathbf I\right),
\label{eq:forward-process}
\end{gather}
with $\alpha(t)=\exp[-\frac{1}{2}\int_0^t\beta(s)\,\mathrm ds]$ and $\sigma^2(t)=1-\exp[-\int_0^t\beta(s)\,\mathrm ds]$. Here, $t$ is uniformly sampled over $[0,1]$, and $w(t)>0$ weights different diffusion times. The Gaussian conditional kernel in \eqnref{forward-process} enables direct sampling of $\bfxx_t$ and provides an analytically tractable conditional score for DSM. With sufficient model capacity and exact optimization, the optimal solution $\theta^*$ satisfies $-\nabla_{\bfxx}u^{\theta^*}_t(\bfxx)=\nabla_{\bfxx}\log p_t(\bfxx)$ for almost every $\bfxx$ and $t$~\citep{vincent2011,song2021}. Therefore, minimizing DSM to its optimum guarantees that the endpoint distribution $p^{\mathrm{SDE},\theta}_0$ generated by reverse diffusion sampling matches the target equilibrium distribution $q^\cg$.

More explicitly, \citet[Theorem~1]{song2021maximum} establish a bound on the Kullback--Leibler (KL) divergence $D_{\rm KL}$ between $q^\cg$ and $p^{{\rm SDE},\theta}_0$. Under likelihood weighting $w(t)=\beta(t)$ and a negligible terminal distribution mismatch, this bound reduces approximately to
\begin{equation}
D_{\rm KL}\!\left(
q^\cg
\,\|\,
p^{\mathrm{SDE},\theta}_0
\right)
\leq
\frac{1}{2}
\left[
\clL_\dsm(\theta)
-
\clL_\dsm(\theta^*)
\right].
\label{eq:methods_reverse_sde_bound}
\end{equation}

In biophysical applications, however, the goal extends beyond generating equilibrium configurations to characterizing thermodynamic properties and conformational dynamics. These properties can be investigated through MD simulations driven by the learned zero-time potential. 
This reflects a fundamental difference in how molecular distributions are generated: reverse diffusion sampling produces the endpoint distribution $p^{\mathrm{SDE},\theta}_0$, whereas CGMD samples the Boltzmann distribution $p^{\cg,\theta}(\bfxx)\propto\exp[-u^\theta_0(\bfxx)]$ induced by the learned potential. Indeed, previous studies have reported substantial discrepancies between $p^{\cg,\theta}$ and $q^\cg$, even when reverse-diffusion sampling accurately reproduces the target ensemble~\citep{plainer2025consistent} (\figref{overview}b, bottom right).

To decipher this discrepancy, we examine how errors along the diffusion path affect the equilibrium distribution induced by the learned potential. A theoretical bound is then established to reveal that the distributional error between $q^\cg$ and $p^{\cg,\theta}$ is jointly controlled by spatial score and temporal energy-derivative errors, with the latter not explicitly constrained by conventional diffusion training.

\begin{theorem}[Spatiotemporal distribution error bound]
\label{thm:distribution_error}
Under the diffusion setting described in \eqnref{forward-process}, assume suitable regularity conditions and terminal matching $p^\theta_1=p_1$. Then the zero-time distribution error satisfies
\begin{equation}
D_{\rm KL}\!\left(
q^\cg\,\|\,p^{\cg,\theta}
\right)
\leq
\clE_{\rm space}+\clE_{\rm time},
\label{eq:methods_endpoint_error_bound}
\end{equation}
where the spatial score error and temporal energy-derivative error are defined as
\begin{align}
\clE_{\rm space}
&=
K\left[
\int_0^1
\bbE_{\bfxx\sim p_t}
\left\|
\nabla_{\bfxx}\log p^\theta_t(\bfxx)
-
\nabla_{\bfxx}\log p_t(\bfxx)
\right\|_2^2
\,\mathrm dt
\right]^{1/2},
\nonumber\\
\clE_{\rm time}
&=
K\left[
\int_0^1
\bbE_{\bfxx\sim p_t}
\left|
\partial_t\log p^\theta_t(\bfxx)
-
\partial_t\log p_t(\bfxx)
\right|^2
\,\mathrm dt
\right]^{1/2}.
\label{eq:methods_spatiotemporal_errors}
\end{align}
Here, $p^\theta_t(\bfxx)\propto\exp[-u^\theta_t(\bfxx)]$ denotes the normalized distribution induced by the learned time-dependent energy, and $K>0$ is a finite constant. The required regularity assumptions are provided in \appxref{method-diffusion}.
\end{theorem}

Therefore, minimizing both $\clE_{\rm space}$ and $\clE_{\rm time}$ to zero guarantees that $p^{\cg,\theta}$ matches the target distribution $q^\cg$, a principle which we term \textbf{spatiotemporal energy consistency}. Under uniform diffusion-time weighting $w(t)=1$, the spatial component of this bound is directly related to DSM,
\begin{equation}
\clL_\dsm(\theta)-\clL_\dsm(\theta^*)
=
K^{-2}\clE_{\rm space}^2.
\label{eq:methods_dsm_spatial_error}
\end{equation}
Minimizing DSM alone therefore does not directly control the full distribution error bound, motivating the complementary temporal matching objective developed in the following subsection. The complete proof is provided in \appxref{si_endpoint_bound}.

\subsection{Tractable Spatiotemporal Energy Matching}
\label{sec:method-sem}

To complement spatial score matching, we seek to reduce $\clE_{\rm time}$ by matching $\partial_tu^\theta_t(\bfxx)$ to $-\partial_t\log p_t(\bfxx)$. This leads to the ideal marginal temporal matching objective $\clL_{\rm time}^{\rm marg}(\theta)=\bbE_{t,\bfxx_t}\left[w_{\rm time}(t)\left(\partial_tu^\theta_t(\bfxx_t)+\partial_t\log p_t(\bfxx_t)\right)^2\right]$, where $w_{\rm time}(t)>0$ weights different diffusion times. 
However, the marginal temporal derivative $\partial_t\log p_t$ is generally intractable. Fortunately, as in DSM, the Gaussian conditional kernel in \eqnref{forward-process} provides an analytically tractable conditional temporal derivative. Replacing this conditional derivative for the intractable marginal derivative yields the tractable temporal matching objective
\begin{equation}
\clL_{\rm time}(\theta)
=
\bbE_{t,\,\bfxx_0\sim p_0,\,\bfxx_t\sim p_t(\cdot\mid\bfxx_0)}
\left[
w_{\rm time}(t)
\left(
\partial_tu^\theta_{t}(\bfxx_t)
+
\partial_t\log p_{t}(\bfxx_t\mid\bfxx_0)
\right)^2
\right].
\label{eq:methods_conditional_time_loss}
\end{equation}

Crucially, the conditional temporal derivative recovers the marginal derivative in expectation,
$\bbE[\partial_t\log p_t(\bfxx_t\allowbreak\mid\bfxx_0)
\mid\bfxx_t,t]
=\allowbreak\partial_t\log p_t(\bfxx_t)$.
Consequently, the conditional and marginal objectives differ only by a term independent of $\theta$ and therefore share the same minimizers. This equivalence is formalized in the following proposition.
\begin{proposition}[Conditional temporal matching]
\label{prop:methods_conditional_time}
Under the diffusion process described in~\eqnref{forward-process}, the marginal and conditional temporal objectives have identical parameter gradients,
$\nabla_\theta\clL_{\rm time}^{\rm marg}(\theta) = \nabla_\theta\clL_{\rm time}(\theta)$.
\end{proposition}

This result allows us to optimize the ideal marginal objective using the tractable conditional target. See the complete proof in~\appxref{method-condition}.
Combining conditional temporal matching with DSM defines the \ourmethod loss (\figref{overview}c),
\begin{equation}
\clL_{\mathrm{STEC}}(\theta)
=
\clL_\dsm(\theta)
+
\lambda_{\rm time}
\clL_{\rm time}(\theta),
\label{eq:methods_stec_loss}
\end{equation}
where $\lambda_{\rm time}$ controls the relative strength of temporal supervision. Both terms use the same perturbed configurations to supervise the spatial and temporal derivatives of a shared scalar energy.

A previous approach enforces the FP equation, $\partial_t p_t(\bfxx)=\frac{\beta(t)}{2}\left[\nabla_{\bfxx}\cdot\left(\bfxx p_t(\bfxx)\right)+\Delta_{\bfxx}p_t(\bfxx)\right]$, through residual regularization of the learned model as adopted in ScoreMD~\citep{plainer2025consistent}. 
However, FP-residual regularization requires higher-order spatial derivatives of the learned energy, particularly its Laplacian, resulting in substantial computational and memory overhead even with the finite-difference approximations used in ScoreMD.
In contrast, STEC directly supervises temporal energy evolution using the analytical conditional target, avoiding these costly spatial derivatives. Our benchmarks demonstrate that STEC substantially reduces training costs relative to FP-based regularization (\figref{overview}c), while improving the thermodynamic fidelity and simulation stability of CGMD across molecular systems (Figs.~\ref{fig:res-iid}--\ref{fig:peptide_generalization}). Further discussion is provided in \appxref{appx_discuss}.

\subsection{Practical Implementation}
\label{sec:method-imple}

Following ScoreMD~\citep{plainer2025consistent}, we employ a graph-transformer architecture to parameterize the molecular energy $u^\theta_{t}(\bfxx)$, adopting the same diffusion process and base training setup with the additional temporal matching objective. For computational efficiency, we approximate the temporal energy derivative using finite differences. Details of the model architecture, diffusion schedules, training procedures, and numerical implementation are provided in \appxref{si_model_training}.
Unless otherwise specified, all molecular ensembles analyzed in the main text are generated by Langevin CGMD using the learned CG force fields. Simulation parameters, initialization procedures, and system configurations are detailed in \appxref{si_sampling}.

\section{Results and Discussions}
\label{sec:results}

\subsection{Thermodynamic Fidelity Across Molecular Systems}
\label{sec:results_equilibrium}

\begin{figure}[!t]
    \centering
     \includegraphics[width=\textwidth]{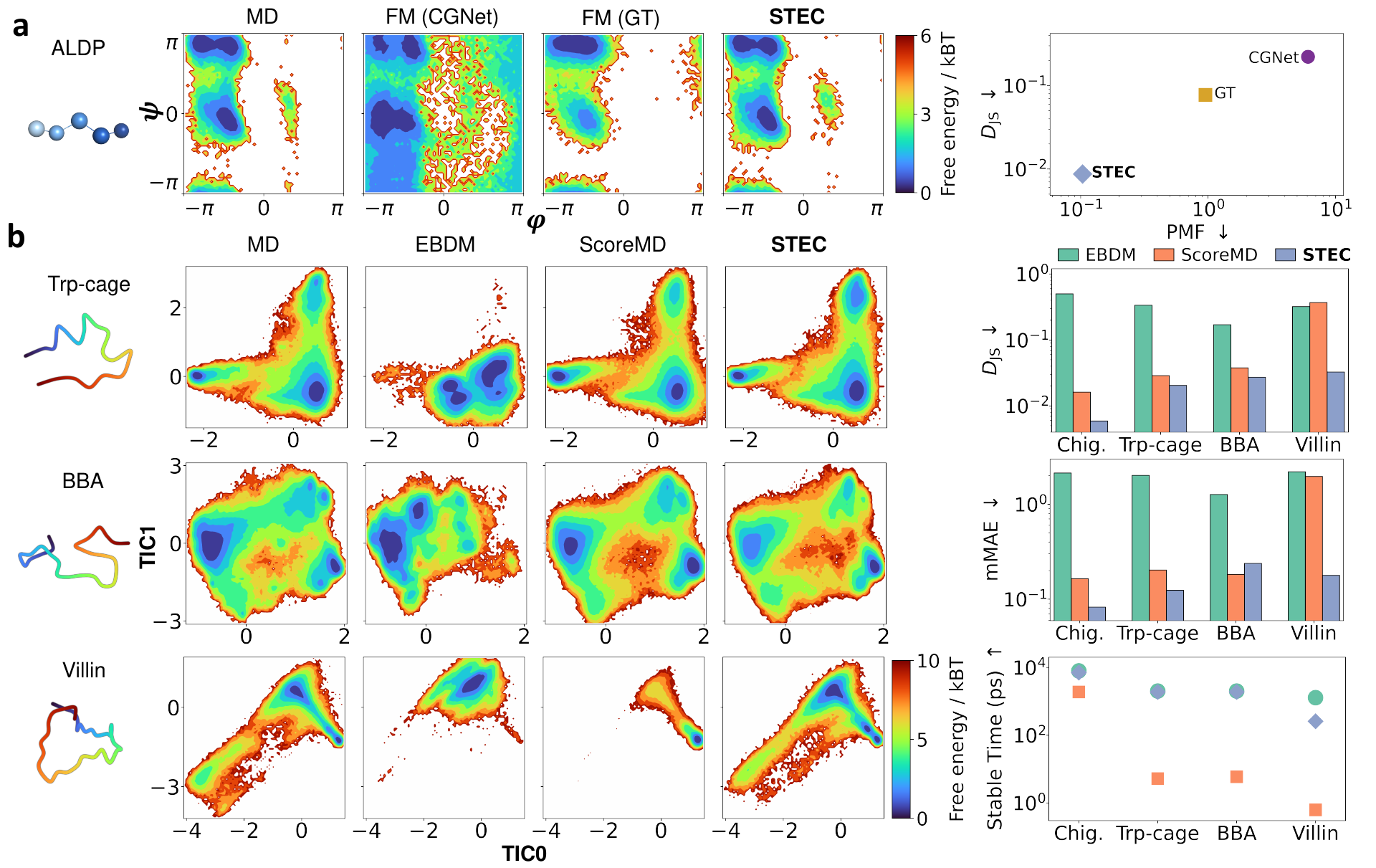}
    \caption{ 
    \textbf{STEC enables thermodynamically faithful and stable CGMD on various molecular systems.}
    All model-induced ensembles are generated by CGMD using the corresponding CG potentials.
    \textbf{(a)} Alanine dipeptide (ALDP). Ramachandran free-energy landscapes projected onto dihedral angles $(\phi,\psi)$ for the mapped atomistic MD reference, FM with CGNet [FM (CGNet)], FM with an architecture-matched graph transformer [FM (GT)], and STEC. The right panel compares PMF error and JS divergence ($D_{\rm JS}$) relative to the reference.
    \textbf{(b)} Fast-folding proteins. Free-energy landscapes projected onto the first two TICs for Trp-cage, BBA, and Villin, comparing the MD reference, standard EBDM, ScoreMD, and STEC. The right panels summarize $D_{\rm JS}$ (top), macrostate free-energy mean absolute error (mMAE; middle), and stable simulation time (bottom) across Chignolin and the three displayed proteins. Stable simulation time is measured until any bonded distance deviates from its equilibrium value by more than $0.5$~\AA. 
    }
    \label{fig:res-iid}
\end{figure}

We first examine whether STEC learns CG potentials that faithfully reproduce atomistic equilibrium ensembles in CGMD (\figref{res-iid}). We begin with alanine dipeptide (ALDP), a widely used benchmark for CG force-field learning~\citep{kohler2021smooth}, using 50,000 atomistic configurations mapped onto five backbone sites. We compare STEC against force matching (FM) with either CGNet~\citep{wang2019machine} or an architecture-matched graph transformer (GT).
As shown in \figref{res-iid}a, STEC closely reproduces the reference Ramachandran free-energy landscape, including the weakly populated metastable basin and the barrier regions separating dominant conformations. In contrast, both FM models exhibit substantial deviations, particularly in low-probability regions. Quantitatively, STEC reduces both the potential of mean force (PMF) error and Jensen--Shannon (JS) divergence $D_{\rm JS}$ by more than 10-fold relative to the best FM baseline. The improvement over FM (GT), which shares the same architecture, further highlights the benefit of distribution-based learning beyond architectural differences. Further details are provided in~\appxref{imple}.

We next evaluate whether this advantage extends to the more complex conformational landscapes of four fast-folding proteins: Chignolin, Trp-cage, BBA, and Villin, spanning 10--35 $\rmC_{\alpha}$ beads. Using long-timescale atomistic trajectories as references~\citep{deshaw2011fastfolding}, we compare STEC against a standard EBDM trained with DSM and the FP-regularized ScoreMD~\citep{plainer2025consistent}. The free-energy landscapes projected onto the first two time-lagged independent components (TICs) reveal that STEC closely recovers the reference conformational distributions across these proteins (\figref{res-iid}b). The distinction is particularly striking for Villin, the largest system considered. Both competing methods fail to adequately sample substantial portions of its reference conformational landscape, whereas STEC recovers the major metastable regions, including an extended basin largely absent from the competing models. This agreement extends beyond the projected energy landscapes, with STEC more accurately reproducing nonlocal contact patterns and intermediate-range structural geometry, as evidenced by contact maps (\figref{si_contact_maps}) and sequence-separated distance distributions (Table~\ref{tab:local_geometry_kl}).

The quantitative comparisons (\figref{res-iid}b, right) further substantiate these observations. STEC achieves chemical accuracy in macrostate free energies (mMAE below $1~\mathrm{kcal/mol}$) and the lowest $D_{\rm JS}$ across all four proteins, with both errors reduced by more than an order of magnitude relative to ScoreMD for Villin (Table~\ref{tab:equilibrium_quality}).
Importantly, STEC maintains stable CGMD trajectories comparable to those of standard EBDM, while extending stable simulation times by two to three orders of magnitude over ScoreMD for Trp-cage, BBA, and Villin. In contrast, standard EBDM can sustain long trajectories despite substantial thermodynamic errors, whereas ScoreMD can improve distributional accuracy but suffers from premature simulation instability. These findings demonstrate that STEC combines accurate equilibrium sampling with the long-time numerical robustness essential for reliable CGMD. 
Additional comparisons with reverse-diffusion sampling and finite-time EBDM force evaluation further support the distributional and structural fidelity of STEC (\figref{si_iid_cgmd}; \figref{si_two_for_one}).

\begin{figure}[!t]
    \centering
     \includegraphics[width=\textwidth]{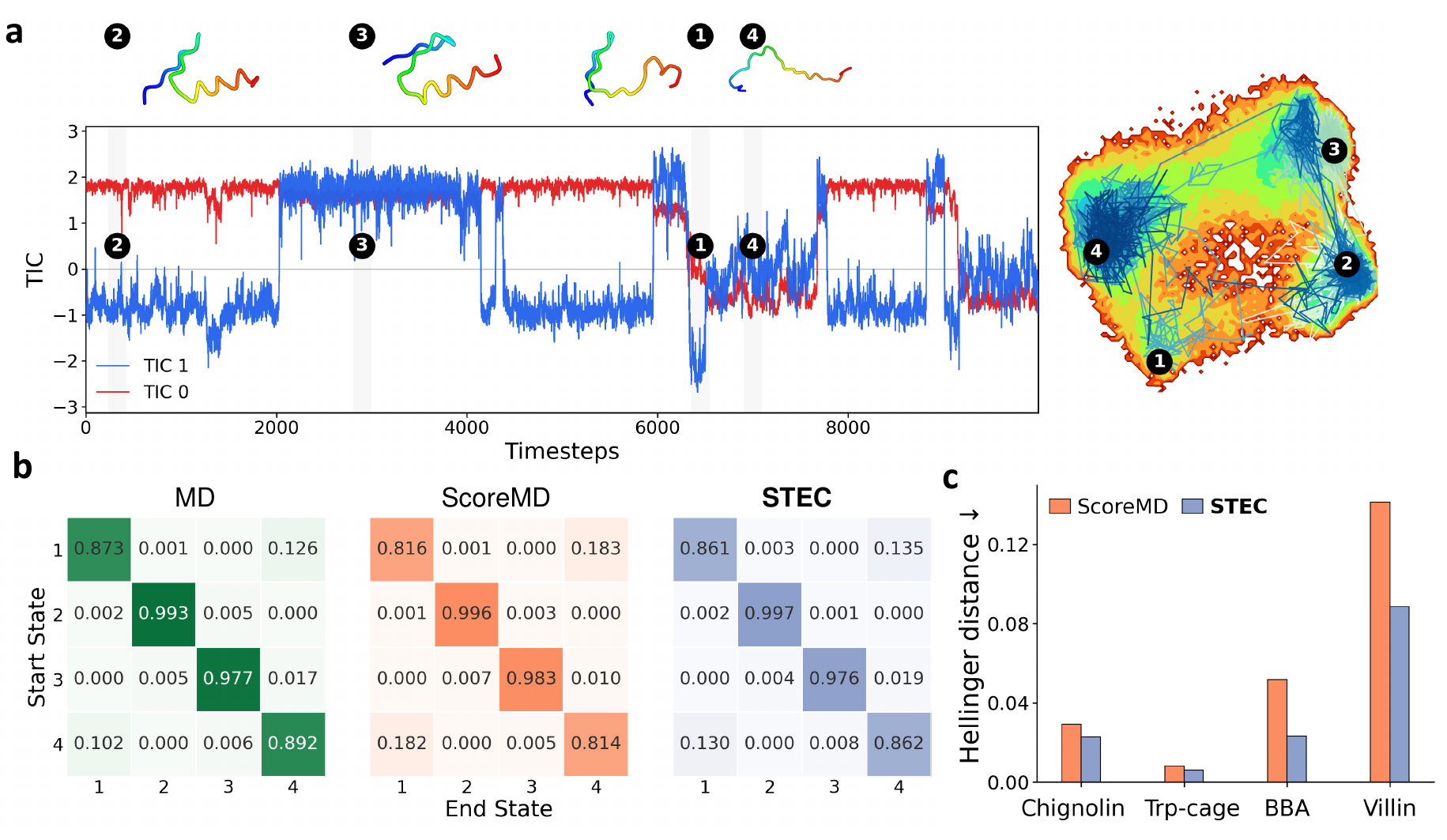}
    \caption{
    \textbf{STEC improves the recovery of metastable transition patterns in CGMD.}
    \textbf{(a)} Representative BBA trajectory generated by STEC-driven CGMD. The first two TICs are shown along the trajectory (left), with selected molecular conformations labeled 1--4 above. The same trajectory is projected onto the atomistic MD reference free-energy landscape (right), illustrating transitions among distinct metastable basins.
    \textbf{(b)} Transition probability matrices for BBA obtained from the atomistic MD reference and CGMD trajectories generated by ScoreMD and STEC. All matrices use the same four-state partition, with rows and columns representing initial and final states, respectively.
    \textbf{(c)} Hellinger distances between the CGMD and atomistic reference transition probability matrices for Chignolin, Trp-cage, BBA, and Villin. 
    }
    \label{fig:trajectory_transition}
\end{figure}

\subsection{Metastable Dynamics in CGMD}
\label{sec:results_dynamics}

Beyond thermodynamic fidelity, a reliable CG potential should accurately capture metastable-state transitions observed in atomistic MD. We assess this capability using CGMD trajectories and their state-to-state transition probabilities. For BBA, a representative STEC-driven trajectory exhibits repeated interconversions between extended and folded conformations, traversing the major basins of the reference free-energy landscape (\figref{trajectory_transition}a). 

To quantify these transitions, we construct the lag-time-dependent transition probability matrix~\citep{jones2021determining} $T_{ij}(\tau)=P(s_{\ell+\tau}=j\mid s_\ell=i)$, where $s_\ell$ denotes the metastable state occupied at physical simulation time $\ell$, $i$ and $j$ index the initial and final states, respectively, and $\tau$ is the lag time. For each protein, the same state partition is used for
the atomistic and CGMD trajectories, with transitions
counted between adjacent frames of the processed trajectories (\appxref{si_evaluation}). As shown in \figref{trajectory_transition}b, STEC more accurately reproduces the reference transition probabilities than ScoreMD, particularly the dominant exchange between states \texttt{1} and \texttt{4} and their associated state persistence. For example, the transition probability from state \texttt{1} to state \texttt{4} is 0.135 for STEC, closely matching the atomistic reference of 0.126, whereas ScoreMD overestimates this transition at 0.183. Similar improvements are observed for the reverse transition and diagonal persistence probabilities. 

This agreement extends across all four fast-folding proteins, as further supported by the transition matrices in \figref{si_transition_matrices}. Quantitatively, STEC yields consistently lower Hellinger distances than ScoreMD, with particularly pronounced improvements for BBA and Villin (\figref{trajectory_transition}c). These results demonstrate that STEC more accurately captures metastable-state persistence and conformational interconversion across diverse protein systems.

\begin{figure}[!t]
    \centering
     \includegraphics[width=\textwidth]{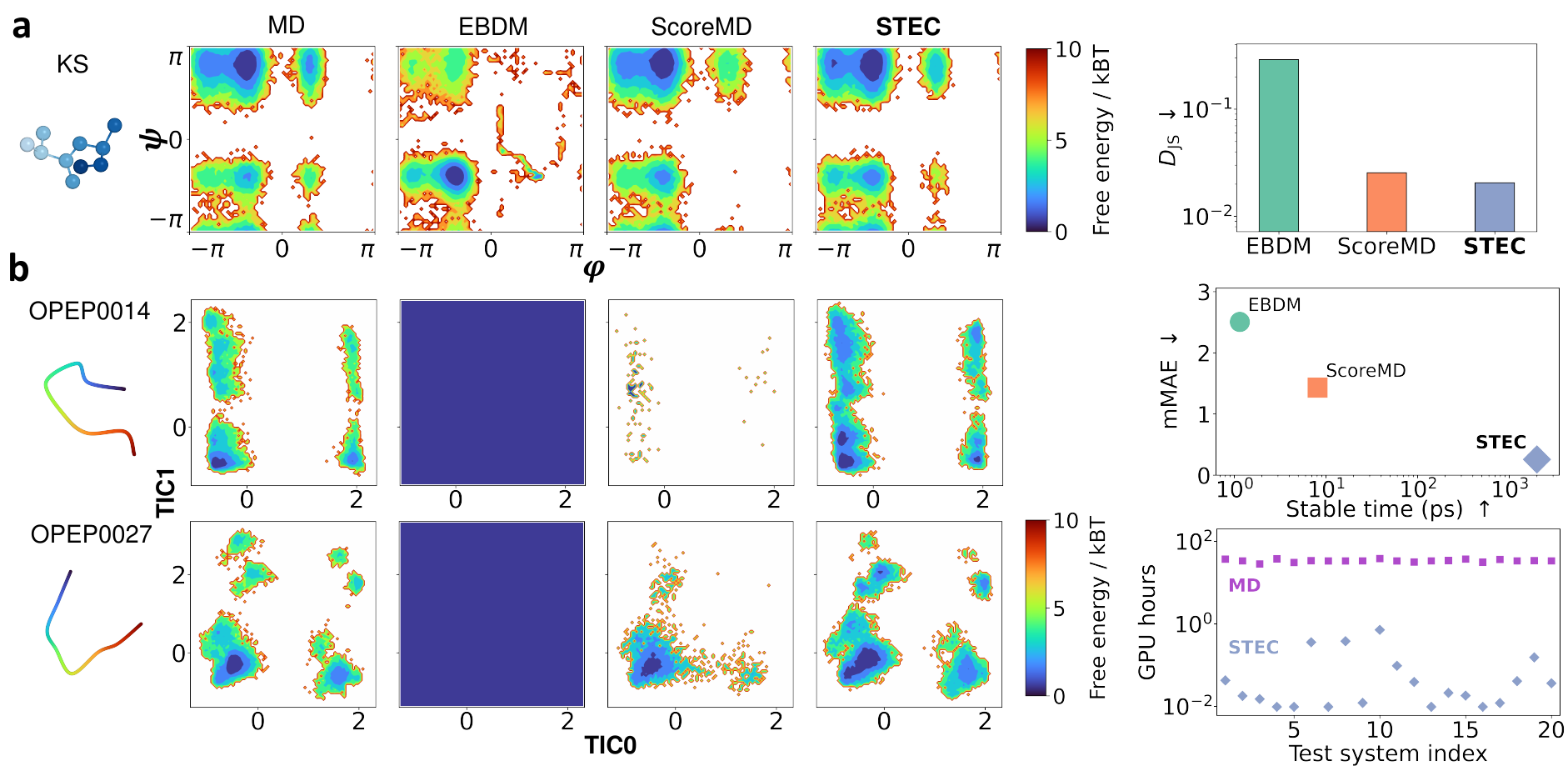}
    \caption{
    \textbf{STEC enables thermodynamically accurate and stable CGMD for unseen peptide systems.}
    \textbf{(a)} Transfer to unseen dipeptides. Ramachandran free-energy landscapes of the representative held-out lysine–serine (KS) dipeptide, projected onto backbone dihedral angles $(\phi,\psi)$, comparing the mapped atomistic MD reference, standard EBDM, ScoreMD, and STEC. The right panel reports the JS divergence ($D_{\rm JS}$) averaged over 15 held-out dipeptides.
    \textbf{(b)} Transfer to unseen octapeptides. Free-energy landscapes projected onto the first two TICs for two representative held-out sequences, OPEP0014 and OPEP0027, comparing the MD reference and the three learned models. The upper-right panel compares free-energy errors (mMAE) and stable simulation time across 20 held-out octapeptides. The lower-right panel compares the production GPU cost of constructing the atomistic reference ensemble with that of STEC-driven CGMD reaching 80\% coverage of the reference low-energy conformational region.
}
    \label{fig:peptide_generalization}
\end{figure}

\subsection{Transferability to Unseen Peptides}
\label{sec:results_generalization}

Having established the thermodynamic and dynamical fidelity of \ourmethod on individual molecular systems, we next assess its transferability to unseen peptide sequences. We first consider a dipeptide benchmark comprising combinations of the 20 standard amino acids~\citep{klein2023timewarp}.
Following ScoreMD~\citep{plainer2025consistent}, we train on 175 dipeptides and evaluate transferability on 15 held-out sequences.
For the representative unseen lysine--serine (KS) dipeptide, \ourmethod faithfully recovers the major conformational basins and their relative populations in the reference Ramachandran free-energy landscape, whereas standard EBDM and ScoreMD fail to capture key conformational regions (\figref{peptide_generalization}a). This advantage extends across 15 evaluated test dipeptides, where \ourmethod achieves markedly lower $D_{\rm JS}$ than both competing methods.

We next examine transferability to larger and more conformationally complex systems using a benchmark of 1,100 octapeptide sequences~\citep{charron2025navigating}, each represented by approximately 35--40 CG beads. We train on 550 octapeptides and evaluate the resulting CG potentials on 20 held-out sequences.
For the representative OPEP0014 and OPEP0027 systems (\figref{peptide_generalization}b), \ourmethod recovers the major low-energy basins of the atomistic reference landscapes. In contrast, standard EBDM fails to reproduce the reference conformational ensembles, while ScoreMD samples restricted or severely distorted conformational regions.

The quantitative comparisons across these 20 octapeptides further demonstrate substantial improvements in thermodynamic fidelity and simulation stability (\figref{peptide_generalization}b, upper right). \ourmethod reduces mMAE by more than 5-fold relative to ScoreMD, while extending stable CGMD simulation times by approximately three and two orders of magnitude relative to standard EBDM and ScoreMD, respectively. Additional free-energy landscapes across the evaluated dipeptide and octapeptide sequences corroborate the recovery of the major reference conformational basins (\figref{si_dipeptide_generalization} and \figref{si_octapeptide_generalization}). Together, these results demonstrate that \ourmethod generalizes thermodynamically faithful CG potentials to unseen peptide sequences while maintaining robust CGMD as molecular size and conformational complexity increase.

\begin{figure}[!t]
    \centering
    \includegraphics[width=0.8\textwidth]{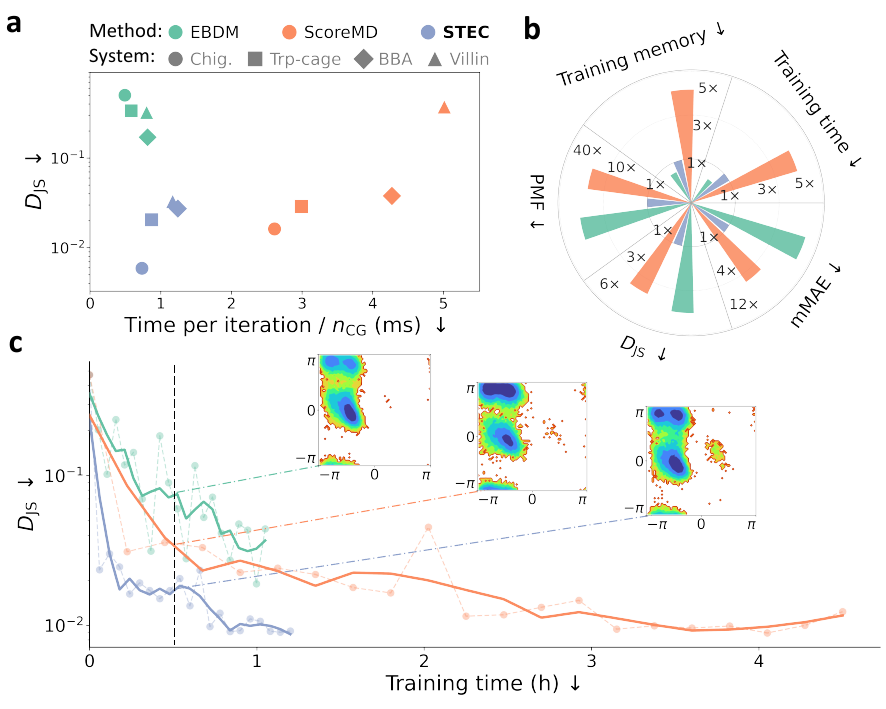}
    \caption{
    \textbf{STEC achieves high thermodynamic fidelity with modest training overhead.}
    Training costs in \textbf{(a)} and \textbf{(b)} are measured on a single NVIDIA A100 GPU, with per-iteration runtimes averaged over three independent runs.
    \textbf{(a)} Trade-off between CGMD distribution error ($D_{\rm JS}$) and per-iteration training time normalized by the number of CG sites ($n_{\rm CG}$) across four fast-folding proteins. Colors denote methods, while marker shapes distinguish proteins.
    \textbf{(b)} Training costs and thermodynamic errors on the octapeptide benchmark, expressed relative to STEC ($1\times$). The mMAE, $D_{\rm JS}$, and PMF error are averaged with equal weights over 20 held-out peptides before normalization. Each metric is displayed on an independent piecewise-linear radial scale.
    \textbf{(c)} CGMD equilibrium distribution error ($D_{\rm JS}$) as a function of elapsed training time on ALDP. The vertical dashed line marks a training budget of 0.5~hour, with the corresponding Ramachandran free-energy landscapes shown as insets.
    }
    \label{fig:efficiency}
\end{figure}

\subsection{Computational Efficiency of \ourmethod}
\label{sec:results_efficiency}

We finally demonstrate the computational efficiency of \ourmethod in learning accurate CG potentials and exploring molecular conformational space. Across the four fast-folding
proteins, we compare CGMD distribution errors against optimization-step
times normalized by the number of CG sites
(\figref{efficiency}a). \ourmethod achieves lower $D_{\rm JS}$ than both
baselines, with only a 46--53\% increase in per-step runtime relative
to standard EBDM (Table~\ref{tab:si_training_efficiency}). For the 20 held-out octapeptides, we jointly compare
training costs and equilibrium accuracy (\figref{efficiency}b).
ScoreMD requires approximately 4.0 times the per-step runtime and
4.2 times the peak device memory of \ourmethod.
Its mean mMAE, $D_{\rm JS}$, and PMF error are also higher by factors
of 5.5, 3.9, and 21.3, respectively.

Beyond per-step efficiency, we examine how rapidly each method learns a CG potential capable of repro-
ducing the target equilibrium distribution.
For ALDP, we evaluate intermediate training checkpoints by running Langevin CGMD and measuring the resulting $D_{\rm JS}$ against the atomistic reference (\figref{efficiency}c). \ourmethod converges substantially faster than both standard EBDM and ScoreMD, reaching a $D_{\rm JS}$ on the order of $10^{-2}$ within just 0.5~hour of training. The corresponding free-energy landscapes further reveal that \ourmethod already captures the major conformational basins and their relative populations, whereas the competing methods exhibit incomplete coverage at the same training budget. This rapid improvement in CGMD equilibrium fidelity demonstrates that \ourmethod produces accurate, simulation-ready potentials with substantially less training time.

We further assess the production cost of conformational sampling using the 20 unseen octapeptides examined above. Taking the low-energy conformational regions of the atomistic reference ensembles as targets, we measure the GPU cost required for \ourmethod-driven CGMD to achieve 80\% structural coverage (\figref{peptide_generalization}b, lower right). Across these systems, \ourmethod achieves a median production-cost reduction of 1749.5-fold compared with generating the corresponding $1~\mu\mathrm{s}$ atomistic reference ensembles, with savings spanning two to four orders of magnitude. The coverage definition and production-cost calculation are detailed in~\appxref{si_computational_efficiency},
with per-peptide results reported in Table~\ref{tab:si_opep_sampling_cost}. Together, these results establish the practical efficiency of \ourmethod in both learning transferable CG potentials and using them for extensive conformational sampling.

\section{Conclusions}

We have developed \ourmethod, an efficient diffusion-based framework for learning thermodynamically faithful CG force fields. Our theoretical characterization of zero-time distribution error reveals a fundamental gap between conventional diffusion training and CG potential learning, highlighting the joint roles of spatial score and temporal energy-derivative errors. To address this gap, STEC combines spatial score matching with a tractable conditional temporal energy-matching objective, avoiding the costly higher-order spatial derivatives required by Fokker--Planck regularization.
Across diverse molecular systems, STEC achieves chemical accuracy in macrostate free energies, reproduces key metastable transition patterns, and enables stable, transferable CGMD. With modest training overhead, it reduces median conformational-sampling production costs by over 1700-fold relative to atomistic reference simulation.

While STEC demonstrates promising transferability to unseen peptides, extending its application to larger proteins and more complex conformational dynamics remains an important direction. Recent advances in diffusion-based protein ensemble modeling, including DiG~\citep{zheng2024predicting} and BioEmu~\citep{lewis2025scalable}, have demonstrated scalable sampling of diverse protein conformations. Combining these advances with spatiotemporal energy supervision and broader training ensembles could enable thermodynamically faithful CGMD of larger biomolecular systems. Extending STEC to condition-dependent CG potentials could further enable the investigation of how temperature and solvent environments reshape protein free-energy landscapes and functional conformational transitions~\citep{charron2025navigating}.

\section*{Data Availability}

The molecular simulation datasets used in this study were obtained from previously published sources. The alanine dipeptide (ALDP) dataset is available through the \texttt{bgmol} repository (\url{https://github.com/noegroup/bgmol})~\citep{kohler2021smooth,arts2023,plainer2025consistent}. The atomistic trajectories of fast-folding proteins, originally reported by~\citet{deshaw2011fastfolding}, are available for noncommercial research upon request from D.~E.~Shaw Research (\url{https://www.deshawresearch.com/}). The dipeptide dataset of~\citet{klein2023timewarp} is hosted on Hugging Face (\url{https://huggingface.co/datasets/microsoft/timewarp}), with a reduced version available from~\citet{klein2024tbg} on OSF (\url{https://osf.io/n8vz3/}). The octapeptide dataset of~\citet{charron2025navigating} is deposited on Zenodo (\url{https://doi.org/10.5281/zenodo.15465782}).

\section*{Acknowledgements}

D.H. is supported by the Shanghai Science and Technology
Innovation Action Plan (23JC1402500). C.L. is supported by the National Natural Science Foundation
of China (NSFC 62676047) and Zhongguancun Academy (C20250506). H.Z. is supported by the National Natural Science Foundation of China (NSFC T2600069).

\printbibliography

\newpage

\appendix
\setcounter{secnumdepth}{2}
\setcounter{table}{0}
\renewcommand{\thetable}{S\arabic{table}}
\setcounter{equation}{0}
\renewcommand{\theequation}{S\arabic{equation}}
\renewcommand{\theHequation}{S\arabic{equation}}
\setcounter{figure}{0}
\renewcommand{\thefigure}{S\arabic{figure}}
\setcounter{section}{0}
\renewcommand{\thesection}{\Alph{section}}
\renewcommand{\thesubsection}{\thesection.\arabic{subsection}}

\section*{Appendix}

\section{Related Work}
\label{appx:related}

In this section, we provide a more detailed discussion of prior work that uses machine learning to approximate effective coarse-grained (CG) potentials for molecular simulation. Given a coarse-graining map $\bfxx=\Xi(\bfrr)$, the atomistic equilibrium ensemble induces the CG distribution
\begin{equation}
q^\cg(\bfxx)
=
\int q^{\mathrm{AT}}(\bfrr)
\delta\!\left(\Xi(\bfrr)-\bfxx\right)
\,\mathrm d\bfrr,
\end{equation}
whose thermodynamically consistent potential is the potential of mean force~\citep{rudzinski2011coarse},
\begin{equation}
U^\cg(\bfxx)
=
-k_{\mathrm B}T\log q^\cg(\bfxx)
+
\mathrm{const}.
\end{equation}
Because the marginalization over unresolved atomistic degrees of freedom is generally intractable, existing bottom-up methods differ primarily in how they obtain supervision for this effective potential. The development of these methods can be viewed as a progression from force-level supervision to distribution-level learning, and more recently to generative models that attempt to provide both equilibrium sampling and simulation-ready energy functions.

\subsection{Force matching}
\label{appx:related-fm}

Variational force matching (FM) exploits the relation between the CG mean force and projected atomistic forces. Under appropriate conditions on the CG mapping, the exact CG force can be written as a conditional expectation~\citep{kalligiannaki2015geometry}
\begin{equation}
-\nabla_{\bfxx}U^\cg(\bfxx)
=
\mathbb E_{q(\bfrr\mid\bfxx)}
\left[
\Xi_F
\left(
-\nabla_{\bfrr}U^{\mathrm{AT}}(\bfrr)
\right)
\right],
\label{eq:si_force_matching_identity}
\end{equation}
where $\Xi_F$ denotes the corresponding force projection operator. This relation motivates supervised regression of a parameterized CG force field against projected atomistic forces
\citep{noid2008multiscale,doerr2021torchmd,wang2019autoencoders,wang2019machine,husic2020coarse,durumeric2026learning,charron2025navigating,majewski2023machine}.
FM avoids sampling from the learned CG model during training and has therefore become one of the dominant approaches for bottom-up CG potential learning.

The projected force associated with an individual atomistic configuration, however, is only one sample of the conditional distribution in Eq.~\eqref{eq:si_force_matching_identity}. Different unresolved atomistic configurations can map to nearly identical CG coordinates while carrying substantially different projected forces. Consequently, the instantaneous force labels can exhibit large fluctuations around the conditional mean, leading to noisy supervision and poor data efficiency~\citep{kramer2023statistically}. The problem is compounded by the sampling distribution of the atomistic trajectories, which typically concentrates data in metastable basins and provides relatively sparse coverage of transition regions. Recent work has sought to alleviate this limitation through improved data generation. Chen et al.~\citep{chen2025enhanced}, for example, use enhanced sampling along CG coordinates and recompute forces with respect to the unbiased atomistic potential. This preserves the target conditional mean force while improving equilibration and coverage of transition regions. Such approaches can substantially improve FM data efficiency, but they retain the need for atomistic force evaluations and force-level supervision.

\subsection{Relative entropy and generative distribution learning}
\label{appx:related-rem}

Relative entropy minimization (REM) approaches the same problem from the distribution level. A parameterized CG potential $U_\theta^\cg$ defines
\begin{equation}
p_\theta^\cg(\bfxx)
\propto
\exp\left[-\frac{U_\theta^\cg(\bfxx)}
{k_{\mathrm B}T}\right],
\end{equation}
and REM minimizes
\begin{equation}
D_{\mathrm{KL}}
\left(
q^\cg\,\|\,p_\theta^\cg
\right).
\end{equation}
This objective directly targets the equilibrium distribution that the learned potential should reproduce and does not require projected atomistic force labels
\citep{shell2008relative,noid2008multiscale,thaler2022deep}.
Its gradient, however, contains an expectation over the current model distribution,
\begin{equation}
\nabla_\theta
D_{\mathrm{KL}}
\left(
q^\cg\,\|\,p_\theta^\cg
\right)
=
(k_{\mathrm B}T)^{-1}
\left[
\mathbb E_{q^\cg}
\left[\nabla_\theta U_\theta^\cg\right]
-
\mathbb E_{p_\theta^\cg}
\left[\nabla_\theta U_\theta^\cg\right]
\right],
\end{equation}
which generally requires repeated CG sampling as the potential evolves during optimization~\citep{chaimovich2011coarse}.

Generative models offer a route to retaining distribution-level learning while reducing this model-in-the-loop sampling cost. K\"ohler et al.~\citep{kohler2023flow} introduced a normalizing-flow-based approach termed Flow-Matching. The method first learns an explicit approximation to the CG equilibrium density and then generates configurations and corresponding force estimates from the learned flow to train a deterministic CG potential. This two-stage construction avoids atomistic force labels and repeated simulations of the evolving CG potential, combining aspects of relative-entropy learning and force matching. Its effectiveness nevertheless depends on learning an expressive density model, and the force field used for molecular simulation is obtained through a subsequent distillation stage rather than directly from the generative model.

\subsection{Diffusion models as coarse-grained force fields}
\label{appx:related-diffusion-cg}

Diffusion models provide a particularly flexible alternative for learning high-dimensional molecular distributions directly from equilibrium configurations. The central connection to coarse-grained force-field learning follows from the score of the noised molecular distribution. For a Boltzmann-distributed CG ensemble,
\begin{equation}
\nabla_{\bfxx}\log q^\cg(\bfxx)
=
\frac{\mathbf F^\cg(\bfxx)}{k_{\mathrm B}T}.
\label{eq:si_score_force_relation}
\end{equation}
At sufficiently small diffusion noise, the learned score therefore approaches the CG mean force up to the thermal factor. Arts et al.~\citep{arts2023} developed this connection in \textit{Two for One}, showing that a diffusion model trained only on equilibrium molecular configurations can provide both independent samples and a Denoising Force Field for CGMD. This removes the requirement for projected force labels while retaining a direct route from equilibrium configurations to molecular simulation. Their experiments also demonstrated the importance of conservative force parameterizations for stable dynamics.

A practical difficulty arises near the clean-data endpoint. Although the exact score converges to the score of the target Boltzmann distribution as the diffusion noise vanishes, estimating this quantity accurately becomes increasingly difficult at very small noise levels. \textit{Two for One} therefore treats the diffusion level used to extract the force as a hyperparameter and evaluates the force at a finite noise level selected according to simulation performance. This improves robustness in practice, but the corresponding score belongs to a noise-smoothed distribution rather than exactly to the target clean CG ensemble. The distinction becomes important when a diffusion model is expected not only to generate equilibrium configurations, but also to define the potential used for long-time CGMD.

\subsubsection{Temporal and spatiotemporal learning of diffusion energies}
\label{appx:related-diffusion-theory}

Beyond spatial score matching~\citep{hyvarinen2005estimation}, a complementary line of work exploits
variation along the corruption or diffusion-time dimension to learn
probability densities and energy functions. Choi et
al.~\citep{choi2022density} introduced time score matching (TSM), which
estimates the temporal log-density derivative
$\partial_t\log p_t(\bfxx)$ along a continuous probability path and
uses its integral for density-ratio estimation. More recently, Yu et
al.~\citep{yu2025density} derived conditional time-score objectives in
which the generally intractable marginal temporal score is replaced by
a tractable conditional target associated with a prescribed probability
path. These results establish a temporal analogue of denoising score
matching and provide efficient supervision of density variation across
noise levels.

Temporal information has also been incorporated directly into
time-dependent energy models. Guth et
al.~\citep{guth2025learning} proposed dual score matching, jointly
constraining the spatial derivative and noise-level derivative of a
learned energy to recover normalized densities consistently across
noise levels. Rather than matching an infinitesimal temporal derivative,
DiffCLF~\citep{ouyang2026diffusive} learns cross-noise-level energy
relations through a classification objective and combines this signal
with conventional score matching to improve global energy estimation.
Recent Spatiotemporal Noise-Contrastive Estimation
(stNCE)~\citep{yu2026spatiotemporal} provides a broader unifying perspective,
organizing energy-learning objectives according to whether they
constrain spatial, temporal, or joint spatiotemporal variations. stNCE also evaluates consistency between
diffusion and Langevin sampling for coarse-grained
molecular systems.

These approaches demonstrate that temporal information can complement
the local spatial supervision provided by score matching. Their primary
objectives, however, are density-ratio estimation, normalized density
learning, or general-purpose energy-based generative modeling. The
question considered here is different: when the endpoint of an
energy-based diffusion model is used as a physical potential for
molecular simulation, what errors determine whether that potential
reproduces the target equilibrium distribution? Our analysis connects
this endpoint error explicitly to spatial score error and temporal
energy-evolution error, providing the motivation for temporal
supervision specifically in simulation-oriented coarse-grained force
fields.

\subsubsection{Energy-based diffusion and sampling--simulation consistency}
\label{appx:related-diffusion-consis}

Energy parameterization makes the connection between diffusion models and molecular force fields more explicit. Rather than predicting an unconstrained vector-valued score, an energy-based diffusion model represents $\mathbf s_\theta(\bfxx,t)
=-\nabla_{\bfxx}u_\theta(\bfxx,t)$,
so the learned score is conservative by construction. The same time-dependent scalar energy can consequently support reverse-diffusion sampling and provide an endpoint potential for molecular simulation. This property has motivated energy-parameterized diffusion models in molecular modeling as well as in broader generative modeling settings.

Importantly, conservativity alone does not guarantee that these two uses of the model are thermodynamically consistent. Plainer et al.~\citep{plainer2025consistent} showed that an EBDM can produce accurate samples through reverse diffusion while the equilibrium distribution obtained from Langevin dynamics under the learned endpoint energy differs substantially from the training distribution. They traced this sampling--simulation inconsistency to errors at small diffusion times and, in particular, to violations of the Fokker--Planck evolution required by the diffusion process. ScoreMD therefore introduces a regularization objective derived from the Fokker--Planck equation. Related Fokker--Planck regularization had previously been studied for improving score-based generative models \citep{lai2023}; ScoreMD adapts this principle to energy-based molecular diffusion with the goal of obtaining a score that is consistent with the learned potential and suitable for simulation.

Fokker--Planck regularization provides a principled solution to the consistency problem, but it also changes the computational requirements of EBDM training. The FP residual couples the diffusion-time derivative of the density to spatial derivatives of the score. Under an energy parameterization, this introduces terms such as the Laplacian of the energy and requires higher-order spatial differentiation through the molecular energy network. Numerical approximations can reduce this burden, but additional model evaluations and derivative computations remain necessary. This becomes increasingly costly for larger molecular systems and expressive geometric neural networks.

\subsubsection{Energy-parameterized diffusion as statistical potentials.}
\label{appx:related-diffusion-potential}

Recent work also points to a broader role for energy-parameterized diffusion models beyond system-specific CG force-field learning. Roney et al.~\citep{roney2025protein} introduced ProteinEBM, which uses an energy-parameterized protein diffusion model as a learned statistical potential. The resulting energy is applied to structure ranking, mutation-effect prediction, conformational sampling, structure prediction, and exploration of folding pathways. This work illustrates an important advantage of embedding an explicit scalar energy within a generative model: the learned function can be reused for downstream physical or statistical tasks that are not accessible from samples alone. Transferable diffusion models such as BioEmu
and DiG generate protein conformational ensembles
at substantially lower sampling cost than direct
atomistic simulation
\citep{lewis2025scalable,zheng2024predicting}.
However, these models do not directly parameterize
a scalar energy whose gradient defines a
conservative CG force field for molecular simulation.

Our work addresses a complementary question. Rather than asking whether a diffusion-derived energy is useful as a general statistical score, we focus on whether the endpoint energy learned from mapped atomistic equilibrium configurations induces the same thermodynamic distribution when it is used as a CG potential. Our endpoint analysis shows that this error is governed jointly by the spatial score error and the diffusion-time energy-derivative error. Standard DSM directly targets the former, while full Fokker--Planck regularization couples both through a governing PDE. Spatiotemporal Energy Consistency (\ourmethod) instead directly supervises the temporal energy derivative through a tractable conditional objective. It therefore retains configuration-only distribution learning and a conservative endpoint force while avoiding the higher-order spatial derivatives required by full FP residual regularization. In this sense, \ourmethod is not an approximation to the Fokker--Planck objective, but a consistency objective derived directly from the error that matters for the endpoint CG distribution.
\section{Technical Details}

\label{appx:method}

In this section, we provide the detailed mathematical formulations,
derivations of the temporal consistency objective, and theoretical
proofs regarding the equivalence of objectives and the endpoint
distribution error bound.

\subsection{Diffusion process and energy-based parameterization}

\label{appx:method-diffusion}

We consider a diffusion process that gradually transforms the target
CG equilibrium distribution toward a simple reference distribution $p_{\rm simple}$.
Let $\bfxx_0\sim q^\cg$ denote an equilibrium CG configuration.
The forward process follows the SDE
\begin{equation}
\mathrm d\bfxx_t
=
f(\bfxx_t,t)\,\mathrm dt
+
g(t)\,\mathrm d\bfww_t,
\qquad t\in[0,1],
\end{equation}
where $f(\bfxx,t)$ and $g(t)$ denote the drift and diffusion
coefficients, respectively, and $\bfww_t$ is a standard Wiener process.
For the variance-preserving (VP) process used in the main text,
\begin{equation}
f(\bfxx_t,t)
=
-\frac{1}{2}\beta(t)\bfxx_t,
\qquad
g(t)=\sqrt{\beta(t)},
\end{equation}
where $\beta(t)>0$ is the noise schedule.
This process defines a continuous perturbation path
$\{p_t\}_{t\in[0,1]}$ with $p_0=q^\cg$
and $p_1= p_{\rm simple}$.

For $t>0$, the corresponding conditional transition kernel
is Gaussian,
\begin{equation}
p_t(\bfxx_t\mid\bfxx_0)
=
\mathcal N\left(
\bfxx_t;\alpha(t)\bfxx_0,\sigma^2(t)\mathbf I
\right),
\label{eq:appx_transition_kernel}
\end{equation}
where $\alpha(t)$ and $\sigma(t)$ denote the signal and noise
scales, respectively.
They are determined by the noise schedule through
\begin{equation}
\alpha(t)
=\exp\left[-\frac{1}{2}\int_0^t\beta(s)\,\mathrm ds
\right],\qquad \sigma^2(t)
=1-\exp\left[-\int_0^t\beta(s)\,\mathrm ds
\right].
\end{equation}
Equivalently, perturbed configurations can be sampled through
the reparameterization
$\bfxx_t=\alpha(t)\bfxx_0+\sigma(t)\bfeps$,
with $\bfeps\sim\mathcal N(0,\mathbf I)$.
The marginal distribution at each diffusion time is obtained
by integrating over the equilibrium ensemble,
\begin{equation}
p_t(\bfxx_t)
=
\int
p_t(\bfxx_t\mid\bfxx_0)
q^\cg(\bfxx_0)\,\mathrm d\bfxx_0.
\end{equation}

To transform samples from $p_{\rm simple}$ toward the target
equilibrium distribution $q^\cg$, the reverse-time SDE is given by
\begin{equation}
\mathrm d\bfxx_t
=
\left[
f(\bfxx_t,t)
-
g^2(t)\nabla_{\bfxx_t}\log p_t(\bfxx_t)
\right]\mathrm dt
+
g(t)\,\mathrm d\bar{\bfww}_t,
\end{equation}
where $\bar{\bfww}_t$ denotes a reverse-time Wiener process.
For generation, this SDE is integrated from $t=1$ to $t=0$,
initialized with samples from $p_{\rm simple}$. The reverse drift depends on the score of the intermediate diffusion marginal,
$\nabla_{\bfxx_t}\log p_{t}(\bfxx_t)$, which is generally unknown and therefore needs to be learned from data. Energy-based diffusion models parameterize the score field using a time-dependent scalar energy function $u^\theta_{t}(\bfxx)$,
$\mathbf s^\theta_{t}(\bfxx)=-\nabla_{\bfxx}
u^\theta_{t}(\bfxx)=\nabla_{\bfxx}\log p^\theta_{t}(\bfxx),$ where the associated normalized model density is
$p^\theta_{t}(\bfxx)\propto \exp[-u^\theta_{t}(\bfxx)].$
This energy parameterization ensures a conservative score field by construction and provides a direct connection to CGMD. At the clean
endpoint $t=0$, the learned energy induces the CG potential $U^{\cg,\theta}(\bfxx)=k_{\mathrm B}T
u^\theta_{0}(\bfxx),$
with the corresponding conservative force
$\mathbf F^{\cg,\theta}(\bfxx)=
-k_{\mathrm B}T\nabla_{\bfxx}u^\theta_{0}(\bfxx).$ The corresponding CGMD equilibrium distribution is
$p^{\cg,\theta}(\bfxx)=p^\theta_0(\bfxx)
\propto\exp[-u^\theta_0(\bfxx)]$.

Since the marginal score $\nabla_{\bfxx_t}\log p_{t}(\bfxx_t)$ is not available explicitly, EBDMs are trained using denoising score
matching. For the Gaussian transition kernel in Eq.~\eqref{eq:appx_transition_kernel}, the conditional score can be computed analytically as
\begin{equation}
\nabla_{\bfxx_t}
\log p_{t}(\bfxx_t\mid\bfxx_0)=
-\frac{\bfxx_t-\alpha(t)\bfxx_0}{\sigma^2(t)}.
\end{equation}
The resulting DSM objective is
\begin{equation}
\mathcal L_{\mathrm{DSM}}(\theta)=\mathbb E_{t,\bfxx_0,\bfxx_t}\left[w(t)\left\|\nabla_{\bfxx_t}
u^\theta_{t}(\bfxx_t)-\frac{\bfxx_t-\alpha(t)\bfxx_0}{\sigma^2(t)}\right\|_2^2\right],
\label{eq:appx_DSM}
\end{equation}
where the expectation is taken over diffusion times, equilibrium configurations $\bfxx_0\sim q^\cg$, and perturbed samples
$\bfxx_t\sim p_{t}(\bfxx_t|\bfxx_0)$.

Besides the stochastic reverse process, the same diffusion path can also be described by a deterministic probability-flow ordinary differential
equation (ODE), whose marginal distributions are identical to those of the forward SDE~\citep{song2021}. The probability-flow ODE is given by
\begin{equation}
\label{eq:appx_prob_flow_ode}
\frac{\mathrm d\bfxx_t}{\mathrm dt}=
\mathbf v(\bfxx_t,t)\triangleq
f(\bfxx_t,t)-\frac{1}{2}g^2(t)\nabla_{\bfxx_t}
\log p_{t}(\bfxx_t).
\end{equation}
The equivalence between the forward SDE and the probability-flow ODE follows from their identical Fokker--Planck equation, which guarantees the same marginal density evolution along the diffusion path. The
probability-flow velocity can be expressed as the posterior expectation of the conditional velocity,
\begin{equation}
\mathbf v(\bfxx_t,t)=
\mathbb E_{\bfxx_0\sim
p(\bfxx_0|\bfxx_t)}
\left[
\mathbf v(\bfxx_t,t|\bfxx_0)
\right].
\end{equation}
Using the Gaussian transition kernel defined above, the conditional velocity is given by
\begin{equation}
\mathbf v(\bfxx_t,t|\bfxx_0)
=
\alpha'(t)\bfxx_0
+
\frac{\sigma'(t)}{\sigma(t)}
\left[
\bfxx_t-\alpha(t)\bfxx_0
\right].
\label{eq:appx_cond_velocity}
\end{equation}
This conditional velocity provides a tractable representation of the probability-flow dynamics associated with the Gaussian perturbation
process and will be used to derive the temporal consistency objective and the endpoint distribution error bound. The theoretical results below
are established under the following regularity assumptions on the diffusion marginals and probability-flow velocity.

\paragraph{Assumptions.}
\label{appx:method-assumptions}

We make the following assumptions throughout the theoretical analysis. These conditions ensure the regularity, integrability, and boundary behavior required for the probability-flow dynamics and the derivation of the distribution error bound. Similar assumptions are commonly adopted in analyses of probability-flow ODEs and score-based diffusion models~\citep{song2021maximum}.

\begin{enumerate}[label=(A\arabic*)]

\item
$p_{t}(\bfxx)$ and $p_{t}^\theta(\bfxx)$ are $C^2$ on
$\mathbb R^D\times[0,1]$, where $D=3n$.

\item

$\mathbb E_{\bfxx_0\sim p_{0}}
\left[
\|\bfxx_0\|_2^2
\right]
<\infty.$

\item
$\alpha\in C^1([0,1])$,
$\sigma\in C([0,1])\cap C^1((0,1])$,
and $g\in C^1([0,1])$, with
$\alpha(t)>0$ for $t\in[0,1]$,
$\sigma(0)=0$, and
$\sigma(t)>0$ for $t\in(0,1]$.

\item

$\exists C>0,\ \forall\,(\bfxx,t)\in\mathbb R^D\times[0,1]:
\quad
\|f(\bfxx,t)\|_2
\le
C(1+\|\bfxx\|_2).$

\item
$\exists C>0,\ \forall\,(\bfxx,t)\in\mathbb R^D\times[0,1]:
\quad
\left\|
\nabla_{\bfxx}\log p_{t}(\bfxx)
\right\|_2
\le
C(1+\|\bfxx\|_2).$

\item

$\exists C>0,\ \forall\,(\bfxx,t)\in\mathbb R^D\times[0,1]:
\quad
\left\|
\nabla_{\bfxx}\log p_{t}^\theta(\bfxx)
\right\|_2
\le
C(1+\|\bfxx\|_2).$

\item

$\mathbb E_{t\sim\mathcal U(0,1),\,\bfxx\sim p_{t}}
\left[
\left\|
\nabla_{\bfxx}\log p_{t}^\theta(\bfxx)
-
\nabla_{\bfxx}\log p_{t}(\bfxx)
\right\|_2^2
\right]
<\infty.$

\item

$\mathbb E_{t\sim\mathcal U(0,1),\,\bfxx\sim p_{t}}
\left[
\left|
\partial_t\log p_{t}^\theta(\bfxx)
-
\partial_t\log p_{t}(\bfxx)
\right|^2
\right]
<\infty.$

\item

$\forall\,t\in[0,1],\ \exists k>0:
p_{t}(\bfxx)
=
O\left(
e^{-\|\bfxx\|_2^k}
\right)$
as
$\|\bfxx\|_2\rightarrow\infty .$

\end{enumerate}

\subsection{Derivation of the Conditional Temporal Matching Objective}
\label{appx:method-condition}

We provide the proof of Proposition~\ref{prop:methods_conditional_time}
from the main text. The derivation starts from the continuity equation
associated with the probability-flow ODE and shows that the marginal
temporal log-density derivative can be expressed as the posterior
expectation of a tractable conditional counterpart. This identity
establishes the equivalence between the marginal and conditional
temporal objectives at the level of parameter gradients.

\begin{proposition}[Continuity equation]
\label{prop:continuity_eq}

Let $p_t(\bfxx)$ be the density transported by the ODE
$\mathrm d\bfxx_t/\mathrm dt=\mathbf v(\bfxx_t,t)$.
Then $p_t$ satisfies
\begin{equation}
\partial_t p_t(\bfxx)
+
\nabla_{\bfxx}\cdot
\left[
p_t(\bfxx)\mathbf v(\bfxx,t)
\right]
=0.
\label{eq:continuity_eq_app}
\end{equation}
Equivalently,
\begin{equation}
\partial_t\log p_t(\bfxx)
=
-\mathbf v(\bfxx,t)\cdot
\nabla_{\bfxx}\log p_t(\bfxx)
-
\nabla_{\bfxx}\cdot\mathbf v(\bfxx,t).
\label{eq:log_continuity_eq_app}
\end{equation}

\end{proposition}

\begin{proof}

Let $\Omega_t$ be an arbitrary region transported by
the velocity field $\mathbf v(\bfxx,t)$.
Conservation of probability mass gives
\begin{equation}
\frac{\mathrm d}{\mathrm dt}
\int_{\Omega_t}
p_t(\bfxx)\,\mathrm d\bfxx
=0.
\end{equation}
By the Reynolds transport theorem,
\begin{equation}
\frac{\mathrm d}{\mathrm dt}
\int_{\Omega_t}
p_t(\bfxx)\,\mathrm d\bfxx
=
\int_{\Omega_t}
\left[
\partial_t p_t(\bfxx)
+
\nabla_{\bfxx}\cdot
\left(
p_t(\bfxx)\mathbf v(\bfxx,t)
\right)
\right]
\,\mathrm d\bfxx.
\end{equation}
Since $\Omega_t$ is arbitrary, the integrand must vanish,
yielding Eq.~\eqref{eq:continuity_eq_app}.
Dividing by $p_t(\bfxx)$ and applying the product rule gives
\begin{align}
\partial_t\log p_t(\bfxx)
&=
-\frac{1}{p_t(\bfxx)}
\nabla_{\bfxx}\cdot
\left[
p_t(\bfxx)\mathbf v(\bfxx,t)
\right]
\nonumber\\
&=
-\nabla_{\bfxx}\cdot\mathbf v(\bfxx,t)
-
\mathbf v(\bfxx,t)\cdot
\nabla_{\bfxx}\log p_t(\bfxx),
\end{align}
which proves Eq.~\eqref{eq:log_continuity_eq_app}.

\end{proof}

\paragraph{Conditional temporal target.}

The marginal target
$\tau_t(\bfxx_t)=\partial_t\log p_t(\bfxx_t)$
is not directly available because it depends on the unknown
diffusion marginal $p_t$. We therefore construct a conditional
target based on the known perturbation kernel. Analogously, for
a fixed clean configuration $\bfxx_0$, we define the conditional
temporal log-density derivative as
\begin{equation}
\tau_{t}^{\mathrm{cond}}
(\bfxx_t\mid\bfxx_0)
:=
-
\nabla_{\bfxx_t}\cdot
\mathbf v(\bfxx_t,t\mid\bfxx_0)
-
\mathbf v(\bfxx_t,t\mid\bfxx_0)
\cdot
\nabla_{\bfxx_t}
\log p_{t}(\bfxx_t\mid\bfxx_0).
\label{eq:appx_conditional_temporal_target_general}
\end{equation}
This is the conditional analogue of
Eq.~\eqref{eq:log_continuity_eq_app}.

For the Gaussian perturbation kernel in
Eq.~\eqref{eq:appx_transition_kernel}, the conditional score is
\begin{equation}
\nabla_{\bfxx_t}
\log p_{t}(\bfxx_t\mid\bfxx_0)
=
-
\frac{
\bfxx_t-\alpha(t)\bfxx_0
}{
\sigma^2(t)
}.
\label{eq:appx_gaussian_conditional_score}
\end{equation}
Using the conditional velocity in Eq.~\eqref{eq:appx_cond_velocity}, its divergence with respect to $\bfxx_t$ is
\begin{equation}
\nabla_{\bfxx_t}\cdot
\mathbf v(\bfxx_t,t\mid\bfxx_0)
=
3n\frac{\sigma'(t)}{\sigma(t)}.
\label{eq:appx_gaussian_conditional_divergence}
\end{equation}
Substituting the conditional velocity and the Gaussian transition kernel into Eq.~\eqref{eq:appx_conditional_temporal_target_general} gives the analytical conditional temporal derivative
\begin{equation}
\tau_{t}^{\mathrm{cond}}(\bfxx_t\mid\bfxx_0)
=\frac{\alpha'(t)}{\sigma(t)}
\bfxx_0\cdot\bfeps
+\frac{\sigma'(t)}{\sigma(t)}
\left(
\|\bfeps\|_2^2-3n
\right),
\label{eq:appx_conditional_temporal_target}
\end{equation}
where $\bfeps=(\bfxx_t-\alpha(t)\bfxx_0)/\sigma(t)$ and $n$ denotes the number of CG beads, such that the dimensionality of the CG coordinate space is $3n$. This conditional temporal target is used in the STEC objective. Throughout this derivation, temporal derivatives are
evaluated at fixed molecular coordinates.

\paragraph{Conditional-to-marginal identity.}

We next show that the conditional temporal target recovers the marginal temporal log-density derivative after averaging over the posterior of
the clean configuration. The marginal probability current admits the decomposition
\begin{equation}
\mathbf v(\bfxx_t,t)p_{t}(\bfxx_t)=
\int
\mathbf v(\bfxx_t,t\mid\bfxx_0)
p_{t}(\bfxx_t\mid\bfxx_0)
p_{0}(\bfxx_0)
\,\mathrm d\bfxx_0.
\label{eq:appx_conditional_current_decomposition}
\end{equation}
Applying $-\nabla_{\bfxx_t}\cdot$ to both sides and using
Eq.~\eqref{eq:continuity_eq_app} yields
\begin{align}
p_{t}(\bfxx_t)\tau_{t}(\bfxx_t)
&=-\nabla_{\bfxx_t}\cdot
\left[\mathbf v(\bfxx_t,t)p_{t}(\bfxx_t)\right]\nonumber\\
&=-\int\nabla_{\bfxx_t}\cdot\left[
\mathbf v(\bfxx_t,t\mid\bfxx_0)
p_{t}(\bfxx_t\mid\bfxx_0)\right]
p_{0}(\bfxx_0)\,\mathrm d\bfxx_0
\nonumber\\
&=\int p_{t}(\bfxx_t\mid\bfxx_0)
\tau_{t}^{\mathrm{cond}}(\bfxx_t\mid\bfxx_0)
p_{0}(\bfxx_0)\,\mathrm d\bfxx_0.
\label{eq:appx_temporal_current_identity}
\end{align}
Dividing by $p_{t}(\bfxx_t)$ gives
\begin{equation}
\tau_{t}(\bfxx_t)=
\mathbb E_{\bfxx_0\sim p(\bfxx_0\mid\bfxx_t)}
\left[\tau_{t}^{\mathrm{cond}}
(\bfxx_t\mid\bfxx_0)\right].
\label{eq:appx_conditional_temporal_expectation_identity}
\end{equation}
Thus, the conditional temporal target is an unbiased conditional representation of the marginal temporal log-density derivative.

\paragraph{Proof of the conditional temporal matching equivalence.}

For compactness, define
$h_\theta(\bfxx_t,t):=\partial_tu^\theta_t(\bfxx_t)$ and $\rho(t):=w_{\rm time}(t)$. The marginal temporal objective is
\begin{equation}
\clL_{\mathrm{time}}^{\mathrm{marg}}(\theta)=
\mathbb E_{t,\bfxx_t\sim p_{t}}
\left[\rho(t)\left(h_\theta(\bfxx_t,t)+
\tau_{t}(\bfxx_t)\right)^2\right].
\label{eq:appx_time_marginal_loss_proof}
\end{equation}
Since $\rho(t)$ and $\tau_{t}$ do not depend on $\theta$, its parameter gradient is
\begin{equation}
\nabla_\theta
\clL_{\mathrm{time}}^{\mathrm{marg}}(\theta)
=2\mathbb E_{t,\bfxx_t\sim p_{t}}
\left[\rho(t)\left(h_\theta(\bfxx_t,t)+
\tau_{t}(\bfxx_t)\right)
\nabla_\theta h_\theta(\bfxx_t,t)\right].
\label{eq:appx_grad_time_marginal}
\end{equation}

The tractable conditional temporal objective is
\begin{equation}
\clL_{\rm time}(\theta)=
\mathbb E_{t,\bfxx_0\sim p_{0},\bfxx_t\sim p_{t}(\bfxx_t|\bfxx_0)}
\left[\rho(t)\left(h_\theta(\bfxx_t,t)+
\tau_{t}^{\mathrm{cond}}
(\bfxx_t|\bfxx_0)\right)^2\right],
\label{eq:appx_conditional_time_loss_proof}
\end{equation}
and its gradient is
\begin{equation}
\nabla_\theta
\clL_{\rm time}(\theta)
=2\mathbb E_{t,\bfxx_0\sim p_{0},\bfxx_t\sim p_{t}(\bfxx_t|\bfxx_0)}
\left[\rho(t)\left(h_\theta(\bfxx_t,t)+
\tau_{t}^{\mathrm{cond}}
(\bfxx_t|\bfxx_0)\right)
\nabla_\theta h_\theta(\bfxx_t,t)\right].
\label{eq:appx_grad_conditional_time}
\end{equation}

The quadratic terms involving
$h_\theta(\bfxx_t,t)\nabla_\theta h_\theta(\bfxx_t,t)$ are identical in Eqs.~\eqref{eq:appx_grad_time_marginal} and
\eqref{eq:appx_grad_conditional_time}, because sampling
$\bfxx_0\sim p_{0}$ followed by
$\bfxx_t\sim p_{t}(\bfxx_t|\bfxx_0)$
induces the marginal distribution $p_{t}(\bfxx_t)$.

For the remaining cross terms, the conditional-to-marginal identity in
Eq.~\eqref{eq:appx_conditional_temporal_expectation_identity} gives
\begin{align}
&
\mathbb E_{\bfxx_0\sim p_{0},\bfxx_t\sim p_{t}(\bfxx_t|\bfxx_0)}
\left[\tau_{t}^{\mathrm{cond}}
(\bfxx_t|\bfxx_0)\nabla_\theta h_\theta(\bfxx_t,t)
\right]\nonumber\\
&=
\mathbb E_{\bfxx_t\sim p_{t}}
\left[\tau_{t}(\bfxx_t)
\nabla_\theta h_\theta(\bfxx_t,t)\right].
\label{eq:appx_temporal_cross_term_identity}
\end{align}

Since $\rho(t)$ depends only on diffusion time, the same equality holds after weighting by $\rho(t)$ and averaging over $t$. Therefore,
\begin{equation}
\nabla_\theta
\clL_{\mathrm{time}}^{\mathrm{marg}}(\theta)
=\nabla_\theta\clL_{\rm time}(\theta).
\label{eq:appx_temporal_gradient_equivalence}
\end{equation}
Thus, the marginal and conditional temporal objectives have identical optimization landscapes up to a parameter-independent constant and share the same minimizers.

\subsection{Boundedness of the Probability-Flow Velocity}
\label{appx:si_velocity_bound}

We next establish that the probability-flow velocity has finite
integrated squared norm along the diffusion path. This property is used
to control the spatial contribution in the proof of the spatiotemporal
distribution error bound.

\begin{proposition}[Finite kinetic energy]
\label{prop:boundedness}

Under Assumptions (A2)--(A5) in \appxref{method-diffusion}, the probability-flow velocity
$\mathbf v(\bfxx_t,t)
=f(\bfxx_t,t)-\frac{1}{2}g^2(t)\nabla_{\bfxx}\log p_{t}(\bfxx_t)$
satisfies
\begin{equation}
\int_0^1 \mathbb E_{\bfxx_t\sim p_{t}}
\left[\left\|\mathbf v(\bfxx_t,t)
\right\|_2^2\right]\,\mathrm dt<\infty.
\label{eq:appx_finite_velocity_energy}
\end{equation}

\end{proposition}

\begin{proof}

Using the Gaussian perturbation representation
$\bfxx_t=\alpha(t)\bfxx_0+\sigma(t)\boldsymbol\epsilon$, with $\bfxx_0\sim p_{0}$ and
$\boldsymbol\epsilon\sim\mathcal N(0,I)$, we have
\begin{align}
\mathbb E_{\bfxx_0,\boldsymbol\epsilon}
\left\|\bfxx_t\right\|_2^2
&=\mathbb E_{\bfxx_0,\boldsymbol\epsilon}
\left\|\alpha(t)\bfxx_0+\sigma(t)\boldsymbol\epsilon\right\|_2^2\nonumber\\
&\le 2\alpha^2(t)\mathbb E_{\bfxx_0}
\left\|\bfxx_0\right\|_2^2+2\sigma^2(t)
\mathbb E_{\boldsymbol\epsilon}
\left\|\boldsymbol\epsilon\right\|_2^2 .
\label{eq:appx_second_moment_xt}
\end{align}
By Assumptions (A2) and (A3), $\alpha(t)$ and $\sigma(t)$ are bounded on $[0,1]$. Therefore, Eq.~\eqref{eq:appx_second_moment_xt} implies that there exists a finite constant $C_0$ such that
\begin{equation}
\sup_{t\in[0,1]}
\mathbb E_{\bfxx_t\sim p_{t}}
\left[\left\|\bfxx_t\right\|_2^2
\right] \le C_0 .
\label{eq:appx_uniform_second_moment}
\end{equation}

From the definition of the probability-flow velocity,
\begin{equation}
\left\|\mathbf v(\bfxx_t,t)
\right\|_2^2\le 2\left\| f(\bfxx_t,t)\right\|_2^2
+\frac{1}{2}g^4(t)\left\|\nabla_{\bfxx}\log p_{t}(\bfxx_t)\right\|_2^2 .
\label{eq:appx_velocity_square_bound}
\end{equation}

Since $g(t)$ is bounded on $[0,1]$, Assumptions (A4) and (A5) give
\[\left\|\mathbf v(\bfxx_t,t)
\right\|_2^2\le C_1(1+\|\bfxx_t\|_2)^2 \]
for some finite constant $C_1$. Taking expectation with respect to $p_{t}$ and using Eq.~\eqref{eq:appx_uniform_second_moment} yields
\begin{equation}
\sup_{t\in[0,1]}
\mathbb E_{\bfxx_t\sim p_{t}}
\left[\left\|\mathbf v(\bfxx_t,t)
\right\|_2^2\right]<\infty .
\end{equation}

Therefore,
\begin{equation}
\int_0^1
\mathbb E_{\bfxx_t\sim p_{t}}
\left[\left\|\mathbf v(\bfxx_t,t)
\right\|_2^2\right]\,\mathrm dt<\infty,
\end{equation}
which proves the result.
\end{proof}

\subsection{Proof of the Spatiotemporal Distribution Error Bound}
\label{appx:si_endpoint_bound}

\begin{proof}[Proof of Theorem~\ref{thm:distribution_error}]

For any $s\in[0,1]$, define
\begin{equation}
\mathrm{KL}(s):=D_{\mathrm{KL}}
\left(p_{s}\,\|\,p^\theta_{s}
\right).
\label{eq:appx_si_kl_s}
\end{equation}
Differentiating with respect to $s$ gives
\begin{align}
\frac{\mathrm d}{\mathrm ds}
\mathrm{KL}(s)={}&\int\partial_s p_{s}(\bfxx)\log\frac{p_{s}(\bfxx)}{
p^\theta_{s}(\bfxx)}
\,\mathrm d\bfxx\nonumber\\
&+\int p_{s}(\bfxx)
\left[\partial_s\log p_{s}(\bfxx)
-\partial_s\log p^\theta_{s}(\bfxx)
\right]\,\mathrm d\bfxx .
\label{eq:appx_si_kl_derivative}
\end{align}

Using the continuity equation from
Proposition~\ref{prop:continuity_eq},
\[\partial_s p_{s}(\bfxx)
=-\nabla_{\bfxx}\cdot\left[p_{s}(\bfxx)
\mathbf v(\bfxx,s)\right],
\]
the first term in Eq.~\eqref{eq:appx_si_kl_derivative} can be rewritten by integration by parts as
\begin{align}
&
\int
\partial_s p_{s}(\bfxx)\log
\frac{p_{s}(\bfxx)}{
p^\theta_{s}(\bfxx)}
\,\mathrm d\bfxx\nonumber\\
&=-\mathbb E_{\bfxx\sim p_{s}}
\left[\mathbf v(\bfxx,s)
\cdot\left(\nabla_{\bfxx}\log p^\theta_{s}(\bfxx)-\nabla_{\bfxx}\log p_{s}(\bfxx)\right)\right].
\label{eq:appx_si_kl_spatial_term}
\end{align}
The boundary term vanishes under Assumptions (A5), (A6), and (A9),
which guarantee the required growth control of the score functions and the decay of the diffusion marginals.

Combining Eqs.~\eqref{eq:appx_si_kl_derivative} and
\eqref{eq:appx_si_kl_spatial_term} yields
\begin{align}
\frac{\mathrm d}{\mathrm ds}
\mathrm{KL}(s)
=-\mathbb E_{\bfxx\sim p_{s}}\Bigg[
&
\mathbf v(\bfxx,s)\cdot\left(
\nabla_{\bfxx}\log p^\theta_{s}(\bfxx)
-\nabla_{\bfxx}\log p_{s}(\bfxx)\right)
\nonumber\\
&+\left(\partial_s\log p^\theta_{s}(\bfxx)-\partial_s\log p_{s}(\bfxx)
\right)\Bigg].
\label{eq:appx_si_kl_derivative_error}
\end{align}
Integrating Eq.~\eqref{eq:appx_si_kl_derivative_error} from an arbitrary $t\in[0,1]$ to the terminal time $1$ gives
\begin{align}
D_{\mathrm{KL}}\left(
p_{t}\,\|\,p^\theta_{t}
\right)
={}&
D_{\mathrm{KL}}\left(
p_{1}\,\|\,p^\theta_{1}
\right)\nonumber\\
&+\int_t^1 \mathbb E_{\bfxx\sim p_{s}}
\Bigg[\mathbf v(\bfxx,s)
\cdot\left(
\nabla_{\bfxx}\log p^\theta_{s}(\bfxx)
-\nabla_{\bfxx}\log p_{s}(\bfxx)
\right)\nonumber\\
&\hspace{2.8cm}
+\left(\partial_s\log p^\theta_{s}(\bfxx)-\partial_s\log p_{s}(\bfxx)
\right)\Bigg]\,\mathrm ds .
\label{eq:appx_si_kl_integrated}
\end{align}
Define
\begin{equation}
K_v:=
\left[
\int_0^1
\mathbb E_{\bfxx\sim p_s}
\left\|\mathbf v(\bfxx,s)\right\|_2^2
\,\mathrm ds
\right]^{1/2},
\qquad
K:=\max\{K_v,1\}.
\label{eq:appx_si_velocity_constant}
\end{equation}
By Proposition~\ref{prop:boundedness}, $K_v<\infty$.
Thus, $K$ is finite and independent of diffusion time $t$.
For the intermediate-time derivation, define
\begin{align*}
\clE_{\rm space}(t)
&=
K\left[
\int_t^1
\bbE_{\bfxx\sim p_s}
\left\|
\nabla_{\bfxx}\log p^\theta_s(\bfxx)
-
\nabla_{\bfxx}\log p_s(\bfxx)
\right\|_2^2
\,\mathrm ds
\right]^{1/2},
\\
\clE_{\rm time}(t)
&=
K\left[
\int_t^1
\bbE_{\bfxx\sim p_s}
\left|
\partial_s\log p^\theta_s(\bfxx)
-
\partial_s\log p_s(\bfxx)
\right|^2
\,\mathrm ds
\right]^{1/2}.
\end{align*}
At $t=0$, these quantities coincide with the errors
defined in Eq.~\eqref{eq:methods_spatiotemporal_errors}.

Applying the Cauchy--Schwarz inequality to the spatial term gives
\begin{align}
&\left|
\int_t^1
\mathbb E_{\bfxx\sim p_s}
\left[
\mathbf v(\bfxx,s)\cdot
\left(
\nabla_{\bfxx}\log p^\theta_s(\bfxx)
-
\nabla_{\bfxx}\log p_s(\bfxx)
\right)
\right]
\,\mathrm ds
\right|
\nonumber\\
&\leq
\left[
\int_t^1
\mathbb E_{\bfxx\sim p_s}
\left\|\mathbf v(\bfxx,s)\right\|_2^2
\,\mathrm ds
\right]^{1/2}
\frac{\clE_{\rm space}(t)}{K}
\label{eq:appx_si_spatial_cs}\\
&\leq
\frac{K_v}{K}\clE_{\rm space}(t),
\label{eq:appx_si_spatial_error_bound}
\end{align}
where the last inequality follows from
$[t,1]\subseteq[0,1]$.

Similarly, the temporal term satisfies
\begin{align}
&\left|
\int_t^1
\mathbb E_{\bfxx\sim p_s}
\left[
\partial_s\log p^\theta_s(\bfxx)
-
\partial_s\log p_s(\bfxx)
\right]
\,\mathrm ds
\right|
\nonumber\\
&\leq
\sqrt{1-t}
\left[
\int_t^1
\mathbb E_{\bfxx\sim p_s}
\left|
\partial_s\log p^\theta_s(\bfxx)
-
\partial_s\log p_s(\bfxx)
\right|^2
\,\mathrm ds
\right]^{1/2}
\nonumber\\
&=
\frac{\sqrt{1-t}}{K}\clE_{\rm time}(t)
\leq
\frac{1}{K}\clE_{\rm time}(t).
\label{eq:appx_si_temporal_error_bound}
\end{align}

Combining Eqs.~\eqref{eq:appx_si_kl_integrated},
\eqref{eq:appx_si_spatial_error_bound}, and
\eqref{eq:appx_si_temporal_error_bound} yields
\begin{equation}
D_{\rm KL}\!\left(p_t\,\|\,p^\theta_t\right)
\leq
D_{\rm KL}\!\left(p_1\,\|\,p^\theta_1\right)
+
\frac{K_v}{K}\clE_{\rm space}(t)
+
\frac{1}{K}\clE_{\rm time}(t).
\label{eq:appx_si_pathwise_intermediate}
\end{equation}
Since $K\geq K_v$ and $K\geq1$, we obtain
\begin{equation}
D_{\rm KL}\!\left(p_t\,\|\,p^\theta_t\right)
\leq
D_{\rm KL}\!\left(p_1\,\|\,p^\theta_1\right)
+
\clE_{\rm space}(t)
+
\clE_{\rm time}(t).
\label{eq:appx_si_pathwise_error_bound}
\end{equation}
Under the terminal matching assumption $p^\theta_1=p_1$
in Theorem~\ref{thm:distribution_error},
the terminal KL term vanishes, giving
\begin{equation*}
D_{\rm KL}\!\left(p_t\,\|\,p^\theta_t\right)
\leq
\clE_{\rm space}(t)+\clE_{\rm time}(t).
\end{equation*}

At the clean endpoint $t=0$, we have $p_0=q^\cg$ and
$p^\theta_0=p^{\cg,\theta}$. The bound therefore reduces to
\begin{equation}
D_{\rm KL}\!\left(
q^\cg\,\|\,p^{\cg,\theta}
\right)
\leq
\clE_{\rm space}+\clE_{\rm time},
\label{eq:appx_si_endpoint_error_bound}
\end{equation}
where $\clE_{\rm space}=\clE_{\rm space}(0)$ and
$\clE_{\rm time}=\clE_{\rm time}(0)$.
This proves Theorem~\ref{thm:distribution_error}.

\end{proof}

\subsection{Discussion}
\label{appx:method-discuss}

\paragraph{Energy normalization along the diffusion path.}

Under exact spatial score matching, the learned energy satisfies
$u^\theta_{t}(\bfxx)
=-\log p_{t}(\bfxx)+c(t)$,
where $c(t)$ is independent of molecular coordinates.
Temporal supervision further constrains the energy evolution by matching
$\partial_tu^\theta_{t}$ to $-\tau_{t}$.
Exact temporal matching therefore gives $c'(t)=0$, so the normalization
constant $Z^\theta_{t}=e^{-c(t)}$ remains unchanged along
the diffusion path.
Consequently,
$-\partial_tu^\theta_{t}
=\partial_t\log p^\theta_{t}$.
The remaining additive constant does not affect the equilibrium
distribution or the conservative force field used for CGMD.

\paragraph{Relation to Fokker--Planck regularization.}
\label{appx:appx_discuss}
An alternative approach to enforcing diffusion-time consistency is to
directly regularize the Fokker--Planck equation governing the evolution
of the probability density along the diffusion path. ScoreMD~\citep{plainer2025consistent}
introduces a Fokker--Planck-based regularization to constrain the
learned energy dynamics. Under the energy-based parameterization
$p^\theta_{t}(\bfxx)\propto
\exp[-u^\theta_{t}(\bfxx)]$, the corresponding Fokker--Planck
residual is given by
\begin{equation}
\begin{aligned}
\mathcal R_{\mathrm{FP}}^\theta(\bfxx_t,t)
={}&
\partial_t u^\theta_{t}(\bfxx_t)
-\frac{1}{2}g^2(t)
\left[
\Delta_{\bfxx_t}u^\theta_{t}(\bfxx_t)
-
\left\|
\nabla_{\bfxx_t}u^\theta_{t}(\bfxx_t)
\right\|_2^2
\right]
\\
&+
f(\bfxx_t,t)\cdot
\nabla_{\bfxx_t}u^\theta_{t}(\bfxx_t)
-
\nabla_{\bfxx_t}\cdot f(\bfxx_t,t).
\end{aligned}
\label{eq:appx_discussion_fp_residual}
\end{equation}
The corresponding FP regularization objective is
\begin{equation}
\mathcal L_{\mathrm{FP}}(\theta)
=
\mathbb E_{t,\bfxx_t\sim p_{t}}
\left[
\frac{w_{\mathrm{FP}}(t)}{(3n)^2}
\left|
\mathcal R_{\mathrm{FP}}^\theta(\bfxx_t,t)
\right|^2
\right],
\label{eq:appx_discussion_fp_loss}
\end{equation}
where $w_{\mathrm{FP}}(t)$ denotes the diffusion-time weighting used in
the FP regularization.

The Fokker--Planck residual provides a straightforward constraint on the
diffusion evolution. However, directly evaluating this residual requires
higher-order spatial derivatives of the learned energy field, including
the score divergence and Laplacian terms. This introduces substantial
computational overhead, especially when gradients with respect to model
parameters are computed.

ScoreMD reduces this cost using a Gaussian-smoothed
weak FP residual~\citep{plainer2025consistent},
\begin{equation}
\widetilde{\mathcal R}_{\rm FP}^\theta(\bfxx_t,t)
=
\mathbb E_{\mathbf z}
\left[
\mathcal R_{\rm FP}^\theta
(\bfxx_t+\eta\mathbf z,t)
\right],
\qquad
\mathbf z\sim\mathcal N(\mathbf 0,\mathbf I).
\end{equation}
The divergence term is estimated using first-order
score evaluations at $\bfxx_t\pm\eta\mathbf z$,
avoiding explicit evaluation of the energy Laplacian.
Two independent residual estimates,
$\widehat{\mathcal R}_1$ and $\widehat{\mathcal R}_2$,
are combined through
\begin{equation}
\mathbb E
\left[
\widehat{\mathcal R}_1
\widehat{\mathcal R}_2
\right]
=
\left|
\widetilde{\mathcal R}_{\rm FP}^\theta(\bfxx_t,t)
\right|^2.
\end{equation}
This requires four additional spatially perturbed
score evaluations.
Temporal derivatives are further approximated using
three-point finite differences, requiring additional
energy evaluations at neighboring diffusion times.

Motivated by Theorem~\ref{thm:distribution_error},
STEC directly supervises temporal energy derivatives
through a tractable conditional objective.
With finite-difference evaluation, this requires energy
evaluations at neighboring diffusion times while keeping
the molecular coordinates fixed.
As summarized in Table~\ref{tab:consistency_derivative_order},
STEC reuses the spatial score computation from DSM
without explicitly evaluating additional second-order
spatial derivatives.

\begin{table}[ht]
\centering
\small
\renewcommand{\arraystretch}{1.3}
\setlength{\tabcolsep}{5pt}
\caption{
Energy evaluations and explicit input derivatives used
by FP regularization, ScoreMD, and STEC.
For ScoreMD, $\mathbf z_1,\mathbf z_2
\sim\mathcal N(\mathbf 0,\mathbf I)$ are independent
perturbation vectors, and $\eta$ is the perturbation scale.
The listed scalar energy evaluations use the variant
that evaluates non-divergence residual terms at
the positive perturbations.
The backward and forward temporal steps are
$h_s$ and $h_d$, respectively.
For STEC, $t_-=\max(t-\Delta t,0)$.
Derivative orders refer to input differentiation before
backpropagation with respect to model parameters.
The entries list required quantities rather than
separate model forward-pass counts.
}
\label{tab:consistency_derivative_order}

\begin{tabular}{lccc}
\hline
Order & Exact FP & ScoreMD & STEC \\
\hline

0th-order
&
\begin{tabular}[c]{@{}c@{}}
$u^\theta_t(\bfxx_t)$
\end{tabular}
&
\begin{tabular}[c]{@{}c@{}}
$u^\theta_{t+h_d}(\bfxx_t+\eta\mathbf z_1)$\\
$u^\theta_t(\bfxx_t+\eta\mathbf z_1)$\\
$u^\theta_{t-h_s}(\bfxx_t+\eta\mathbf z_1)$\\
$u^\theta_{t+h_d}(\bfxx_t+\eta\mathbf z_2)$\\
$u^\theta_t(\bfxx_t+\eta\mathbf z_2)$\\
$u^\theta_{t-h_s}(\bfxx_t+\eta\mathbf z_2)$
\end{tabular}
&
\begin{tabular}[c]{@{}c@{}}
$u^\theta_t(\bfxx_t)$\\
$u^\theta_{t_-}(\bfxx_t)$
\end{tabular}
\\
\hline

1st-order
&
\begin{tabular}[c]{@{}c@{}}
$\nabla_{\bfxx_t}u^\theta_t(\bfxx_t)$\\
$\partial_tu^\theta_t(\bfxx_t)$
\end{tabular}
&
\begin{tabular}[c]{@{}c@{}}
$\nabla_{\bfxx_t}u^\theta_t(\bfxx_t)$\\
$\nabla_{\bfxx_t}u^\theta_t
(\bfxx_t+\eta\mathbf z_1)$\\
$\nabla_{\bfxx_t}u^\theta_t
(\bfxx_t-\eta\mathbf z_1)$\\
$\nabla_{\bfxx_t}u^\theta_t
(\bfxx_t+\eta\mathbf z_2)$\\
$\nabla_{\bfxx_t}u^\theta_t
(\bfxx_t-\eta\mathbf z_2)$
\end{tabular}
&
\begin{tabular}[c]{@{}c@{}}
$\nabla_{\bfxx_t}u^\theta_t(\bfxx_t)$
\end{tabular}
\\
\hline

2nd-order
&
$\Delta_{\bfxx_t}u^\theta_t(\bfxx_t)$
&
\begin{tabular}[c]{@{}c@{}}
None
\end{tabular}
&
\begin{tabular}[c]{@{}c@{}}
None
\end{tabular}
\\
\hline
\end{tabular}
\end{table}

\paragraph{Why temporal energy consistency improves endpoint potentials.}

The effectiveness of STEC originates from the complementary roles of
spatial and temporal supervision in controlling the learned energy along
the diffusion path. The DSM objective optimizes an averaged denoising
error over diffusion times and constrains the spatial score
$-\nabla_{\bfxx}u^\theta_{t}$ throughout the perturbation
trajectory. However, a small value of this averaged objective does not
necessarily imply uniformly small score errors near the clean endpoint
$t=0$. In particular, the low-noise regime is intrinsically challenging
for diffusion models, where the score field becomes increasingly
sensitive to local distribution details
\citep{kim2022softtruncation,Yang2023LipschitzSI}. Consequently, DSM
may allow localized score deviations near the endpoint while maintaining
a small overall training loss.

STEC provides a complementary constraint by supervising the temporal
evolution of the same energy field. Unlike spatial score matching, which
directly constrains the local geometry of the energy landscape at each
diffusion time, temporal supervision restricts how the energy changes
along the diffusion trajectory. A deviation in the temporal derivative
affects the endpoint energy through its accumulation over diffusion time,
\begin{equation}
u^\theta_{0}(\bfxx)
-
u^\theta_{t}(\bfxx)
=
-\int_0^t
\partial_s u^\theta_{s}(\bfxx)
\,\mathrm ds .
\end{equation}
Therefore, temporal matching suppresses endpoint distortions by enforcing
consistency of the energy evolution over the diffusion path, rather than
requiring the endpoint score to be accurately recovered solely from local
low-noise observations.

A natural question is whether the endpoint accuracy can be improved by
simply increasing the emphasis on the low-noise regime in DSM training.
However, ScoreMD~\citep{plainer2025consistent} has explored this strategy
by introducing a low-noise-focused objective for EBDMs, where additional
supervision is applied to diffusion times close to the clean endpoint.
Nevertheless, accurately recovering the endpoint energy landscape remains
challenging, suggesting that stronger spatial score supervision alone may
not fully resolve the inconsistency accumulated along the diffusion path.
STEC addresses this limitation from a different perspective by directly
constraining the temporal evolution of the energy field. Temporal
supervision provides complementary information by restricting the
accumulated change of the energy along the diffusion trajectory, rather
than further emphasizing local score estimation. Therefore, STEC
complements DSM by controlling the missing temporal consistency of the
learned energy field.

\section{Implementation Details}
\label{appx:imple}
\subsection{Molecular Systems and Dataset Splits}
\label{appx:si_datasets}

We consider molecular benchmarks spanning both system-specific and
transferable coarse-grained force-field learning. In the
system-specific setting, a separate model is trained for each molecular
system and evaluated against the corresponding reference ensemble. In
the transferable setting, a single model is trained across multiple
peptide identities and evaluated on molecules that are entirely absent
from model fitting. ALDP and the fast-folding proteins are used for the thermodynamic
evaluations in \figref{res-iid}, with the protein datasets
additionally used for the dynamical analyses in
\figref{trajectory_transition}.
The dipeptide and octapeptide benchmarks are used for the
transferability evaluations in \figref{peptide_generalization}. Table~\ref{tab:si_dataset_summary} summarizes the
molecular representations, simulation temperatures, and data
partitions used throughout the study.

\paragraph{Alanine dipeptide.}
We use the public alanine dipeptide (ALDP) dataset distributed with
\texttt{bgmol}, following previous diffusion-based coarse-graining
studies~\citep{kohler2021smooth,arts2023,plainer2025consistent}.
The dataset consists of four independent 500~ns atomistic trajectories
generated in implicit solvent at 300~K using the Amber ff99SB-ILDN force
field and openMM~\citep{eastman2017openmm}, with 250,000 configurations
saved from each trajectory. We randomly select 50,000 configurations
from the aggregated ensemble for model training. Each configuration is
mapped to five backbone sites,
$[\mathrm{C},\mathrm{N},\mathrm{C}_{\alpha},\mathrm{C},\mathrm{N}]$,
which retain the backbone degrees of freedom required to resolve the
major conformational states in the $(\phi,\psi)$ dihedral space.
Projected atomistic forces are used only for the force-matching
baselines in the controlled ALDP comparison.

\paragraph{Fast-folding proteins.}
Chignolin, Trp-cage, BBA, and Villin are taken from the fast-folding
trajectories of Lindorff-Larsen et al.\
\citep{deshaw2011fastfolding}. Each protein is represented by one
$\mathrm{C}_{\alpha}$ bead per residue, giving 10, 20, 28, and 35 CG
sites, respectively. The reference trajectories span 106~$\mu$s for
Chignolin, 208~$\mu$s for Trp-cage, 223~$\mu$s for BBA, and
125~$\mu$s for Villin, at the temperatures listed in
Table~\ref{tab:si_dataset_summary}. Each mapped trajectory is divided
at the frame level into 80\% training, 10\% validation, and 10\% test
sets. Separate energy models are trained for the four proteins.

\paragraph{Dipeptides.}
We use the two-amino-acid (2AA) dipeptide dataset introduced by Klein et al.~\citep{klein2023timewarp}. The dataset contains implicit-solvent molecular dynamics simulations of diverse dipeptide identities constructed from combinations of the standard amino acids. Each dipeptide contains 49,000 configurations, with training and validation trajectories simulated for 50~ns and test trajectories simulated for 1~$\mu$s using the Amber14 force field at 310~K with openMM~\citep{eastman2017openmm}. Each residue is represented by five CG sites,
$[\mathrm{N},\mathrm{C}_{\alpha},\mathrm{C}_{\beta},\mathrm{C},\mathrm{O}]$,
resulting in up to ten CG sites per dipeptide. Dipeptides containing glycine are excluded because glycine lacks a $\mathrm{C}_{\beta}$ atom.

The dataset is split by molecular identity, with 175 dipeptides assigned to training, 75 to validation, and 92 to testing. All configurations from the same dipeptide are assigned to a single split to prevent information leakage across molecular identities. For the transferability evaluation in
\figref{peptide_generalization}a, we use the same 15 held-out test dipeptides as ScoreMD~\citep{plainer2025consistent}:
AC, AP, AT, ET, HP, HT, IM, KQ, KS, LW, NF, NY, RL, RV, and TD.

\paragraph{Octapeptides.}
We use the octapeptide dataset introduced by Charron et al.~\citep{charron2025navigating}. The dataset contains 1,100 reference octapeptide systems generated from diverse peptide sequences, with approximately $1~\mu$s of aggregate atomistic sampling for each system. We exclude the associated decoy groups and use only the reference peptide ensembles for training and evaluation. The retained dataset contains 22,657,940 configurations. We use the CG coordinates and bead labels provided with the dataset, with each octapeptide containing 35--40 CG beads depending on sequence. Systems are split by molecular identity into 550 training, 220 validation, and 330 test peptides. Variable-size systems are padded to 40 nodes for batch processing, with a node mask applied to exclude padded nodes during model operations.

For the quantitative transferability comparisons in
\figref{peptide_generalization}b, we use a fixed subset consisting of the first 20 peptides under the fixed ordering of the held-out test split.
OPEP0014 and OPEP0027 are shown in the same figure as representative free-energy landscapes.

\begin{table}[t]
\centering
\caption{
Summary of the molecular benchmarks used in this work. The
system-specific trajectory datasets are partitioned or sampled at the
frame level, whereas the dipeptide and octapeptide benchmarks are
partitioned by molecular identity. ALDP uses 50,000 trajectory frames
for system-specific training.
}
\label{tab:si_dataset_summary}
\small
\setlength{\tabcolsep}{5pt}
\begin{tabular}{lcccl}
\toprule
Benchmark
& CG sites
& Temperature (K)
& Split unit
& Data partition \\
\midrule
ALDP
& 5
& 300
& Frames
& 50,000 training frames \\

Chignolin
& 10
& 340
& Frames
& 80\% / 10\% / 10\% \\

Trp-cage
& 20
& 290
& Frames
& 80\% / 10\% / 10\% \\

BBA
& 28
& 325
& Frames
& 80\% / 10\% / 10\% \\

Villin
& 35
& 360
& Frames
& 80\% / 10\% / 10\% \\

Dipeptides
& 10
& 310
& Systems
& 175 / 75 / 92 \\

Octapeptides
& 35--40
& 300
& Systems
& 550 / 220 / 330 \\
\bottomrule
\end{tabular}

\vspace{2pt}
\begin{minipage}{0.94\linewidth}
\footnotesize
For entries with three values, the order is training / validation /
test.
\end{minipage}
\end{table}

\subsection{Model Architecture and Training}
\label{appx:si_model_training}

This subsection provides details of the graph-transformer energy
architecture and training setup outlined in section
\nameref{sec:method-imple}. Within each benchmark, standard EBDM, ScoreMD, and STEC use the same
molecular representation and conservative energy architecture. The
methods differ primarily in the objective used to constrain the learned
diffusion path. This matched design allows differences in the resulting
CG force fields to be attributed to the training objective rather than
to model capacity. Unless otherwise stated, the models are optimized
with AdamW and cosine learning-rate decay.

\paragraph{Molecular representation and energy model.}
A molecular configuration is represented as
$\bfxx=\{\bfxx_i\}_{i=1}^{n}$, where
$\bfxx_i\in\mathbb{R}^{3}$ denotes the Cartesian coordinate of CG
site $i$. Each site corresponds to one graph node. The initial node
representation combines a diffusion-time embedding with the categorical
descriptors available for the corresponding molecular representation,
\begin{equation}
h_i^{(0)}
=
\operatorname{concat}
\left[
E_{\mathrm{type}}(z_i),
E_{\mathrm{index}}(b_i),
E_{\mathrm{res}}(r_i),
E_{\mathrm{aa}}(a_i),
\gamma(t)
\right],
\label{eq:appx_node_features}
\end{equation}
where $z_i$ denotes the bead or atom type, $b_i$ the bead index,
$r_i$ the residue index, $a_i$ the amino-acid identity, and
$\gamma(t)$ the diffusion-time embedding. Features that are not
available or required for a particular benchmark are omitted. The
dipeptide model uses atom identity, atom index, residue
index, and amino-acid identity. The octapeptide model uses bead type, bead
index, and residue index, together with the node mask introduced for
the padded molecular representation.

We use the graph-transformer energy architecture adopted in previous
diffusion-based CG models
\citep{arts2023,shi2021,plainer2025consistent}. Molecular geometry is
encoded using relative coordinate vectors
\begin{equation}
\mathbf{r}_{ij}
=
\bfxx_i-\bfxx_j,
\label{eq:appx_relative_displacement}
\end{equation}
which remove dependence on the absolute molecular position. The node
representations are updated through $L$ graph-transformer blocks,
\begin{equation}
h^{(\ell+1)}
=
\mathcal{G}_{\ell}
\left(
h^{(\ell)},\{\mathbf{r}_{ij}\}
\right),
\qquad
\ell=0,\ldots,L-1,
\label{eq:appx_graph_transformer}
\end{equation}
where $\mathcal{G}_{\ell}$ denotes one graph-transformer block.

A shared decoder maps the final representation of each physical node to
a scalar contribution. The dimensionless molecular energy is obtained
by summation,
\begin{equation}
u_\theta(\bfxx,t)
=
\sum_{i=1}^{n}
\psi
\left(
h_i^{(L)}
\right).
\label{eq:appx_scalar_output}
\end{equation}
For variable-size systems such as octapeptides,
\begin{equation}
u_\theta(\bfxx,t)
=
\sum_{i=1}^{n_{\max}}
m_i
\psi
\left(
h_i^{(L)}
\right),
\label{eq:appx_masked_output}
\end{equation}
where $m_i=1$ for a physical node and $m_i=0$ for padding. The mask is
also applied to attention and coordinate derivatives so that padded
nodes do not contribute to the learned energy or force.

The spatial score is obtained by differentiating the scalar energy,
\begin{equation}
\mathbf{s}_\theta(\bfxx,t)
=
-\nabla_{\bfxx}
u_\theta(\bfxx,t).
\label{eq:appx_conservative_score}
\end{equation}
The resulting score field is therefore conservative by construction.
Random global rotations are applied during training because the graph
transformer is not explicitly $\mathrm{SO}(3)$ equivariant.

For octapeptides, bead type is represented by a learned
16-dimensional embedding, while bead and residue indices use
16-dimensional positional embeddings. Diffusion time is represented
by an embedding with 4 channels. Pair features additionally include
the relative residue-index offset together with the relative
coordinate vector. EBDM, ScoreMD, and STEC use the same graph
transformer with 4 blocks, hidden dimension 256, and 8 attention heads.

\paragraph{Training objectives and optimization.}
Standard EBDM uses only the denoising score matching objective
$\mathcal{L}_{\mathrm{DSM}}$ defined in
Eq.~\eqref{eq:methods_dsm}.
We retrain ScoreMD~\citep{plainer2025consistent} using DSM augmented
with its Fokker--Planck consistency objective.
STEC supervises the diffusion-time evolution of the learned energy
using the conditional temporal objective derived
in~\appxref{method-condition}. Combining this temporal objective with DSM gives the STEC loss
defined in Eq.~\eqref{eq:methods_stec_loss},
$\mathcal{L}_{\mathrm{STEC}}
=
\mathcal{L}_{\mathrm{DSM}}
+
\lambda_{\mathrm{time}}
\mathcal{L}_{\mathrm{time}}.$

Within each benchmark, the three methods use the same training data,
dataset splits, molecular representation, and graph-transformer
architecture. Separate models are trained for ALDP and each
fast-folding protein, whereas each peptide benchmark uses a single
transferable model per objective.
All models use AdamW~\citep{loshchilov2019adamw} with the same cosine learning-rate decay
schedule within each benchmark. The graph transformers use
8 attention heads and a 16-dimensional feature embedding.

For ALDP, Chignolin, BBA, and dipeptides, we reproduce the published
ScoreMD setup using the same dataset splits and training
hyperparameters, including the FP regularization weights.
Since the original ScoreMD study did not consider Trp-cage, Villin or
octapeptides, we adopt the FP regularization weight from BBA
($\lambda_{\mathrm{FP}}=10^{-2}$), the closest protein benchmark in the number of CG sites.
Architecture and training settings are summarized
in Table~\ref{tab:si_training_summary}.

\paragraph{Temporal-derivative evaluation.}
The temporal energy derivative is evaluated at fixed molecular
coordinates using a first-order backward finite difference,
\begin{equation}
\partial_tu_\theta(\bfxx_t,t)
\approx
\frac{
u_\theta(\bfxx_t,t)
-
u_\theta(\bfxx_t,t_-)
}{
\Delta t_{\mathrm{eff}}
},
\qquad
t_-=\max(t-\Delta t,0),
\qquad
\Delta t_{\mathrm{eff}}=t-t_-,
\label{eq:si_stec_dt}
\end{equation}
where the nominal step is $\Delta t=10^{-3}$.
For $t<\Delta t$, the earlier time is clipped to zero and the
denominator is reduced accordingly to $\Delta t_{\mathrm{eff}}=t$,
avoiding evaluations at negative diffusion times.
In the implementation, the denominator is additionally bounded
below by $10^{-12}$ for numerical safety.
The molecular coordinates $\bfxx_t$ remain unchanged in both
energy evaluations.

We compare finite-difference evaluation with automatic
differentiation during STEC training on BBA. Both runs use the
same energy model, training data, and initialization.
The two methods yield comparable PMF errors and
Jensen--Shannon divergences for both reverse-diffusion samples
and Langevin CGMD ensembles.
In a separate single-GPU benchmark for this BBA case on an
NVIDIA A100 with batch size 32, finite-difference evaluation
reduces the mean time per optimization step by 10.8\% and the
baseline-subtracted peak device memory by 23.6\% relative to
automatic differentiation.

\begin{table}[t]
\centering
\caption{
Architecture and training settings for the diffusion-based models.
}
\label{tab:si_training_summary}
\small
\setlength{\tabcolsep}{4pt}
\renewcommand{\arraystretch}{1.15}

\begin{tabular*}{\linewidth}{@{\extracolsep{\fill}}lccccccc@{}}
\toprule
& \multicolumn{2}{c}{Architecture}
& \multicolumn{3}{c}{Optimization}
& \multicolumn{2}{c}{Loss weights} \\
\cmidrule(lr){2-3}
\cmidrule(lr){4-6}
\cmidrule(lr){7-8}
Benchmark
& Hidden dim.
& Blocks
& Epochs
& Batch size
& Learning rate
& ScoreMD $\lambda_{\mathrm{FP}}$
& STEC $\lambda_{\mathrm{time}}$ \\
\midrule
ALDP
& 96 & 2 & 10,000 & 1,024
& $3\times10^{-4}$ & $5\times10^{-4}$ & $0.1$ \\

Chignolin
& 128 & 3 & 2,000 & 1,024
& $10^{-4}$ & $2\times10^{-3}$ & $0.1$ \\

Trp-cage
& 128 & 3 & 1,500 & 1,024
& $10^{-4}$ & $10^{-2}$ & $0.5$ \\

BBA
& 128 & 3 & 1,500 & 1,024
& $10^{-4}$ & $10^{-2}$ & $0.5$ \\

Villin
& 128 & 3 & 1,500 & 1,024
& $10^{-4}$ & $10^{-2}$ & $0.5$ \\

\addlinespace[3pt]
Dipeptides
& 128 & 3 & 120 & 1,024
& $3\times10^{-4}$ & $5\times10^{-4}$ & $0.1$ \\

Octapeptides
& 256 & 4 & 200 & 1,024\textsuperscript{*}
& $10^{-4}$ & $10^{-2}$ & $0.5$ \\
\bottomrule
\end{tabular*}

\medskip
\begin{minipage}{\linewidth}
\footnotesize
\textsuperscript{*}For octapeptides, EBDM and STEC use a global batch
size of 1,024. ScoreMD uses 256 because of the higher memory
requirement of the FP objective.
\end{minipage}
\end{table}

\paragraph{Diffusion schedule.}
Under the Gaussian perturbation kernel defined in
Eq.~\eqref{eq:forward-process}, noisy configurations are sampled as
$\bfxx_t=\alpha(t)\bfxx_0+\sigma(t)\boldsymbol\epsilon$,
with $\boldsymbol\epsilon\sim\mathcal N(0,\mathbf I)$.
The functions $\alpha(t)$ and $\sigma(t)$ control the signal and
noise scales, respectively.
For all diffusion-based models, we use the same variance-preserving
schedule,
\begin{equation}
\alpha(t)
=
\exp\left(-0.05t-4.975t^2\right),
\qquad
\sigma(t)
=
\sqrt{1-\alpha^2(t)}.
\label{eq:si_vp_schedule}
\end{equation}

These scales can equivalently be expressed through the linear
noise schedule $\beta(t)$,
\begin{equation}
\begin{aligned}
\beta(t)
&=
0.1+19.9t,
\\
\alpha(t)
&=
\exp\left[
-\frac{1}{2}\int_0^t\beta(s)\,\mathrm ds
\right],
\\
\sigma(t)
&=
\sqrt{
1-\exp\left[-\int_0^t\beta(s)\,\mathrm ds\right]
}.
\end{aligned}
\label{eq:appx_vp_beta_relation}
\end{equation}
Substituting these scales into the general relations for the forward
SDE gives the drift and diffusion coefficients
\begin{equation}
\begin{aligned}
f(\bfxx,t)
&=
\frac{\alpha'(t)}{\alpha(t)}\bfxx
=
-\frac{1}{2}\beta(t)\bfxx,
\\
g(t)
&=
\left[
\frac{\mathrm d\sigma^2(t)}{\mathrm dt}
-
2\frac{\alpha'(t)}{\alpha(t)}\sigma^2(t)
\right]^{1/2}
=
\sqrt{\beta(t)}.
\end{aligned}
\label{eq:appx_vp_sde}
\end{equation}

\paragraph{Diffusion-time weighting.}
For the DSM and conditional temporal objectives defined in
Eqs.~\eqref{eq:methods_dsm}
and~\eqref{eq:methods_conditional_time_loss},
we use the same diffusion-time dependence, with an additional
dimensional normalization in the temporal objective:
\begin{equation}
w(t)
=
9n^2\,w_{\mathrm{time}}(t)
=
\sigma^2(t)
\begin{cases}
\dfrac{5}{3}, & 0 \le t \le 0.3, \\[6pt]
\dfrac{5}{7}, & 0.3 < t \le 1,
\end{cases}
\label{eq:si_time_weight}
\end{equation}
where $n$ is the number of CG sites.
The piecewise prefactors are inversely proportional to the
interval lengths and each integrate to $1/2$ over the
corresponding interval, compensating for their unequal
lengths before modulation by $\sigma^2(t)$.

In the implementation, both residuals are multiplied by
$\sigma(t)$ before squaring, and the resulting squared errors
are weighted by the corresponding piecewise prefactors.
The temporal squared error is additionally divided by
$(3n)^2=9n^2$, with this normalization included in
$w_{\mathrm{time}}(t)$.
These operations yield the weights in
Eq.~\eqref{eq:si_time_weight}.
The overall coefficient $\lambda_{\rm time}$ separately
controls the relative strength of temporal supervision.

\paragraph{Force-matching baselines.}
For the controlled ALDP comparison in \figref{res-iid}a, we train CGNet~\citep{wang2019machine} and an
architecture-matched graph-transformer force-matching model as baselines. Both baselines are trained using projected
atomistic forces, in contrast to the configuration-only supervision used by the diffusion models.

For CGNet, we adopt the ALDP force-matching implementation and baseline setup provided with the Flow-Matching study~\citep{kohler2023flow}. The model is trained using the same
50,000 mapped configurations employed by the diffusion models, with
10,000 configurations used for validation.

The architecture-matched force-matching baseline uses the same
five-site ALDP representation and the same conservative
graph-transformer model family as the diffusion models. It contains two
transformer blocks with hidden dimension 96 and feature dimension 16,
matching the corresponding ALDP energy architecture. The model is
trained for 5,000 epochs with a batch size of 512 and an initial
learning rate of $3\times10^{-4}$. Training uses random rotational
augmentation, net-force removal, and the harmonic priors adopted in the
ALDP force-matching setup. This baseline is denoted FM (GT) in the main
text.

\subsection{Sampling and Simulation Protocols}
\label{appx:si_sampling}

After training, each diffusion-based model is evaluated through two sampling routes: reverse-diffusion sampling and Langevin CGMD driven by the learned endpoint force field.
The CGMD ensembles are used for the thermodynamic, dynamical, and transferability evaluations in
\figref{res-iid}, \figref{trajectory_transition}, and \figref{peptide_generalization}, respectively. Within each benchmark, standard EBDM, ScoreMD, and STEC use identical sampling and simulation protocols; only the learned energy model differs. We use $t$ exclusively to denote diffusion time and denote the molecular-dynamics integration timestep by $\Delta t_{\mathrm{MD}}$. 

\paragraph{Reverse-diffusion sampling.}
Approximately independent samples are generated from the trained diffusion models by integrating the reverse-time VP-SDE using the
Euler--Maruyama solver with 1,000 diffusion-time steps. Within each benchmark, EBDM, ScoreMD, and STEC use the same reverse-diffusion protocol.
For each method and evaluation run, we generate 100,000 samples for ALDP, 427,794 for Chignolin, 1,044,000 for Trp-cage, 1,300,156 for BBA, and 502,326 for Villin. For the dipeptide benchmark, we generate 49,000 samples per held-out peptide.
For the octapeptide benchmark, we generate 20,000 samples per peptide for the fixed 20-peptide evaluation subset specified in ~\appxref{si_datasets}.
The comparison between reverse-diffusion and CGMD ensembles for ALDP and the fast-folding proteins is reported in~\appxref{si_sampling_simulation}.

\paragraph{Langevin integrator.}
We perform NVT dynamics with the Langevin integrator as implemented in OpenMM~\citep{eastman2017openmm}. Let $\bfxx_n$ and $\mathbf v_n$ denote the CG coordinates and velocities at simulation step $n$, respectively. Writing $a=\exp(-\gamma\Delta t_{\mathrm{MD}})$, the integration update is
\begin{align}
\mathbf v_{n+1}
&=
a\mathbf v_n
+
\frac{1-a}{\gamma}
\mathbf M^{-1}
\mathbf F_\theta^\cg(\bfxx_n)
+
\sqrt{k_{\mathrm B}T(1-a^2)}
\,\mathbf M^{-1/2}\mathbf R_n,
\nonumber\\
\bfxx_{n+1}
&=
\bfxx_n
+
\Delta t_{\mathrm{MD}}
\mathbf v_{n+1},
\label{eq:si_langevin_update}
\end{align}
where $\gamma$ is the friction coefficient, $\mathbf M$ is the diagonal mass matrix, $\mathbf R_n\sim\mathcal N(\mathbf 0,I)$, $k_{\mathrm B}$ is the Boltzmann constant, and $T$ is the simulation temperature. To match the force-evaluation setting of
ScoreMD~\citep{plainer2025consistent}, we use
$t_{\mathrm{eval}}=10^{-5}$ for standard EBDM, ScoreMD, and STEC.
The forces are computed as
$\mathbf F_\theta^\cg(\bfxx)
=
-k_{\mathrm B}T
\nabla_{\bfxx}u_\theta(\bfxx,t_{\mathrm{eval}}).$
In our tests, direct evaluation at $t=0$ yields nearly identical
results, with no numerical instability observed. For the Two for One comparison
in~\appxref{si_two_for_one}, the EBDM force field is evaluated
at $t_{\mathrm{eval}}=0.005$,
following finite-time force extraction~\citep{arts2023}.

\paragraph{Alanine dipeptide and fast-folding proteins.}
For alanine dipeptide and the four fast-folding proteins, Langevin CGMD is performed using $\Delta t_{\mathrm{MD}}=2~\mathrm{fs}$ with 100 parallel trajectories. Initial conformations are sampled from the corresponding reference CG ensemble. The simulation temperature is matched to the reference dataset: 300~K for alanine dipeptide, 340~K for Chignolin, 290~K for Trp-cage, 325~K for BBA, and 360~K for Villin. Across the parallel trajectories, the aggregate generated simulation time is $1.2~\mu\mathrm{s}$ for alanine dipeptide, $855.6~\mathrm{ns}$ for Chignolin, and $200~\mathrm{ns}$ for each of Trp-cage, BBA, and Villin.

For alanine dipeptide, the friction coefficient and particle masses are taken from the openMM integrator and system definition associated with the reference dataset. For the fast-folding proteins, we use $\gamma=1~\mathrm{ps}^{-1}$, and the bead masses are assigned from the atomic masses of the retained $C_\alpha$ atoms in the corresponding topology. For these system-specific benchmarks, reported uncertainty is obtained from three independent evaluations of the same trained model using different inference seeds. The seeds independently affect reverse-diffusion sampling and Langevin simulation, while model parameters are held fixed across evaluations.

\paragraph{Dipeptides.}
For each held-out dipeptide, we perform Langevin CGMD for 30~ns using a molecular timestep of $\Delta t_{\mathrm{MD}}=0.5~\mathrm{fs}$. Simulations are initialized from 10 randomly selected conformations from the corresponding reference ensemble and are subsequently downsampled to 49,000 configurations for evaluation. We use a friction coefficient of $\gamma=1~\mathrm{ps}^{-1}$, and masses are assigned according to the retained atom identities $[\mathrm N,\mathrm C_\alpha,\mathrm C_\beta,\mathrm C,\mathrm O]$ using their corresponding atomic masses. Unlike the single-system benchmarks, uncertainty for the dipeptide results reflects variation across the 15 held-out molecular systems rather than repeated inference seeds.

\paragraph{Octapeptides.}
For each of the first 20 peptides under the fixed ordering of the held-out  test split, we run 20 parallel Langevin trajectories at 300~K. Initial coordinates are selected uniformly without replacement from 20 configurations of the corresponding reference dataset, and initial velocities are sampled from the Maxwell--Boltzmann distribution. Each trajectory is propagated with $\Delta t_{\mathrm{MD}}=2~\mathrm{fs}$ and $\gamma=1~\mathrm{ps}^{-1}$. One configuration is saved every 20 integration steps, corresponding to a sampling interval of $0.04~\mathrm{ps}$. We retain 50,000 saved configurations per trajectory, corresponding to 2~ns of simulated time per trajectory and an aggregate duration of 40~ns per peptide.
The octapeptide simulations use $k_{\mathrm B}T=2.4943~\mathrm{kJ\,mol^{-1}}$ and assign unit mass to every retained CG site. Reported octapeptide summary statistics reflect variation across this fixed 20-peptide evaluation subset rather than repeated sampling seeds.

\subsection{Evaluation and Analysis Protocols}
\label{appx:si_evaluation}

Unless otherwise stated, equilibrium and structural metrics are
computed from the reference ensembles and Langevin CGMD trajectories
generated by the learned endpoint force fields.
For each metric and molecular system, all methods use the same
projection, discretization, and state definition.
Reverse-diffusion samples are additionally evaluated in the
sampling--simulation consistency analysis
in~\appxref{si_sampling_simulation}.

\paragraph{Equilibrium projections and distribution metrics.}
For the projected equilibrium comparisons in
\figref{res-iid} and \figref{peptide_generalization},
we use backbone dihedral coordinates $(\phi,\psi)$
for ALDP and dipeptides.
Fast-folding protein and octapeptide ensembles are projected
onto the first two time-lagged independent components (TICs).
For each metric and molecular system, the reference and CGMD
ensembles are histogrammed using the same coordinate ranges
and bin boundaries.
The normalized histograms define the reference and model
distributions $p_{\mathrm{ref}}$ and $p_{\mathrm{model}}$,
respectively.

Agreement between these distributions is quantified using
the Jensen--Shannon (JS) divergence~\citep{lin1991divergence},
\begin{equation}
\begin{aligned}
D_{\mathrm{JS}}
\left(p_{\mathrm{ref}}\,\|\,p_{\mathrm{model}}\right)
&=
\frac{1}{2}
D_{\mathrm{KL}}
\left(p_{\mathrm{ref}}\,\|\,p_{\mathrm{mix}}\right)
+
\frac{1}{2}
D_{\mathrm{KL}}
\left(p_{\mathrm{model}}\,\|\,p_{\mathrm{mix}}\right),
\\
p_{\mathrm{mix}}
&=
\frac{1}{2}
\left(p_{\mathrm{ref}}+p_{\mathrm{model}}\right).
\end{aligned}
\label{eq:si_js}
\end{equation}
For any two discrete distributions $p_X$ and $p_Y$,
the Kullback--Leibler (KL) divergence is
\begin{equation}
D_{\mathrm{KL}}(p_X\,\|\,p_Y)
=
\sum_b
p_{X,b}
\log\frac{p_{X,b}}{p_{Y,b}},
\label{eq:si_kl}
\end{equation}
where $b$ indexes histogram bins.
Natural logarithms are used, and terms with $p_{X,b}=0$
contribute zero.
Lower $D_{\mathrm{JS}}$ indicates closer agreement
with the reference equilibrium distribution.

\paragraph{PMF.}
PMF error complements the probability-level comparison
by measuring differences in the dimensionless free energies
of the reference and model distributions~\citep{plainer2025consistent}.
This metric is reported in \figref{res-iid}a,
\figref{efficiency}b, and~\appxref{si_sampling_simulation}.

Let $\mathcal B$ denote the union of occupied bins
in $p_{\mathrm{ref}}$ and $p_{\mathrm{model}}$
on the PMF grid.
Within $\mathcal B$, zero probabilities are replaced
by $10^{-6}$, yielding
$\tilde p_{\mathrm{ref},b}$ and
$\tilde p_{\mathrm{model},b}$.
The PMF error is
\begin{equation}
\begin{aligned}
E_{\mathrm{PMF}}
&=
\sum_{b\in\mathcal B}
w_b
\left[
\log\tilde p_{\mathrm{model},b}
-
\log\tilde p_{\mathrm{ref},b}
\right]^2,
\\
w_b
&=
\frac{
\tilde p_{\mathrm{ref},b}
+
\tilde p_{\mathrm{model},b}
}{
\sum_{c\in\mathcal B}
\left(
\tilde p_{\mathrm{ref},c}
+
\tilde p_{\mathrm{model},c}
\right)
}.
\end{aligned}
\label{eq:si_pmf}
\end{equation}
This dimensionless squared error uses the unshifted
log probabilities, without a final square root.

PMF grids contain 64 bins per axis for ALDP and dipeptides,
100 for fast-folding proteins, and 80 for octapeptides.
These totals include one overflow bin at each end
of each axis, retaining samples outside the central ranges.
The separate discretization used for conformational coverage
is described in~\appxref{si_computational_efficiency}.

\paragraph{Conformational projections and TICA.}
For ALDP, the fast-folding proteins, and dipeptides,
we follow the projection protocols of previous diffusion-based
CG studies~\citep{arts2023,plainer2025consistent}.
ALDP and dipeptides are analyzed in backbone dihedral space,
while fast-folding proteins are analyzed using
TICA~\citep{molgedey1994separation,perezhernandez2013identification,schwantes2013improvements}.

For octapeptides, we follow the TICA-based conformational
analysis of \citet{charron2025navigating}.
The input combines pairwise $\mathrm{C}_{\alpha}$ distances
with sine and cosine encodings of the backbone
$(\phi,\psi)$ angles, yielding a 56-dimensional feature vector.
For each protein and octapeptide, a separate TICA transformation
is fitted to the corresponding atomistic reference ensemble
and applied unchanged to all CGMD ensembles.
The first two TICs provide shared coordinates for the
landscapes in \figref{res-iid}b
and \figref{peptide_generalization}b. TICA is implemented using the \texttt{deeptime} library~\citep{hoffmann2021deeptime}.

\paragraph{Macrostate MAE.}
We use the macrostate mean absolute error
(mMAE)~\citep{lewis2025scalable}
to quantify discrepancies in relative macrostate free energies,
as reported in \figref{res-iid}b
and \figref{peptide_generalization}b.
Macrostates are defined using 3-state hidden Markov models
(HMMs)~\citep{noe2013projected} fitted to the reference ensembles.
The fitted models are held fixed when assigning reference
and generated configurations to macrostates,
providing a common state definition for all compared methods.

Macrostate probabilities $p_{\mathrm{ref},i}$ and
$p_{\mathrm{model},i}$ are estimated from sample counts,
with one pseudocount added to each macrostate
in the generated ensembles.
The relative macrostate free energies are
\begin{equation}
G_i^{X}
=
-k_{\mathrm B}T\log p_{X,i}+C_X,
\qquad
X\in\{\mathrm{ref},\mathrm{model}\},
\label{eq:si_macrostate_free_energy}
\end{equation}
where $C_X$ is chosen independently for each ensemble
such that $\min_i G_i^X=0$.
The mMAE is then
\begin{equation}
\mathrm{mMAE}
=
\frac{1}{K}
\sum_{i=1}^{K}
\left|
G_i^{\mathrm{model}}-G_i^{\mathrm{ref}}
\right|,
\qquad K=3.
\label{eq:si_mmae}
\end{equation}
Free energies are evaluated at the reference simulation
temperature $T$ of each system, and mMAE values are reported
in $\mathrm{kcal/mol}$.

\paragraph{Simulation stability.}
Simulation stability in \figref{res-iid}b and
\figref{si_two_for_one}c is evaluated using the
bonded-distance criterion of Fu et al.~\citep{Fu2022ForcesAN}.
A CGMD trajectory is considered unstable when any bonded
distance deviates from its corresponding equilibrium value
by more than $0.5$~\AA.
The stable simulation time is the elapsed simulation time
before the first occurrence of this criterion.
It characterizes the onset of local bond-length instability
rather than a molecular relaxation or kinetic timescale.

\paragraph{Transition probabilities and Hellinger distance.}
For the transition-matrix comparisons in
\figref{trajectory_transition}
and~\appxref{si_transition_matrices},
configurations are assigned to their nearest fixed center
in two-dimensional TICA space.
Chignolin, Trp-cage, and Villin use 3 states,
whereas BBA uses 4 states.
The same centers are used for the reference and model
ensembles within each protein.
These partitions are distinct from the HMM macrostates
used to evaluate mMAE.

For trajectory-derived matrices, transitions between adjacent
entries of the processed state sequences are counted.
Following Two for One~\citep{arts2023} and
ScoreMD~\citep{plainer2025consistent},
we estimate transition probabilities by normalizing
the outgoing counts for each state:
\begin{equation}
T_{ij}
=
\frac{N_{ij}}{\sum_k N_{ik}},
\label{eq:si_transition_matrix}
\end{equation}
where $N_{ij}$ is the number of observed transitions
from state $i$ to state $j$.
Rows without outgoing counts are set to zero.
For Trp-cage, the comparison uses the recorded matrices
underlying the figure.

The full matrices, including diagonal entries, are compared
using the Hellinger distance.
Its square-root transformation increases the relative
contribution of discrepancies in low-probability transitions
compared with squared differences between raw probabilities.
We add $10^{-12}$ to every matrix entry and normalize
each complete matrix to unit sum.
The resulting reference and model distributions are denoted
by $p_{\mathrm{ref}}$ and $p_{\mathrm{model}}$,
with entries indexed by $(i,j)$.
Their Hellinger distance is
\begin{equation}
H\left(p_{\mathrm{ref}},p_{\mathrm{model}}\right)
=
\frac{1}{\sqrt{2}}
\left[
\sum_{i,j}
\left(
\sqrt{p_{\mathrm{ref},ij}}
-
\sqrt{p_{\mathrm{model},ij}}
\right)^2
\right]^{1/2}.
\label{eq:si_hellinger}
\end{equation}
Lower Hellinger distance indicates closer agreement
with the reference transition-matrix structure.

\paragraph{Pairwise contacts and local-geometry statistics.}
For the structural comparisons
in~\appxref{si_structural_fidelity},
the contact probability $C_{ij}$ is the fraction of
configurations in which sites $i$ and $j$ are separated
by less than $1.0~\mathrm{nm}$.
Contact maps include all site pairs and display
$\log(C_{ij}+10^{-4})$.

Local geometry is assessed using $\mathrm{C}_{\alpha}$
distances at sequence offsets $k=1$, $2$, and $4$.
For each offset, distances are pooled across residue pairs
and 50,000 configurations per ensemble.
Histograms use 120 equal-width bins over
$3.2$--$4.4$~\AA,
$3.0$--$12.5$~\AA, and
$4.0$--$17.5$~\AA, respectively.
Out-of-range distances are excluded, and a pseudocount
of $10^{-12}$ is added to each bin before normalization.
The resulting model and reference distributions,
$p_{\mathrm{model}}^{(k)}$ and $p_{\mathrm{ref}}^{(k)}$,
are compared using
\begin{equation}
D_{\mathrm{KL}}^{(k)}
=
D_{\mathrm{KL}}
\left(
p_{\mathrm{model}}^{(k)}
\,\|\,
p_{\mathrm{ref}}^{(k)}
\right),
\label{eq:si_local_geometry_kl}
\end{equation}
with the KL divergence defined in Eq.~\eqref{eq:si_kl}.

\paragraph{Statistical aggregation and uncertainty.}
Where seed-based uncertainties are reported for ALDP
and fast-folding proteins, means and standard deviations
are obtained from three evaluations with different
inference seeds.
The trained model parameters remain fixed across evaluations,
as described in~\appxref{si_sampling}.
For dipeptides and octapeptides, aggregate statistics
summarize the 15 and 20 held-out molecular systems,
respectively.
Their reported spread reflects variation across molecular
identities rather than repeated inference seeds.
The evaluation subsets are specified
in~\appxref{si_datasets}.
\section{Additional Empirical Results}
\label{appx:add-results}

\subsection{Sampling--Simulation Consistency and Endpoint Equilibrium Fidelity}
\label{appx:si_sampling_simulation}

To complement the endpoint equilibrium comparisons
in~\figref{res-iid}, we examine agreement between
reverse-diffusion sampling and Langevin CGMD under
the same trained energy model.
Using the protocols in~\appxref{si_sampling},
we compare the resulting ensembles for ALDP
and the four fast-folding proteins.

\begin{figure}[t]
\centering
\includegraphics[width=\textwidth]{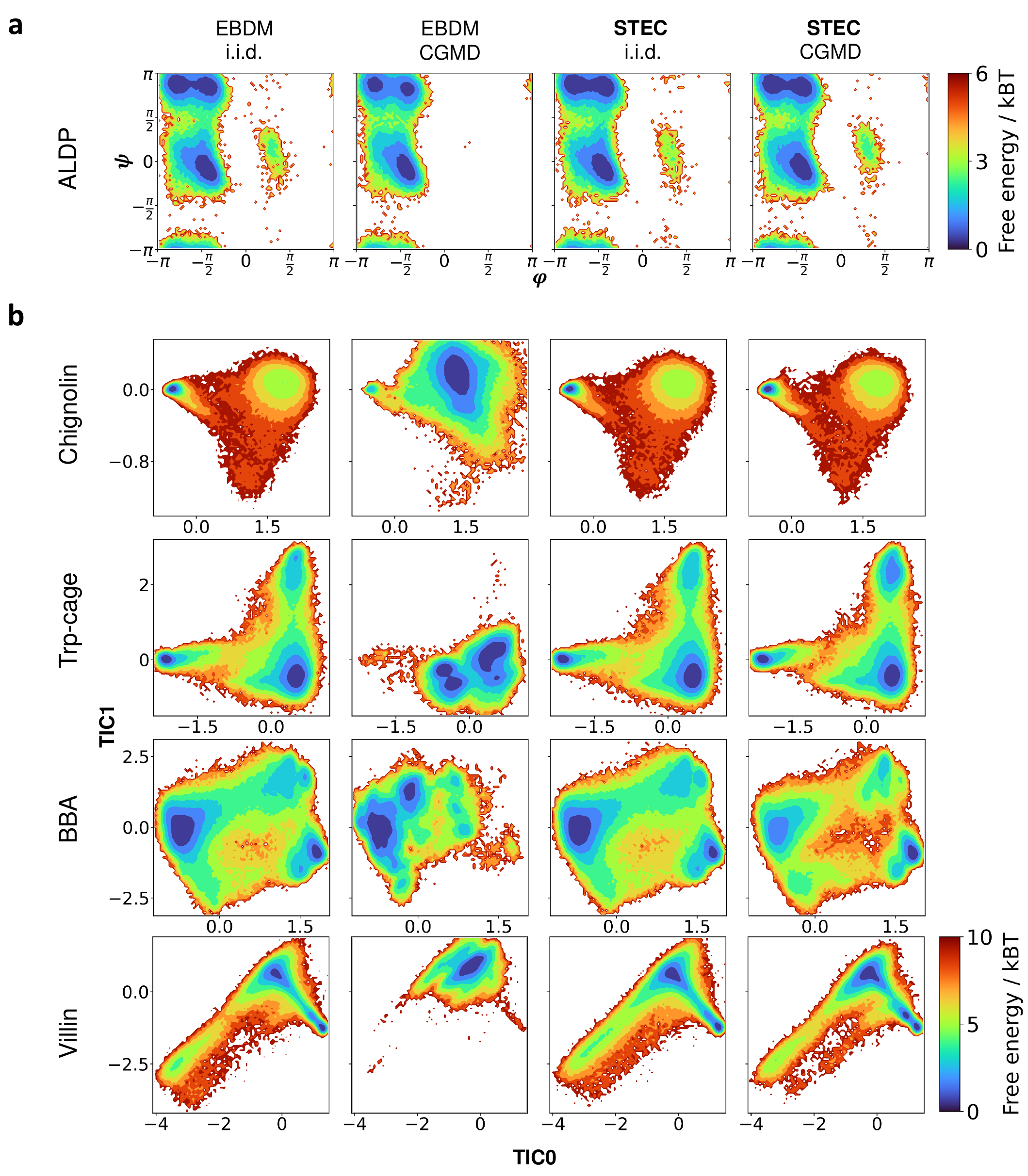}
\caption{
\textbf{STEC improves agreement between reverse-diffusion
and endpoint-CGMD ensembles.}
Free-energy landscapes obtained from approximately
independent reverse-diffusion samples and Langevin CGMD
driven by the corresponding endpoint force fields,
comparing standard EBDM and STEC.
Rows show ALDP, Chignolin, Trp-cage, BBA, and Villin.
ALDP ensembles are projected onto the backbone dihedral
coordinates $(\phi,\psi)$, while protein ensembles
are projected onto the first two TICs.
The same projection is used for both sampling routes
within each system.
Free energies are reported in units of $k_{\mathrm B}T$.
}
\label{fig:si_iid_cgmd}
\end{figure}

Standard EBDM exhibits pronounced differences between
the two sampling routes for the protein systems
(\figref{si_iid_cgmd}).
Reverse-diffusion samples cover broad conformational regions,
whereas endpoint CGMD substantially redistributes
their populations or restricts conformational coverage.
STEC yields closer agreement between the two ensembles,
preserving more similar basin populations and conformational
coverage, including for BBA and Villin.
These comparisons illustrate the distinction between
sampling through the diffusion model and sampling
under its endpoint potential.

We next assess endpoint-CGMD fidelity relative to the
reference ensembles using the metrics defined
in~\appxref{si_evaluation}
(Table~\ref{tab:equilibrium_quality}).
STEC achieves the lowest mean PMF error and
$D_{\mathrm{JS}}$ across all four fast-folding proteins.
The improvement is particularly pronounced for Villin,
where the mean errors are approximately 22-fold and
11-fold lower than those of ScoreMD, respectively.
ScoreMD reduces PMF error across the four proteins,
although its $D_{\mathrm{JS}}$ for Villin exceeds
that of standard EBDM.
For ALDP, ScoreMD gives the lowest mean errors,
with STEC achieving comparable values.
Together with \figref{si_iid_cgmd},
these results show that STEC improves both
sampling--simulation consistency and the fidelity
of endpoint-CGMD protein ensembles to the reference.

\begin{table*}[t]
\centering
\caption{
\textbf{STEC reduces CGMD distribution errors
across the fast-folding proteins.}
PMF error and $D_{\mathrm{JS}}$
relative to the reference ensembles are reported
for standard EBDM, ScoreMD, and STEC.
Metrics are evaluated in backbone dihedral space
for ALDP and two-dimensional TICA space
for the fast-folding proteins.
Values are mean $\pm$ standard deviation across
three CGMD evaluations with different sampling seeds,
using fixed trained model parameters.
Lower values indicate closer agreement with the reference,
and the lowest mean for each system and metric is bold.
}
\label{tab:equilibrium_quality}
\small
\setlength{\tabcolsep}{4.5pt}
\renewcommand{\arraystretch}{1.15}

\begin{tabular}{lcccccc}
\toprule
& \multicolumn{2}{c}{EBDM}
& \multicolumn{2}{c}{ScoreMD}
& \multicolumn{2}{c}{STEC} \\
\cmidrule(lr){2-3}
\cmidrule(lr){4-5}
\cmidrule(lr){6-7}
System
& PMF  $\downarrow$
& $D_{\mathrm{JS}}\downarrow$
& PMF  $\downarrow$
& $D_{\mathrm{JS}}\downarrow$
& PMF  $\downarrow$
& $D_{\mathrm{JS}}\downarrow$ \\
\midrule

ALDP
& $0.210 \pm 0.002$
& $0.0176 \pm 0.0001$
& $\mathbf{0.072 \pm 0.003}$
& $\mathbf{0.0066 \pm 0.0002}$
& $0.074 \pm 0.004$
& $0.0071 \pm 0.0003$ \\

Chignolin
& $15.693 \pm 0.070$
& $0.5011 \pm 0.0002$
& $0.138 \pm 0.018$
& $0.0162 \pm 0.0020$
& $\mathbf{0.047 \pm 0.001}$
& $\mathbf{0.0059 \pm 0.0001}$ \\

Trp-cage
& $10.147 \pm 0.452$
& $0.3369 \pm 0.0217$
& $0.282 \pm 0.009$
& $0.0288 \pm 0.0013$
& $\mathbf{0.200 \pm 0.056}$
& $\mathbf{0.0205 \pm 0.0037}$ \\

BBA
& $5.076 \pm 0.231$
& $0.1705 \pm 0.0069$
& $0.357 \pm 0.047$
& $0.0378 \pm 0.0051$
& $\mathbf{0.288 \pm 0.045}$
& $\mathbf{0.0273 \pm 0.0045}$ \\

Villin
& $12.275 \pm 0.396$
& $0.3215 \pm 0.0059$
& $6.648 \pm 0.381$
& $0.3701 \pm 0.0059$
& $\mathbf{0.298 \pm 0.022}$
& $\mathbf{0.0327 \pm 0.0023}$ \\

\bottomrule
\end{tabular}
\end{table*}

\subsection{Structural Fidelity Beyond Low-Dimensional Projections}
\label{appx:si_structural_fidelity}

The equilibrium comparisons in \figref{res-iid}
and~\appxref{si_sampling_simulation} characterize
conformational populations through low-dimensional projections.
We next examine the underlying intramolecular structure
using pairwise contact maps and sequence-separated
$\mathrm{C}_{\alpha}$ distance distributions,
following the definitions in~\appxref{si_evaluation}.

\begin{figure}[t]
\centering
\includegraphics[width=\textwidth]{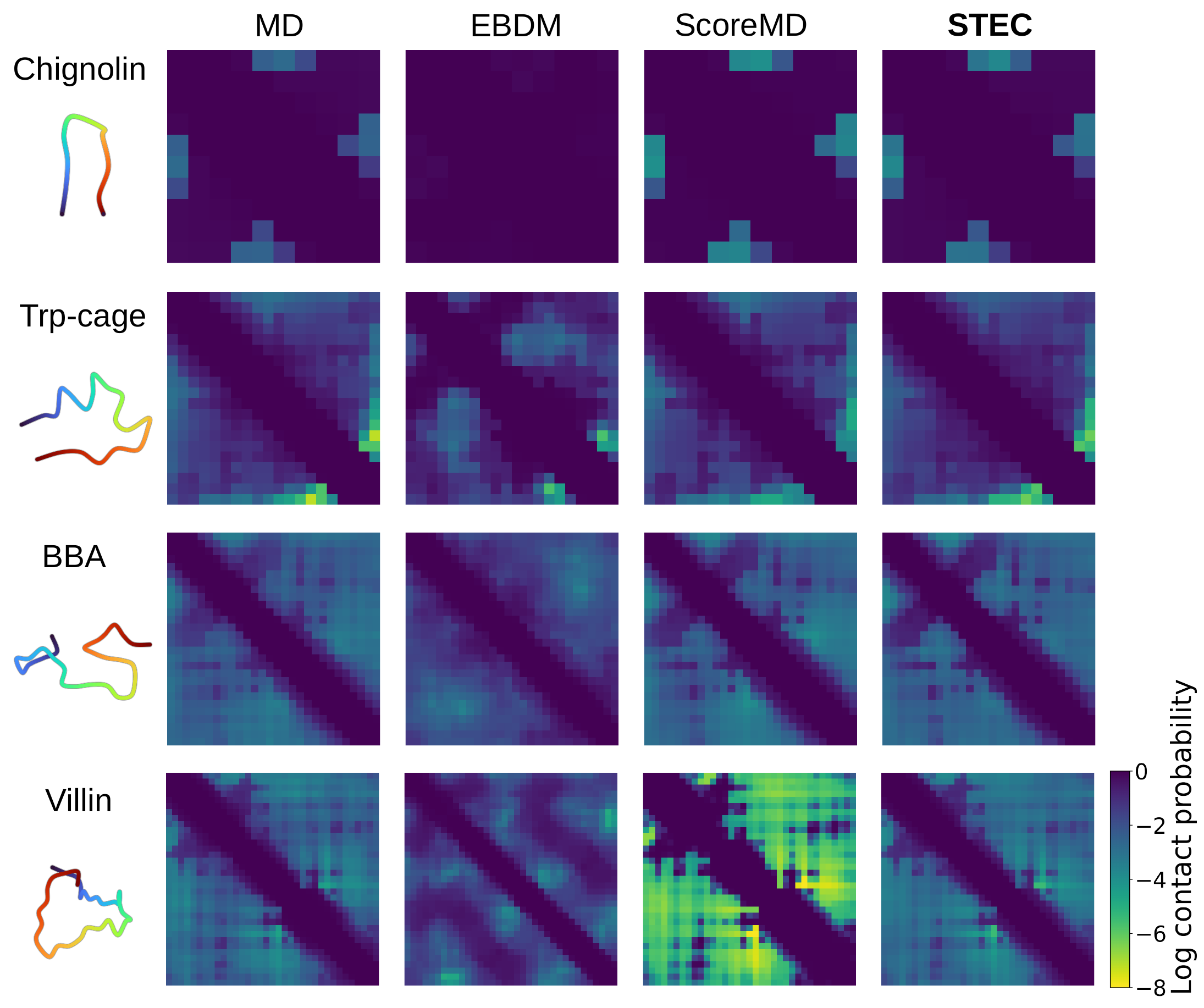}
\caption{
\textbf{STEC more faithfully recovers local and nonlocal
contact patterns in the fast-folding proteins.}
Contact-probability maps for Chignolin, Trp-cage,
BBA, and Villin, comparing the mapped atomistic MD reference
with CGMD ensembles generated by standard EBDM,
ScoreMD, and STEC.
Contacts are defined by a CG-site separation below
$1.0~\mathrm{nm}$.
Maps display $\log(C_{ij}+10^{-4})$,
where $C_{ij}$ is the contact probability
between sites $i$ and $j$.
}
\label{fig:si_contact_maps}
\end{figure}

\paragraph{Pairwise contact statistics.}
STEC more closely reproduces the reference contact patterns
across the four fast-folding proteins
(\figref{si_contact_maps}).
Standard EBDM captures some dominant local contacts
but exhibits substantial deviations in nonlocal contact
probabilities, consistent with its distorted TICA landscapes
in \figref{res-iid}b.
ScoreMD improves many of these patterns,
although discrepancies remain for the larger proteins.
STEC recovers both strong local contacts and weaker
long-range contacts more faithfully,
supporting the structural fidelity of its CGMD ensembles
beyond the projected equilibrium distributions.

\paragraph{Sequence-separated distance distributions.}
We further assess local and intermediate-range geometry
using $\mathrm{C}_{\alpha}$ distances at sequence offsets
$k=1$, $2$, and $4$.
Model and reference distributions are compared using
the KL divergence defined
in Eq.~\eqref{eq:si_local_geometry_kl},
with the sampling and histogram settings specified
in~\appxref{si_evaluation}.

STEC achieves the lowest offset-2 KL divergence
across all four proteins
(Table~\ref{tab:local_geometry_kl}).
It also gives the lowest offset-4 divergence
for Chignolin, Trp-cage, and Villin,
whereas ScoreMD performs best for BBA.
For adjacent distances, standard EBDM remains closest
to the reference for Trp-cage, BBA, and Villin,
despite its larger equilibrium-distribution errors.
Thus, accurate adjacent-distance statistics alone
do not ensure faithful conformational populations.
The structural comparisons complement the transition-matrix
analysis in~\appxref{si_transition_matrices},
which examines transitions between conformational states.

\begin{table}[t]
\centering
\caption{
\textbf{STEC consistently improves offset-2 distance
distributions across the fast-folding proteins.}
KL divergences
$D_{\mathrm{KL}}
(p_{\mathrm{model}}^{(k)}\,\|\,p_{\mathrm{ref}}^{(k)})$
are reported for $\mathrm{C}_{\alpha}$ distance distributions
at sequence offsets $k=1$, $2$, and $4$.
Distributions are normalized within the distance ranges
specified in~\appxref{si_evaluation}.
Lower values indicate closer agreement with the reference,
and the lowest value for each protein and offset is bold.
}
\label{tab:local_geometry_kl}
\small
\setlength{\tabcolsep}{7pt}
\renewcommand{\arraystretch}{1.10}

\begin{tabular}{llccc}
\toprule
System
& Method
& Adjacent
& Offset-2
& Offset-4 \\
&
&
$k=1$
&
$k=2$
&
$k=4$ \\
\midrule

Chignolin
& EBDM
& $0.0548$
& $0.1548$
& $0.9573$ \\
& ScoreMD
& $0.3289$
& $0.0060$
& $0.0078$ \\
& STEC
& $\mathbf{0.0388}$
& $\mathbf{0.0019}$
& $\mathbf{0.0040}$ \\

\midrule

Trp-cage
& EBDM
& $\mathbf{0.0478}$
& $0.0252$
& $0.0511$ \\
& ScoreMD
& $2.9605$
& $0.1749$
& $0.0093$ \\
& STEC
& $0.1268$
& $\mathbf{0.0036}$
& $\mathbf{0.0028}$ \\

\midrule

BBA
& EBDM
& $\mathbf{0.0650}$
& $0.0727$
& $0.0731$ \\
& ScoreMD
& $2.4999$
& $0.0664$
& $\mathbf{0.0133}$ \\
& STEC
& $0.1503$
& $\mathbf{0.0041}$
& $0.0256$ \\

\midrule

Villin
& EBDM
& $\mathbf{0.6066}$
& $1.0466$
& $0.9893$ \\
& ScoreMD
& $4.1621$
& $0.0923$
& $0.1777$ \\
& STEC
& $1.0156$
& $\mathbf{0.0186}$
& $\mathbf{0.0216}$ \\

\bottomrule
\end{tabular}
\end{table}

\subsection{Comparison with Two for One}
\label{appx:si_two_for_one}

We further compare \ourmethod with standard EBDM models whose
force fields are evaluated at a finite diffusion time,
following Two for One~\citep{arts2023}.
The EBDM models are trained using DSM alone, with the protein
architectures and training checkpoints described
in~\appxref{si_model_training}.
We use $t_{\mathrm{eval}}=0.005$ for all four proteins.
This finite-time force-evaluation setting is denoted
Two for One throughout this comparison.
STEC forces are evaluated at $t_{\mathrm{eval}}=10^{-5}$,
following~\appxref{si_sampling}.
The comparison therefore uses the respective force-evaluation
settings of the two methods.

\begin{figure}[t]
    \centering
    \includegraphics[width=\textwidth]{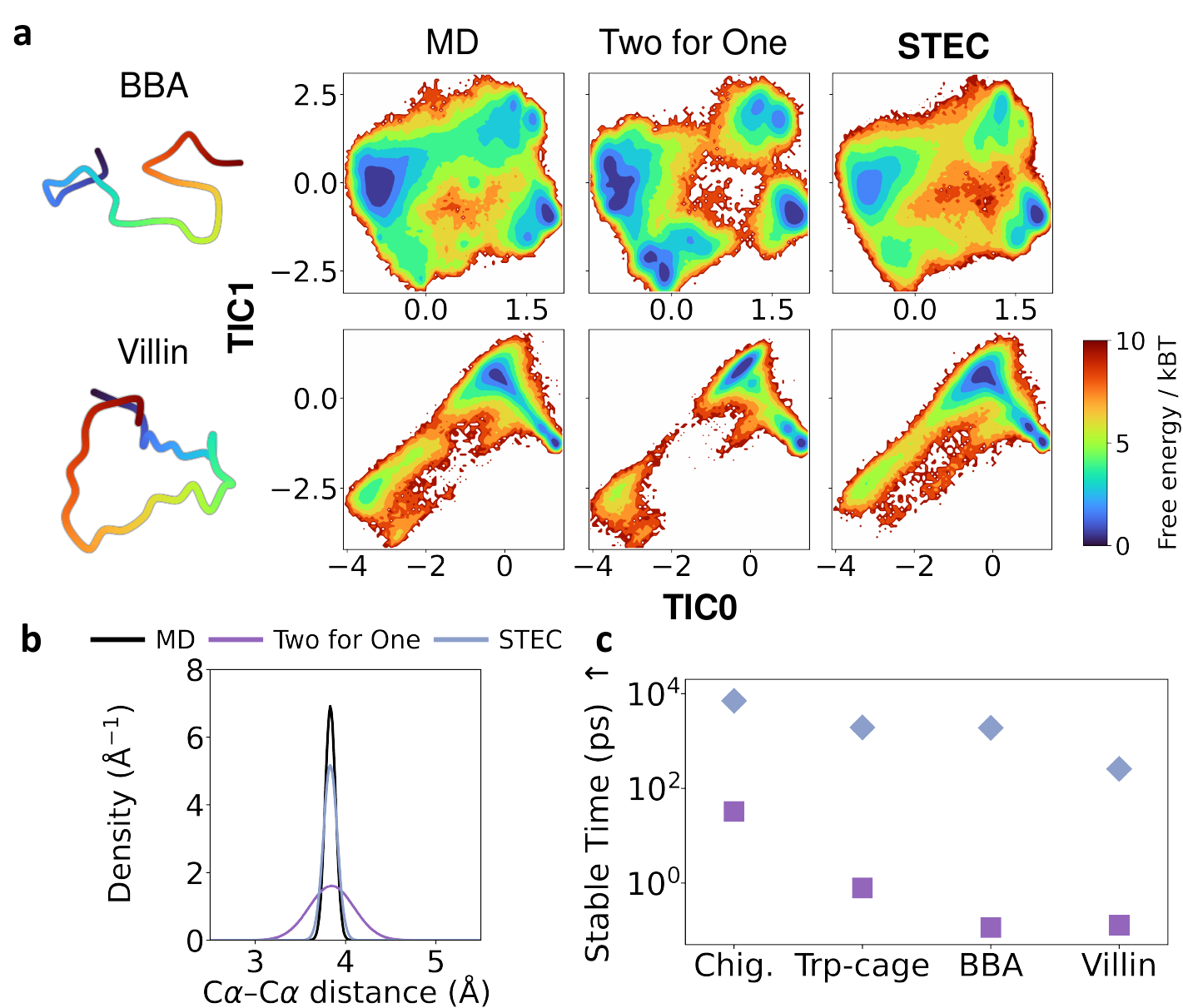}
    \caption{
    \textbf{STEC improves structural fidelity and simulation stability relative to Two for One.}
    Two for One denotes standard EBDM evaluated at a finite diffusion time.
    \textbf{(a)}
    Free-energy landscapes projected onto the first two TICs
    for BBA and Villin, comparing the MD reference,
    Two for One, and STEC.
    STEC more closely recovers the reference conformational
    basins and their relative populations.
    \textbf{(b)}
    Adjacent $C_\alpha$--$C_\alpha$ distance distributions for BBA.
    STEC more faithfully reproduces the narrow reference
    distribution, whereas Two for One produces substantially
    broader bond-length fluctuations.
    \textbf{(c)}
    Mean stable simulation times for Chignolin, Trp-cage,
    BBA, and Villin.
    STEC substantially delays the onset of bond-length
    instability across all four proteins.
    }
    \label{fig:si_two_for_one}
\end{figure}

The TICA-projected landscapes show that \ourmethod more closely
recovers the reference conformational basins of BBA and Villin,
whereas Two for One produces distorted conformational populations
(\figref{si_two_for_one}a).
The adjacent $C_\alpha$ distance distribution provides a
complementary assessment of local backbone geometry
(\figref{si_two_for_one}b).
For BBA, Two for One retains a mean bond length close to the
atomistic reference but produces a substantially broader
distribution.
The bond-length standard deviation is approximately
$0.249$~\AA, compared with $0.058$~\AA\ for the reference
and $0.078$~\AA\ for STEC.
Thus, agreement in the mean bond length does not ensure
accurate recovery of its fluctuations.

We quantify simulation stability using the bonded-distance
criterion defined in~\appxref{si_evaluation}
(\figref{si_two_for_one}c).
Here, reference distances are the bond-specific medians
in the atomistic ensemble, and non-finite distances also
trigger instability.
Within each protein, both methods share the same observation
window.
Trajectories that remain stable throughout this window
contribute the window length when calculating the mean
stable simulation time.
Across the four proteins, STEC yields mean stable simulation
times approximately $2.2\times10^2$--$1.6\times10^4$ times
those of Two for One.
For BBA, the mean stable simulation time increases
from approximately $0.115$~ps for Two for One
to $1.89\times10^3$~ps for STEC.

These observations illustrate the limitation of finite-time
force extraction discussed in the Introduction.
At positive diffusion time, the EBDM energy is learned for
a noise-perturbed marginal rather than the clean target
equilibrium distribution.
The extracted force field therefore departs from the clean
zero-time potential, which can affect both conformational
populations and local structural observables.
Under the settings tested here, STEC improves structural
fidelity and simulation stability while retaining force
evaluation near the clean-data limit.

\subsection{Transition-Probability Structure Across Fast-Folding Proteins}
\label{appx:si_transition_matrices}

Complementing the structural comparisons
in~\appxref{si_structural_fidelity},
we examine the complete transition-probability matrices
for the four fast-folding proteins.
This analysis extends the BBA comparison
in \figref{trajectory_transition}b,
using the shared state partitions and transition-counting
protocols described in~\appxref{si_evaluation}.

\begin{figure}[t]
\centering
\includegraphics[width=\textwidth]{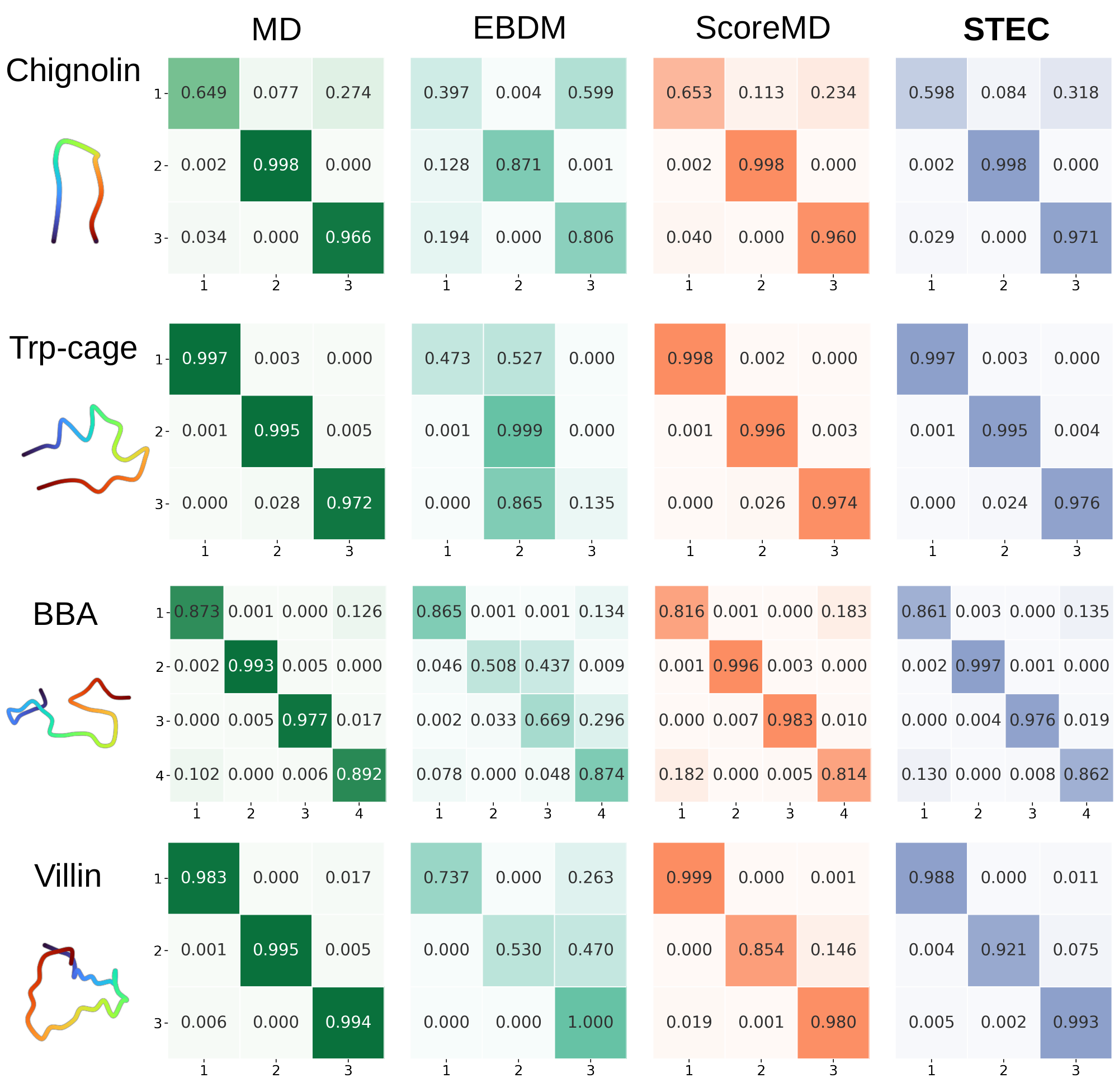}
\caption{
\textbf{STEC more closely reproduces reference
transition-probability patterns across the fast-folding proteins.}
Row-normalized transition matrices for Chignolin,
Trp-cage, BBA, and Villin, comparing the atomistic MD
reference with CGMD trajectories generated by
standard EBDM, ScoreMD, and STEC.
Chignolin, Trp-cage, and Villin use 3 states,
while BBA uses 4 states.
The same state partition is used for all methods
within each protein.
Rows denote starting states and columns denote
destination states; numerical entries give
the corresponding transition probabilities.
The complete matrices, including diagonal entries,
are used for the Hellinger-distance comparison
in \figref{trajectory_transition}c.
}
\label{fig:si_transition_matrices}
\end{figure}

Standard EBDM exhibits substantial deviations from the
reference matrices, affecting both diagonal and
off-diagonal probabilities
(\figref{si_transition_matrices}).
ScoreMD recovers much of the reference structure,
although discrepancies remain for selected transitions.
STEC more closely reproduces the dominant diagonal
entries and principal off-diagonal probabilities.
These comparisons reveal the individual transitions
underlying the Hellinger-distance results
in \figref{trajectory_transition}c.

\subsection{Additional Transferability Results on Held-Out Peptides}
\label{appx:si_peptide_generalization}

We provide conformational landscapes for the complete
evaluation subsets underlying \figref{peptide_generalization}:
15 held-out dipeptides and 20 held-out octapeptides,
as specified in~\appxref{si_datasets}.
Within each benchmark, all test peptides are simulated
using the same transferable STEC model,
following the Langevin CGMD protocols
in~\appxref{si_sampling}.

\begin{figure}[t]
\centering
\includegraphics[width=\textwidth]{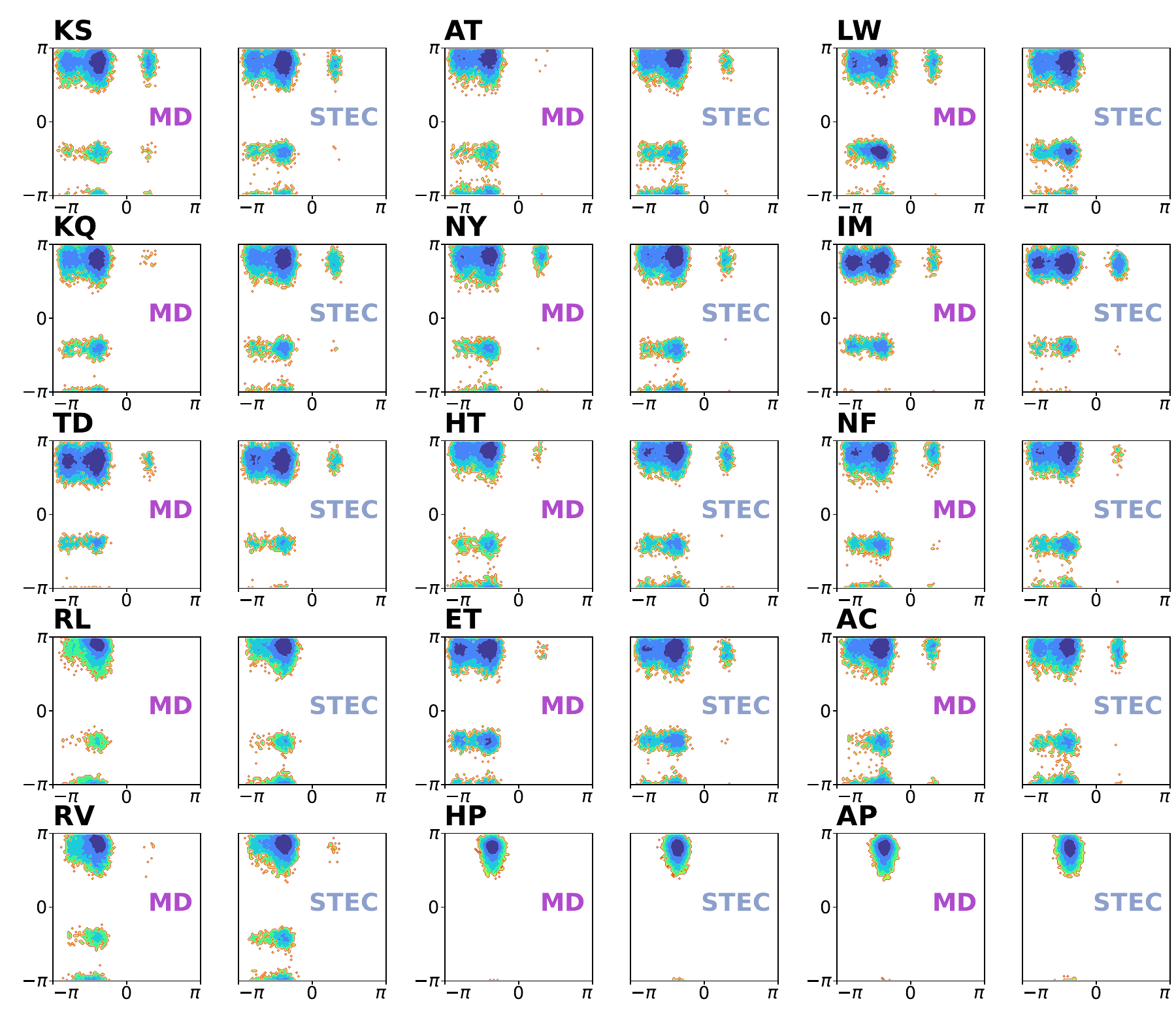}
\caption{
\textbf{STEC reproduces sequence-dependent backbone
conformational distributions in held-out dipeptides.}
Ramachandran landscapes for the 15 held-out dipeptides
evaluated in \figref{peptide_generalization}a.
Each pair compares the mapped atomistic MD reference
(left) with STEC-driven CGMD (right).
Horizontal and vertical axes represent the backbone
dihedral angles $\phi$ and $\psi$, respectively.
All STEC ensembles are generated using the same model.
}
\label{fig:si_dipeptide_generalization}
\end{figure}

For dipeptides, STEC closely reproduces the dominant
reference basins across the evaluated peptide identities
(\figref{si_dipeptide_generalization}).
The landscapes retain sequence-dependent differences
in backbone conformational preferences,
including the concentrated distributions of HP and AP.
These comparisons extend the representative KS landscape
in \figref{peptide_generalization}a
to the full 15-peptide evaluation subset.

\begin{figure}[t]
\centering
\includegraphics[width=\textwidth]{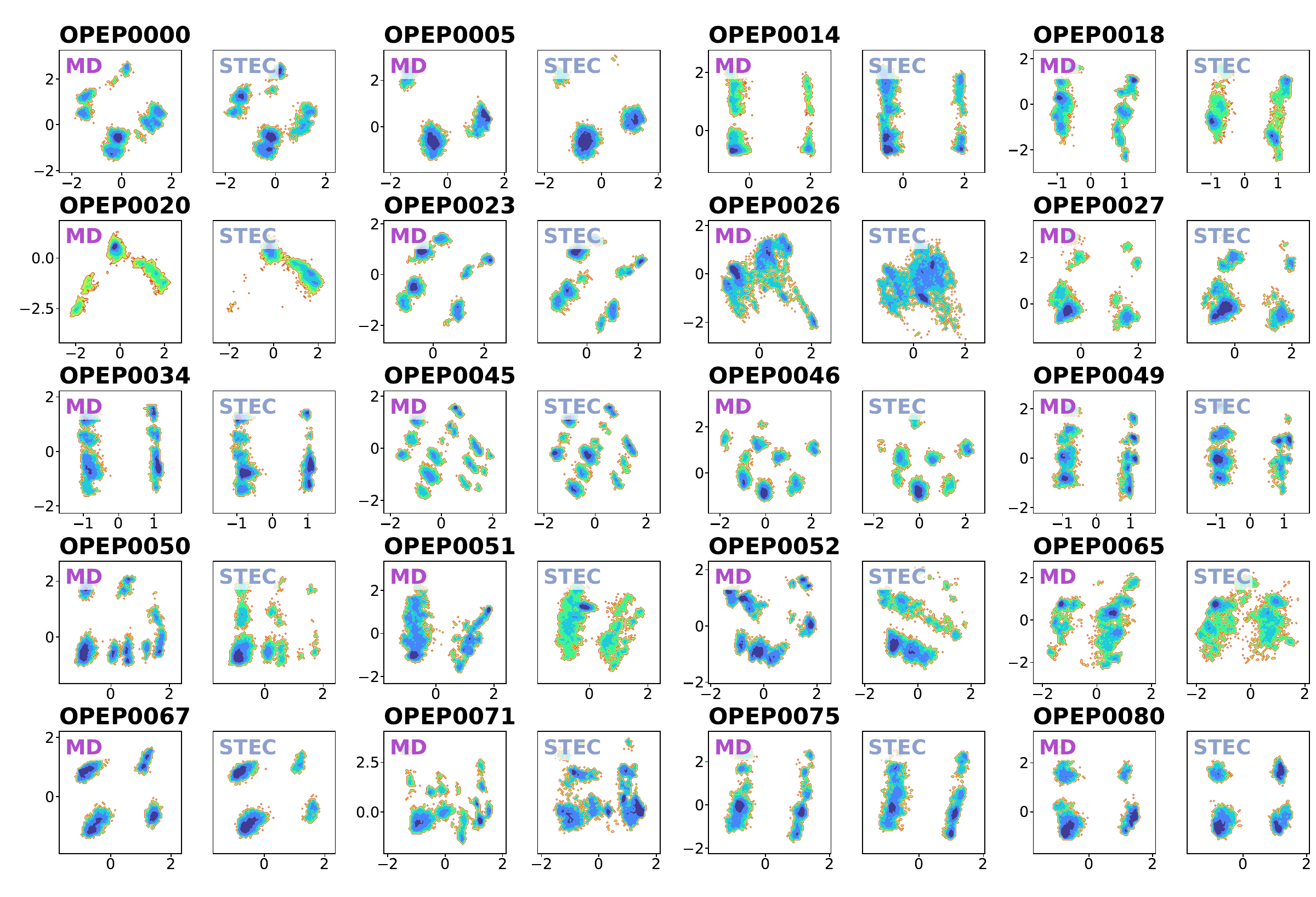}
\caption{
\textbf{STEC captures major reference conformational
basins across held-out octapeptides.}
Conformational landscapes for the 20 held-out octapeptides
evaluated in \figref{peptide_generalization}b.
Each pair compares the mapped atomistic MD reference
(left) with STEC-driven CGMD (right).
Horizontal and vertical axes show the first and second
TICs, respectively.
For each peptide, both ensembles use the same
reference-fitted TICA transformation,
as described in~\appxref{si_evaluation}.
All STEC ensembles are generated using the same model.
Systems follow the order
in Table~\ref{tab:si_opep_sampling_cost}.
}
\label{fig:si_octapeptide_generalization}
\end{figure}

For octapeptides, STEC captures major reference basins
across a range of conformational landscapes,
including the separated basins of OPEP0000 and OPEP0005
(\figref{si_octapeptide_generalization}).
Agreement varies across peptide identities,
with residual differences in some basin shapes
and relative populations.
These comparisons extend the representative OPEP0014
and OPEP0027 landscapes in \figref{peptide_generalization}b
and complement the aggregate mMAE results.

\subsection{Computational Efficiency and Resource Usage}
\label{appx:si_computational_efficiency}

\paragraph{Training time and memory.}
We quantify the resource costs underlying the comparisons
in \figref{efficiency}a,b under controlled single-GPU conditions.
Within each benchmark, EBDM, ScoreMD, and STEC use the same
model architecture, profiling data, and batch size.
Batch sizes are 128 for the fast-folding proteins and
dipeptides, and 64 for octapeptides, without gradient accumulation.
The architectures and full model-training settings are given
in~\appxref{si_model_training}
and Table~\ref{tab:si_training_summary}.

All benchmarks are profiled on a single NVIDIA A100 GPU.
Each method is run in three separate processes for 10 epochs,
with the first 2 discarded as warm-up.
For each process, mean time per optimizer update is calculated
from the synchronized training time over the remaining 8 epochs.
Runtime is reported as the mean and sample standard deviation
of these three process-averaged times.
Device memory is sampled every 0.2~s.
The sampled peak from each process is corrected
for a 4~MiB idle baseline, then averaged across processes.

STEC incurs substantially lower runtime and memory costs
than ScoreMD across the profiling benchmarks
(Table~\ref{tab:si_training_efficiency}).
Relative to standard EBDM, STEC requires
$1.46$--$1.53\times$ the per-update runtime and
$1.19$--$1.34\times$ the sampled peak device memory.
ScoreMD requires $5.13$--$6.26\times$ the runtime and
$3.44$--$5.71\times$ the memory.
This larger overhead is consistent with the additional
energy and spatial-score evaluations required
by its finite-difference FP objective.

Relative costs in Table~\ref{tab:si_training_efficiency}
are normalized to standard EBDM.
For \figref{efficiency}a, protein runtimes are divided
by the number of CG sites.
The octapeptide training costs in \figref{efficiency}b
are instead normalized to STEC.

\begin{table}[!t]
\centering
\caption{
\textbf{STEC requires substantially less training time
and device memory than ScoreMD.}
All methods are profiled on a single NVIDIA A100 GPU
using matched architectures, profiling data, and batch sizes
within each benchmark.
Batch sizes are 128 for the fast-folding proteins
and dipeptides, and 64 for octapeptides.
Runtime values are mean $\pm$ sample standard deviation
across three separate profiling processes.
Memory values are the mean sampled peak device-memory usage,
after subtracting a 4~MiB idle baseline.
Parentheses report costs relative to standard EBDM.
}
\label{tab:si_training_efficiency}
\small
\setlength{\tabcolsep}{8pt}
\renewcommand{\arraystretch}{1.12}

\begin{tabular}{lccc}
\toprule
System & EBDM & STEC & ScoreMD \\
\midrule
\multicolumn{4}{l}{\textit{Time per optimizer update (ms)}} \\
\addlinespace[2pt]

Chignolin
& $4.92 \pm 0.03$
& $7.35 \pm 0.02$ ($1.49\times$)
& $26.11 \pm 0.13$ ($5.31\times$) \\

Trp-cage
& $11.67 \pm 0.03$
& $17.41 \pm 0.11$ ($1.49\times$)
& $59.83 \pm 0.31$ ($5.13\times$) \\

BBA
& $22.81 \pm 0.03$
& $34.80 \pm 0.47$ ($1.53\times$)
& $119.62 \pm 0.09$ ($5.24\times$) \\

Villin
& $28.04 \pm 0.05$
& $40.91 \pm 0.07$ ($1.46\times$)
& $175.47 \pm 1.24$ ($6.26\times$) \\

Dipeptides
& $4.95 \pm 0.06$
& $7.33 \pm 0.04$ ($1.48\times$)
& $26.31 \pm 0.07$ ($5.32\times$) \\

Octapeptides
& $29.39 \pm 0.03$
& $44.22 \pm 0.16$ ($1.50\times$)
& $178.65 \pm 0.73$ ($6.08\times$) \\

\midrule
\multicolumn{4}{l}{\textit{Sampled peak device memory (GiB)}} \\
\addlinespace[2pt]

Chignolin
& 1.11
& 1.32 ($1.19\times$)
& 3.82 ($3.44\times$) \\

Trp-cage
& 2.30
& 2.86 ($1.24\times$)
& 11.03 ($4.80\times$) \\

BBA
& 4.01
& 5.05 ($1.26\times$)
& 19.65 ($4.90\times$) \\

Villin
& 5.25
& 7.04 ($1.34\times$)
& 29.96 ($5.71\times$) \\

Dipeptides
& 1.11
& 1.33 ($1.20\times$)
& 3.83 ($3.45\times$) \\

Octapeptides
& 5.39
& 6.90 ($1.28\times$)
& 28.72 ($5.33\times$) \\

\bottomrule
\end{tabular}
\end{table}

\paragraph{Conformational-coverage cost for octapeptides.}
Beyond training overhead, we assess the post-training
production cost of exploring reference conformational
landscapes, as reported in \figref{peptide_generalization}b.
We use the fixed 20-peptide evaluation subset defined
in~\appxref{si_datasets}, with both atomistic MD
and STEC evaluated on a single NVIDIA A100 GPU.
The comparison measures the cost of constructing
a fixed atomistic reference ensemble against the STEC cost
of reaching 80\% coverage of its low-energy region.

For each octapeptide, the atomistic reference comprises
100 independent 10-ns trajectories,
giving $1000$~ns of aggregate simulation time.
Atomistic simulations use OpenMM with a 4-fs integration
timestep, propagating one trajectory per GPU at a time.
Let $v^{\mathrm{AT}}$ denote the measured single-trajectory
throughput in ns/day.
The corresponding production cost is
\begin{equation}
C^{\mathrm{AT}}
=
\frac{1000}{v^{\mathrm{AT}}}\times 24,
\label{eq:si_opep_at_cost}
\end{equation}
reported in GPU hours.

For STEC, 20 Langevin CGMD trajectories are propagated
concurrently as a single PyTorch batch,
using the simulation settings in~\appxref{si_sampling}.
Initial configurations are drawn from the atomistic
reference ensemble.
We evaluate trajectory prefixes of
20, 40, 100, 200, 400, and 2000~ps.
At each prefix length, configurations from all trajectories
are pooled and uniformly subsampled to 10,000 frames.
The reference ensemble is independently subsampled
to the same number of frames.

Reference and STEC configurations are projected using
the reference-fitted TICA transformation described
in~\appxref{si_evaluation} and discretized
on a common $50\times50$ grid.
Let $p_{\mathrm{ref},b}$ and $p_{\mathrm{model},b}$
denote the normalized reference and STEC probabilities
in grid cell $b$.
The reference free-energy surface is
\begin{equation}
F_{\mathrm{ref},b}
=
-k_{\mathrm B}T
\log\left(p_{\mathrm{ref},b}+10^{-12}\right)+C,
\label{eq:si_opep_reference_fes}
\end{equation}
where $C$ sets its minimum to zero.
The reference low-energy region is
\begin{equation}
\mathcal L
=
\left\{
b:
F_{\mathrm{ref},b}
\le 4~\mathrm{kcal\,mol^{-1}}
\right\},
\label{eq:si_opep_low_energy_region}
\end{equation}
and STEC coverage is
\begin{equation}
\mathrm{Coverage}
=
\frac{
\left|
\left\{
b\in\mathcal L:
p_{\mathrm{model},b}>0
\right\}
\right|
}{
|\mathcal L|
}.
\label{eq:si_opep_coverage}
\end{equation}
Each target cell contributes equally.
Coverage therefore measures the fraction of the
reference low-energy region visited,
rather than its equilibrium probability mass.

For each system, $t^{\mathrm{CG}}_{\mathrm{cov}}$
is the estimated per-trajectory duration required
to reach 80\% coverage.
We estimate it by interpolation between the first
pair of adjacent evaluated prefixes that bracket
the threshold.
If the threshold is already reached at 20~ps,
this earliest prefix provides an upper bound
on the required duration.

The aggregate CGMD duration is
$20t^{\mathrm{CG}}_{\mathrm{cov}}$.
Let $v^{\mathrm{CG}}_{\mathrm{batch}}$ denote
the measured aggregate throughput of the 20-trajectory
batch in ns/day.
The STEC production cost is
\begin{equation}
C^{\mathrm{CG}}
=
\frac{
20t^{\mathrm{CG}}_{\mathrm{cov}}/1000
}{
v^{\mathrm{CG}}_{\mathrm{batch}}
}
\times 24,
\label{eq:si_opep_cg_cost}
\end{equation}
where $t^{\mathrm{CG}}_{\mathrm{cov}}$ is expressed in ps
and the factor $1000$ converts ps to ns.
The relative production cost is summarized by
\begin{equation}
S
=
\frac{C^{\mathrm{AT}}}{C^{\mathrm{CG}}}.
\label{eq:si_opep_cost_ratio}
\end{equation}

This ratio combines the sampling duration required
for the specified task with the measured sampling throughput.
The throughput comparison includes 20 concurrent CGMD
trajectories versus one atomistic trajectory under
the implementations used here.
Accordingly, $S$ reflects both sampling requirements
and implementation-dependent trajectory batching.

Across the 20 held-out octapeptides,
atomistic reference construction costs
28.4--38.4 GPU hours,
while STEC requires 0.0097--0.715 GPU hours
to reach the coverage target
(Table~\ref{tab:si_opep_sampling_cost}).
The reported cost ratios range from
$53.7\times$ to $3907.7\times$,
with a median of $1749.5\times$.
The table row order defines the test-system indices
in \figref{peptide_generalization}b.

\begin{table*}[t]
\centering
\caption{
\textbf{STEC reaches 80\% reference conformational coverage
at substantially lower production cost than atomistic
reference construction.}
Measured throughputs and derived production costs
for the 20 held-out octapeptides on a single NVIDIA A100 GPU.
Each atomistic reference contains $1000$~ns
of aggregate simulation time.
The CGMD duration is the estimated aggregate time
across 20 concurrent STEC trajectories required
to reach 80\% coverage of the reference low-energy TICA region.
$v^{\mathrm{AT}}$ is the single-trajectory atomistic
throughput, and $v^{\mathrm{CG}}_{\mathrm{batch}}$
is the aggregate throughput of the STEC batch.
$C^{\mathrm{AT}}$ and $C^{\mathrm{CG}}$ are the corresponding
production costs, with $S=C^{\mathrm{AT}}/C^{\mathrm{CG}}$.
Asterisks mark systems that reach the threshold
at the earliest evaluated prefix of 20~ps.
Their STEC costs are upper bounds,
and their cost ratios are lower bounds.
Rows follow the system indices
in \figref{peptide_generalization}b.
}
\label{tab:si_opep_sampling_cost}
\small
\setlength{\tabcolsep}{4.5pt}
\renewcommand{\arraystretch}{1.08}

\begin{tabular}{lrrrrrr}
\toprule
System
& \multicolumn{1}{c}{CGMD time}
& $v^{\mathrm{AT}}$
& $v^{\mathrm{CG}}_{\mathrm{batch}}$
& $C^{\mathrm{AT}}$
& $C^{\mathrm{CG}}$
& $S$ \\
& \multicolumn{1}{c}{to 80\% coverage}
& (ns/day)
& (ns/day)
& (GPU h)
& (GPU h)
& \\
& (aggregate ns)
&
&
&
&
& \\
\midrule

OPEP0000 & 1.748  & 646.62 & 988.09 & 37.116 & 0.0425 & $874.1\times$ \\
OPEP0005 & 0.737  & 715.41 & 989.66 & 33.547 & 0.0179 & $1877.8\times$ \\
OPEP0014 & 0.619  & 844.95 & 989.78 & 28.404 & 0.0150 & $1892.3\times$ \\
OPEP0018$^{*}$ & 0.400  & 633.45 & 990.13 & 37.888 & 0.0097 & $3907.7\times$ \\
OPEP0020$^{*}$ & 0.400  & 780.09 & 990.12 & 30.766 & 0.0097 & $3173.1\times$ \\
OPEP0023 & 14.749 & 707.46 & 990.21 & 33.924 & 0.3575 & $94.9\times$ \\
OPEP0026$^{*}$ & 0.400  & 716.56 & 990.13 & 33.493 & 0.0097 & $3454.5\times$ \\
OPEP0027 & 15.649 & 716.61 & 990.35 & 33.491 & 0.3792 & $88.3\times$ \\
OPEP0034 & 0.499  & 715.17 & 990.52 & 33.558 & 0.0121 & $2776.6\times$ \\
OPEP0045 & 29.433 & 625.70 & 988.06 & 38.357 & 0.7149 & $53.7\times$ \\
OPEP0046 & 3.956  & 715.53 & 988.51 & 33.542 & 0.0961 & $349.3\times$ \\
OPEP0049 & 1.605  & 757.59 & 990.13 & 31.679 & 0.0389 & $814.1\times$ \\
OPEP0050$^{*}$ & 0.400  & 715.04 & 989.65 & 33.564 & 0.0097 & $3460.1\times$ \\
OPEP0051 & 0.871  & 700.52 & 989.67 & 34.260 & 0.0211 & $1621.1\times$ \\
OPEP0052 & 0.749  & 643.73 & 989.67 & 37.283 & 0.0182 & $2051.9\times$ \\
OPEP0065$^{*}$ & 0.400  & 766.64 & 990.22 & 31.306 & 0.0097 & $3229.1\times$ \\
OPEP0067 & 0.495  & 658.94 & 989.67 & 36.422 & 0.0120 & $3037.0\times$ \\
OPEP0071 & 1.680  & 713.89 & 990.00 & 33.619 & 0.0407 & $825.5\times$ \\
OPEP0075 & 6.329  & 703.64 & 990.22 & 34.108 & 0.1534 & $222.4\times$ \\
OPEP0080 & 1.492  & 715.45 & 990.10 & 33.545 & 0.0362 & $927.7\times$ \\

\bottomrule
\end{tabular}
\end{table*}

\end{document}